\documentclass[10pt,letterpaper]{article}
\usepackage[top=0.85in,left=1.85in,footskip=1in,marginparwidth=1.75in]{geometry}
\usepackage{amsmath}
\usepackage[version=4]{mhchem}
\usepackage{authblk}
\usepackage{array}
\usepackage[export]{adjustbox}
\usepackage{tabularx}
\usepackage{rotating}
\usepackage{multirow}
\usepackage{float}

\usepackage[table]{xcolor}
\usepackage{booktabs}
\usepackage[super,comma,sort&compress]{natbib}
\usepackage{changepage}
\usepackage{seqsplit}
\usepackage{titlesec}

\titleformat{\section}{\raggedright\large\bfseries}{\thesection}{1em}{}

\usepackage[T2A,T1]{fontenc}
\usepackage[ukrainian,english]{babel} 

\usepackage{nameref,hyperref}

\usepackage[aboveskip=1pt, labelfont=bf, labelsep=period, singlelinecheck=off, margin=1cm]{caption}

\makeatletter
\renewcommand{\@biblabel}[1]{\quad#1.}
\makeatother

\usepackage{lastpage,fancyhdr,graphicx}
\fancyheadoffset[L]{0.75in}
\fancyfootoffset[L]{0.75in}
\fancyheadoffset[R]{0.25in}
\fancyfootoffset[R]{0.25in}

\usepackage{color}

\definecolor{Gray}{gray}{.25}

\usepackage{graphicx}

\usepackage{sidecap}

\usepackage{wrapfig}
\usepackage[pscoord]{eso-pic}
\usepackage[fulladjust]{marginnote}
\reversemarginpar

\begin{document}


\vspace*{0.35in}

\begin{flushleft}

{\Large{\textbf{Broadening the temperature range of blue phases using \textit{azo} compounds of various molecular geometries assembled from modular ``LEGO'' molecular units}}}
\newline
\\
Igor A.~Gvozdovskyy\textsuperscript{1,}\footnote[1]{igvozd@gmail.com},
Vitalii O.~Chornous\textsuperscript{2,}\footnote[2]{chornous.vitalij@bmsu.edu.ua},
Halyna V.~Bogatyryova\textsuperscript{1},
Oleksandr M.~Samoilov\textsuperscript{3},
Longin N.~Lisetski\textsuperscript{3,}\footnote[3]{lcsciencefox@gmail.com},
Serhiy V.~Ryabukhin\textsuperscript{4,5},
Yurii V.~Dmytriv\textsuperscript{5,6},
Mykhaylo V.~Vovk\textsuperscript{7}
\\
\bigskip
\textit{\textsuperscript{{1}} Department of Optical Quantum Electronics Institute of Physics of the Natl. Acad. Sci. of Ukraine Kyiv, 03028, Ukraine} 
\\
\textit{\textsuperscript{{2}} Department of Medical and Pharmaceutical Chemistry Bukovinian State Medical University, Chernivtsi, 58002, Ukraine} 
\\
\textit{\textsuperscript{{3}} Department of Nanostructured Materials Institute for Scintillation Materials of the Natl. Acad. Sci. of Ukraine Kharkiv, 61072, Ukraine} 
\\
\textit{\textsuperscript{{4}} Department of Supramolecular Chemistry, Institute of High Technology Taras Shevchenko National University of Kyiv, Kyiv, 03022, Ukraine}
\\
\textit{\textsuperscript{{5}} Enamine Ltd, Kyiv, 02094, Ukraine}
\\
\textit{\textsuperscript{{6}} Department of Chemical Technology, National Technical University of Ukraine ”Igor Sikorsky Kyiv Polytechnic Institute”, Kyiv, 03056, Ukraine}
\\
\textit{\textsuperscript{{7}} Department of Mechanisms of Organic Reactions, Institute of Organic Chemistry of the Natl. Acad. Sci. of Ukraine, Kyiv, 02660, Ukraine}
\\
\bigskip

\end{flushleft}


\section*{Abstract}
The temperature range of the blue phases (BPs) formed in highly chiral mixtures based on cholesteryl oleyl carbonate (COC) and the nematic liquid crystal E7 was studied in the presence of various chemical structures. The \textit{azo} compounds used were of both chiral and achiral nature, and their molecular geometry was modified by substitution of modular "LEGO" molecular units of varying alkyl chain lengths and types of bridging groups, which could substantially affect the mesomorphic properties of the matrix mixture. It was shown that in many cases these dopants effectively broadened the BP temperature range. This effect depends on both the variation in the molecular geometry of the \textit{azo} compounds and the increase in the \textit{cis}-isomer concentration under UV irradiation. The presence of the \textit{cis}-isomers formed have a stronger impact on broadening the BP temperature range than the initial \textit{trans}-isomers. These results demonstrate that the temperature range of BPs can be precisely controlled via a combination of molecular engineering and \textit{trans}---\textit{cis} photo-isomerization.


\section{Introduction}
The chirality of molecules is a main key to create the helicoidal supramolecular structure in liquid crystalline (LC) media characterized by the Bragg diffraction of light (BRL) in the visible spectral range\cite{Kitzerow2001}. The phenomenon of the BRL, which is a characteristic feature of cholesteric liquid crystals (CLC), including blue phases (BPs) and oblique helicoidal structures (Ch\textsubscript{OH}) in alternating applied field\cite{Lavrentovich2022}, makes these LC systems attractive for different applications.
The chiral compounds can form an intrinsic cholesteric phase (Ch or $\text{N}^*$), like, \textit{e.g.}, cholesterol esters, or can induce the $\text{N}^*$ phase upon their dissolution in nematics (N), serving in this case as chiral dopants (ChDs).\cite{ChilayaLisetski1986} Depending on the nature of chiral molecules, the right-handed or left-handed helix can be formed in the $\text{N}^*$ phase. The well-known elegant classical experiments carried out by O. Lehmann\cite{Lehmann1900} and other authors\cite{Tabe2003,Kausar2011,GvozdTeren2002,GvozdLis2007,GvozdLis2008} visually demonstrate the handedness of the helix formed in the $\text{N}^*$ phase. The emerging pitch \textit{P} of the cholesteric helix depends on both the helical twisting power (HTP, $\beta$) and the concentration \textit{C} of ChD in the nematic matrix as follows:\cite{Kitzerow2001,ChilayaLisetski1986} $\text{HTP} = \frac{1}{P \times C}$. It is known that helical pitch \textit{P} of CLC systems can be changed by various external factors, namely temperature,\cite{Ennulat1974,Zhang2021} electric and magnetic fields,\cite{Helfrich1971,Hurault1973,Niggemann1989,Huh2007} irradiation\cite{Feringa2003,Li2016,Bunning2007,ChornousGvozd2022} and other external stimuli\cite{Simoni1987,Han2010,Chen2024}.
For CLCs under conventional measurement conditions, \textit{i.e.}, for a layer of $\sim 5$--$50~\mu\text{m}$ thickness in the planar texture (helix axis normal to the cell boundaries), the wavelength $\lambda_{\text{max}}$ of BRL and the helix pitch \textit{P} are interconnected as follows:\cite{Kitzerow2001,ChilayaLisetski1986}
\begin{equation}
  \lambda_{\text{max}} = P \times \bar{n} \times \cos(\theta)
\end{equation}
where $\bar{n}$ is the average refractive index of the LC medium and $\theta$ is the angle of incidence of light with respect to the normal direction of the quasi-nematic layers (\textit{i.e.}, cholesteric helix axis). In this case the wavelength can be changed continuously within wide range of visible spectrum by changing the external excitations.
\begin{figure*}
\begin{adjustwidth}{-1in}{0in}
 \centering
 \includegraphics[width=\textwidth]{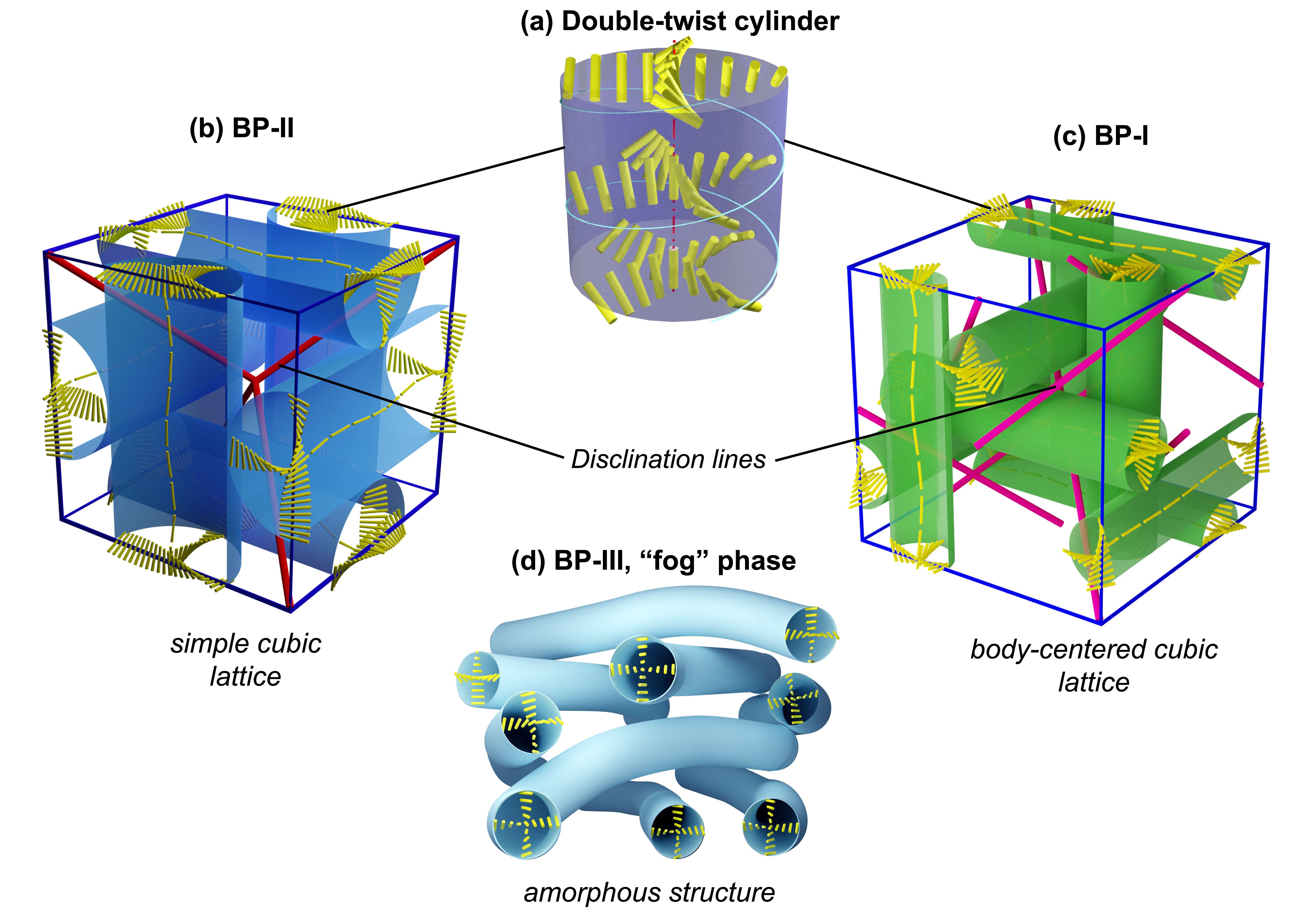}
 \caption{Schematic presentation of blue phases consisting of double-twist cylinders (a). 3D crystal unit cell formed by a network of disclination lines for: (b) simple cubic-–-BP-II and (c) body-centered cubic–--BP-I lattices. (d) An amorphous structure of high temperature BP-III (``fog'' phase).}
 \label{fgr:Figure1}
 \end{adjustwidth}
\end{figure*}
It is known that if the initial cholesteric helix pitch \textit{P} of certain CLCs is less than $\sim 500$~nm\cite{BelyakovDmitrienko1985}, then at temperatures close to the isotropic (Iso) phase there often exists a set of thermodynamically distinct mesophases within narrow temperature ranges (of the orders of several K or less), \textit{i.e.}, so-called blue phases (BPs)\cite{McKinnon1981,MarcusGoodby1982}. The blue phases, as distinct examples of the highly chiral media, are characterized by the presence of double-twist cylinders emerging as local low-energy structures (Fig.~1(a))\cite{Sethna1985,Wright1989,OswaldPieranski2005}, which can be self-assembled into 3D photonic crystals. The 3D crystal unit cell (\textit{i.e.}, an element of the quasi-crystal lattice) is characterized by a network of disclination lines of simple cubic (\textit{i.e.}, BP-II) or body-centered cubic (\textit{i.e.}, BP-I) lattices\cite{OswaldPieranski2005}, as schematically shown in Fig.~1(b) and (c), respectively. Alongside BP-I and BP-II, at higher temperatures closer to isotropic phase (Iso), there can exist another (amorphous) structure of BP (\textit{i.e.}, BP-III or so-called ``fog'' BP) (Fig.~1(d)). Only for BP-II and BP-I the phenomenon of the selective BRL at specific reflected wavelengths can be observed. The reflected wavelength of light, according to Bragg’s law, can be written as follows using Miller indices [\textit{h}, \textit{k}, \textit{l}] (\textit{h} + \textit{k} + \textit{l} is an even number) describing the orientation of lattice planes\cite{OswaldPieranski2005}:

\begin{equation}
\lambda_{h,k,l} = \frac{2 \times \bar{n} \times a}{\sqrt{h^2 + k^2 + l^2}} \cdot \cos(\theta)
\end{equation}

where \textit{a} is the lattice constant of BP, and $\theta$ is, as in the Eq.~(1), the angle of incidence of the light with respect to the normal direction of a Miller plane.

 \begin{figure*}
 \begin{adjustwidth}{-1in}{0in}
 \centering
 \includegraphics[width=\textwidth]{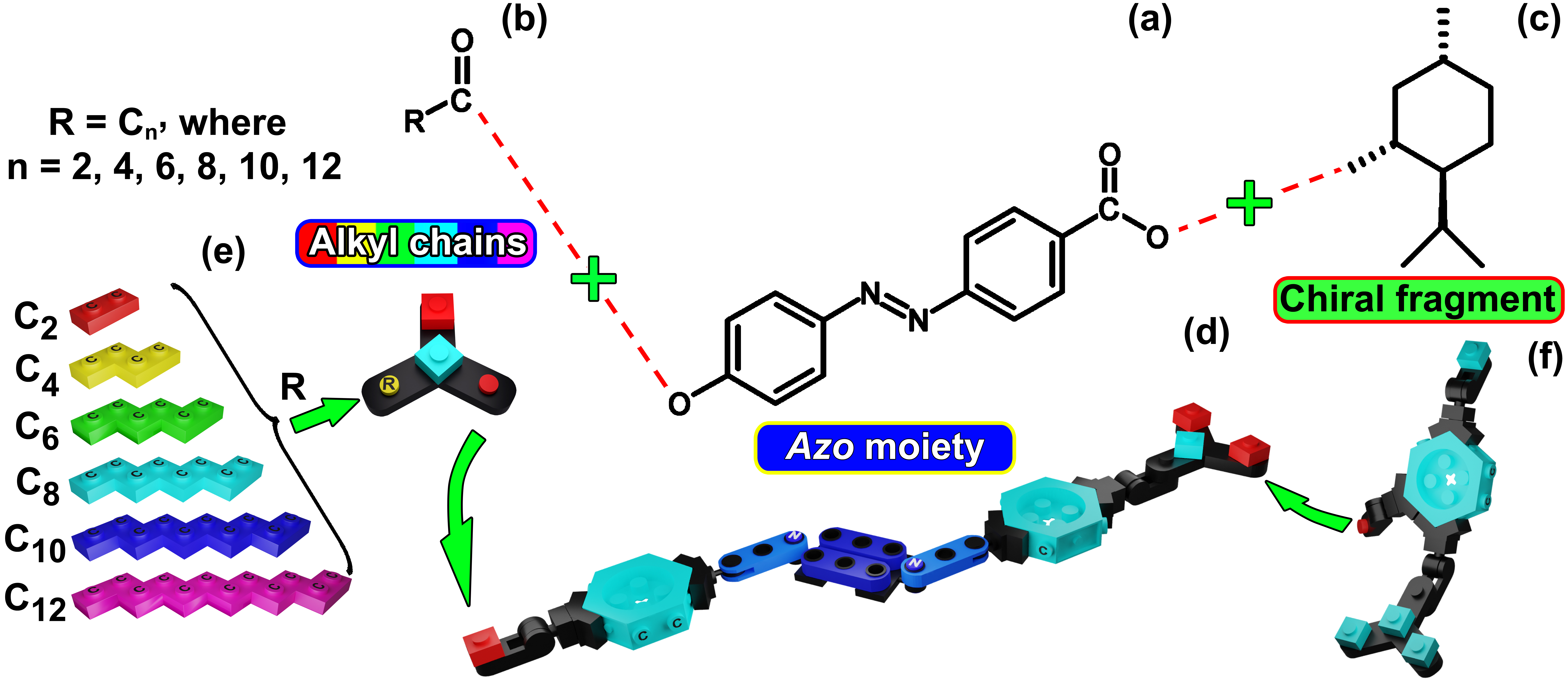}
 \caption{Chemical structure of the chiral \textit{azo} compounds characterized by: (a) \textit{azo} moiety, (b) alkyl chains of various lengths (\textit{i.e.}, different number of carbon atoms) and (c) chiral fragment (\textit{l}-menthol). A schematical build-up of \textit{azo} compounds by means of ``LEGO'' molecular units: (d) \textit{azo} moiety, (e) alkyl chain and (f) chiral fragment.}
 \label{fgr:Figure2}
 \end{adjustwidth}
\end{figure*}

The main advantages of BP are based on their optical and electro-optical properties that can be used for different applications\cite{Yoshizawa2013,Choi2019,Khoo2016,Yang2025,Cho2021}. However, there exist at least two substantial disadvantages. Blue phases typically exist within narrow temperature ranges; furthermore, the random orientation of 3D unit cells in the BP medium leads to the appearance of the polycrystalline (or platelet) texture of BP,\cite{Dierking2003}, which limits their practical application. To achieve an essential broadening of the BP thermal stability (\textit{i.e.}, to increase the temperature range of the BP existence), addition of dopants, \textit{e.g.}, various nanoparticles (NPs)\cite{Muševič2010,Dierking2012,Wang2012,Lavrič2013,Dierking2020,Oton2022} and polymers\cite{ColesPivnenko2005,Kikuchi2002,Kikuchi2021,PabloGonzalez2025,Gvozd2016} to the liquid-crystalline medium was proposed. In the case of BP-II and BP-I, the available network of disclinations serves as a certain trap where nanoparticles or polymers are accumulated thereon, and a significant broadening of the temperature range ($\Delta{T}_{\text{BP}}$) of BP (\textit{i.e.}, blue phases stability) is observed.\cite{Muševič2010,Dierking2012,Wang2012,Lavrič2013,Dierking2020,Oton2022,ColesPivnenko2005,Kikuchi2002,Kikuchi2021,PabloGonzalez2025} In addition, the thermal stability also occurs for BP containing molecules having different shapes, \textit{e.g.}, bimesogenic\cite{WangYang2023}, T-shaped\cite{YoshizawaSato2005}, bent-core\cite{TakezoeJakli2010} and hydrogen-bonded\cite{GuoYang2013,Gvozd2015} molecules. The influence of chirality and parameters such as elastic constant ${\text{K}}_{22}$, molecular length of nematics (\textit{i.e.}, various number of carbon atoms in the alkyl chains including the odd-even effect), and orientation order parameter on the temperature range of BPs was studied both theoretically and experimentally\cite{MillerGlesson1993}. For the practical application of BPs as the medium for various optical devices based on Bragg reflection, the uniform orientation of lattice planes is essential. With this aim, the use of different aligning layers\cite{FukudaŽumer2020,GleesonGoodby2015,Ozaki2018,OtonOton2020} and patterned surfaces\cite{GonzalezPablo2017,ParkPablo2021,NakajimaOzaki2022} was reported to favor the appearance of sharp diffraction lines (\textit{i.e.}, so-called K\"{o}ssel diagrams\cite{Kossel1935,JérómePieranski1989}) of reflected light from uniformly oriented 3D unit cells.
In general, the wavelength of BRL of the BP can be controlled by means of the temperature change that leads to the deformation of the 3D crystal unit cell, but in this case the rate of deformation is relatively slow\cite{OswaldPieranski2005}. However, the BP lattice and its orientation can also be rapidly deformed when a high electric field is applied, resulting in changes of the K\"{o}ssel diagrams, the Kerr effect, electrostriction and electric field-induced phases\cite{Kitzerow2001,OswaldPieranski2005,Yoshizawa2013,Kossel1935,JérómePieranski1989}. In addition, owing to the reversible \textit{trans-cis} photo-isomerization of the constituent molecules, the reversible control of 3D crystal unit cells deformation can also be reached by irradiation of blue phases with light of appropriate wavelengths, with LC systems containing achiral molecules (\textit{e.g.}, nematic LCs\cite{ChilayaCollings2005}, photochromic molecules,\cite{LiuWang2010} acids\cite{MaKim2025}) and the chiral\cite{LuiCheng2012,LinLee2014,LinQLi2013,HeCao2017} \textit{azo} compounds. Therefore, the use of photosensitive compounds (in particular, \textit{azo} compounds\cite{HeCao2017,MenonRoy2025}), possessing reversible \textit{trans--cis} photo-isomerization in a highly chiral liquid crystalline medium that is additionally characterized by BP, sparks interest from a scientific perspective and possible applications as a promising way to study and control their phase transitions and optical features, especially the BRL in the visible range of spectrum.
   
In this work, we will focus our studies on BPs of enhanced smectic phase (EPS), based on cholesteryl oleyl carbonate (COC) and nematic liquid-crystalline mixture E7 in a certain weight ratio\cite{KasianGvozd2025}, doped with both chiral (ChD) and achiral (aChD) \textit{azo} compounds used as photosensitive additives. For the obtained blue phase mixtures (BPMs), the temperature range of the existence of BPs both before and after irradiation of mixtures will be determined to allow comparison demonstrating the influence of \textit{trans--cis} photo-isomerization on the properties of BPs. To obtain photosensitive \textit{azo} compounds, the \textit{azo} coupling of diazonium salts\cite{MenonRoy2025} with the activated aromatic system\cite{ChornousGvozd2022,ChornousGvozd2023} was used in this work as standard synthesis. Analogously to ``LEGO'' elements in children's games (\textit{i.e.}, the concept of reticular synthesis, proposed by O. Yaghi\cite{Yaghi2003}, which involves the assembly of new materials from molecular building blocks), the modification of the base molecule (Fig.~2(a)) was carried out by several ways, \textit{e.g.}, changing of the alkyl chain length (Fig.~2(b)) or adding of chiral fragment (Fig.~2(c)), which, in general, allowed us to obtain various compounds having different properties, from the viewpoints of their features as chiral dopants (Fig.~2) and their influence on general characteristics of BPMs, in particular the BP temperature range ($\Delta{T}_{\text{BP}}$). Special attention will be focused on how the BP temperature range is affected by: (i) the presence of the chiral fragment, (ii) the flexibility provided by means of the various bridges (\textit{e.g.}, \ce{-CH2-}), and (iii) variations in alkyl chain length within the \textit{azo}-based ``LEGO'' molecular units.

In general, we propose the following structure of this article. First, we demonstrate the influence of \textit{cis}-isomers formed under UV irradiation on the broadening of the BP temperature range of the highly chiral mixture (HChM)\cite{KasianGvozd2025}, into which various chiral and achiral \textit{azo} compounds were added. Second subsection considers the influence of a specifically shaped chiral fragment (\textit{e.g.}, \textit{l}-menthol, Figs.~2(c) and (f)) modifying the base \textit{azo} compounds (Figs.~2(a) and (d)) with different lengths of alkyl chain and the same achiral analogs. The third subsection examines the influence of flexibility of both the alkyl chain and chiral fragment linked to \textit{azo}~moiety via the ether (\ce{-C-O-C-}), ester (\ce{-COO-}) and carboxymethylene (\ce{-COO-CH2-}, \textit{i.e.}, \ce{-COO-} and a methylene bridge \ce{-CH2-} linked together) bridges. In the fourth subsection, we focus on the effects of modifying the base \textit{azo} compound (Fig.~2(a) and (d)) by varying the length of alkyl chain (number of carbon atoms \ce{C_n}, Figs.~2(b) and (e)) and the impact of UV irradiation (\textit{i.e.}, increasing of concentration of \textit{cis}-isomer) on the change of the BP temperature range.

\section{Experimental}
\subsection{Materials}
To study the broadening of the BP temperature range, we used the base highly chiral mixture (HChM), consisting of 65~wt\%  cholesteryl oleyl carbonate (COC) and 35~wt\% nematic liquid crystal E7. COC as a chiral dopant (Fig.~3(a)) was obtained from Sigma-Aldrich (St. Louis, MO, USA). COC is characterized by various liquid-crystalline phases, namely, the smectic A phase (SmA) and the intrinsic cholesteric phase ($\text{N}^*$). The temperatures of the phase transitions of COC during both the cooling (Iso $\xrightarrow{36^{\circ}\text{C}}$ $\text{N}^*$ $\xrightarrow{22^{\circ}\text{C}}$ SmA $\xrightarrow{0^{\circ}\text{C}}$ Cr) and heating (Cr $\xrightarrow{22^{\circ}\text{C}}$ $\text{N}^*$ $\xrightarrow{40^{\circ}\text{C}}$ Iso) processes were presented in our earlier paper\cite{LebovkaSoskin2013}. The nematic liquid crystal E7 (a four-component mixture, containing cyanobiphenyl and cyanoterphenyl compounds, Fig.~3(b)), was obtained from Licrystal, Merck (Darmstadt, Germany). As distinct from the recently described cholesteric mixtures based on COC and nematic 5CB\cite{LebovkaSoskin2013}, with similar mixtures containing the nematic E7, which has higher temperature of the nematic-isotropic (N--Iso) transition ($T_{\text{Iso}} = 58^{\circ}\text{C}$\cite{LCbrochure1994}) as compared with 5CB (${T}_\text{Iso}\approx35.5^\circ\text{C}$\cite{LCbrochure1994}), we could observe the appearance of BP at the concentration of COC within 50 to 75~wt\% range.

\begin{figure}[h]
\begin{adjustwidth}{-1in}{0in}
\centering
  \includegraphics[width=\textwidth]{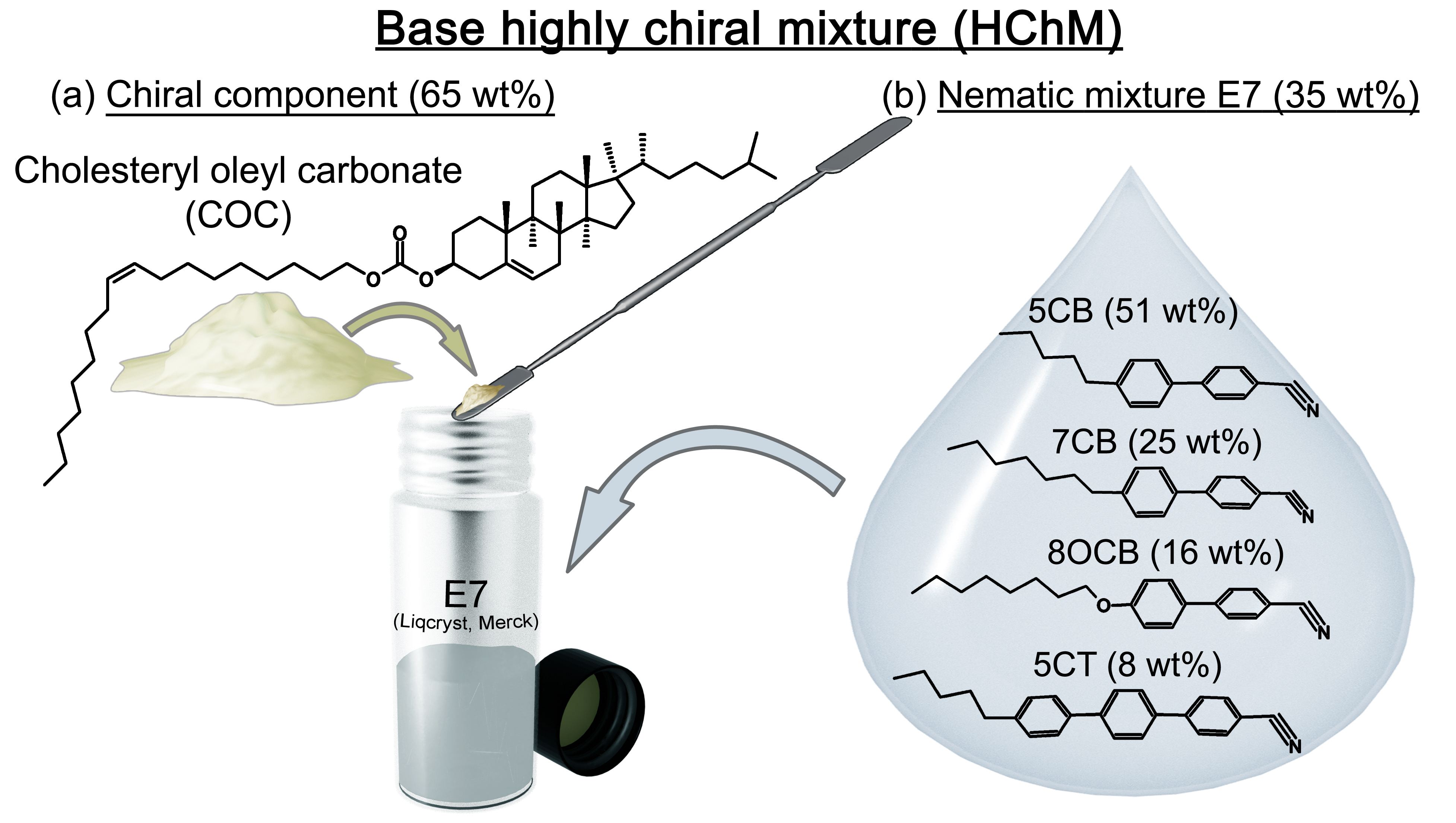}
  \caption{Chemical structures of compounds and their weight ratio in the base highly chiral mixture formed by: (a) 65~wt\% cholesteryl oleyl carbonate (COC) as chiral component and (b) 35~wt\% nematic liquid crystal E7, which is a mixture with 51~wt\% 4-cyano-4'-n-pentyl-biphenyl (5CB), 25~wt\% 4-cyano-4'-n-heptyl-biphenyl (7CB), 16~wt\% 4-cyano-4'-n-oxyoctyl-biphenyl (8OCB) and 8~wt\% 4-pentyl-4''-n-pentyl-p-terphenyl (5CT).}
  \label{fgr:Figure3}
  \end{adjustwidth}
\end{figure}

The achiral compounds used consisted of \textit{azo} moiety and alkyl chains of various lengths (\textit{i.e.}, different numbers of carbon atoms), while chiral \textit{azo} compounds are additionally characterized by the presence of a chiral fragment (\textit{i.e.}, \textit{l}-menthol). \textit{L}-menthol, with a specific rotation $[\alpha]^{20^{\circ}\text{C}}_\text{D} = -50^{\circ}$, was obtained from Shanghai Linsai Trade Co., Ltd. (Shanghai, China). The synthesis of all \textit{azo} compounds used is presented in Section~S1 of the SI\footnote[4]{SI---Supplementary Information}.

To check the concentration dependence of BP temperature range broadening in BPMs consisting of \textit{azo} compounds and the base HChM, we also used the left-handed chiral dopant S-811 (Merck, Darmstadt, Germany), which lacks an \textit{azo} moiety.

To obtain planar alignment of chiral liquid-crystalline mixtures, we used the PI2555 polyimide precursor obtained from HD MicroSystems (Parlin, NJ, USA)\cite{PI2555}.

\subsection{Methods}
To determine the helical twisting power (HTP, $\beta$) of various chiral \textit{azo} compounds, we measured the helical pitch \textit{P} of our mixtures using the Grandjean-Cano method, as described elsewhere in detail\cite{ChornousGvozd2022}. The wedge-like cell was assembled with glass substrates covered by PI2555 film and characterized by thickness of the thin end ${d}_\text{0}$ and thick end \textit{d} of the wedge, with \textit{d} varying from 18 to 30~$\mu$m. The length \textit{P} of cholesteric helix is related to the number ${N}_\text{C}$ of Grandjean-Cano zones indicating the half pitch \textit{P}/2 as follows\cite{ChornousGvozd2022}:

\begin{equation}
P = \frac{2 \times (d - d_0)}{N_C}
\end{equation}

By knowing the dependence of the cholesteric helical pitch \textit{P} on the concentration \textit{C} of ChD, the HTP values of various ChDs in the nematic host E7 were determined from the linear part of the 1/\textit{P}(\textit{C}) plot, which should always pass through the origin of coordinates\cite{ChornousGvozd2022}.

To determine the handedness of the cholesteric helix, we used the rotation effect, \textit{i.e.}, observing the direction of rotation of the ChD solid microcrystals during dissolution at the top of the nematic E7 droplet using a polarizing optical microscope (POM), as described previously in detail\cite{GvozdTeren2002,GvozdLis2007,GvozdLis2008}. The clockwise rotation in this case corresponds to the left-handed cholesteric helix, and vice versa. The data thus obtained are in good agreement with the Grandjean-Cano method traditionally used for determining of the helix screw sense\cite{Gerber1980}.
Planar LC cells of thickness $\sim 20~\mu\text{m}$, used for other experiments, were assembled as described in Ref.\cite{KasianGvozd2025}

The blue phase mixtures (BPMs), consisting of the HChM (which itself exhibited the BPs close to the isotropic transition\cite{KasianGvozd2025}) and various dopants with an \textit{azo} moiety, were used to study the effects of \textit{trans}--\textit{cis} photo-isomerization on the BP temperature range. These dopants, presented in Sections~S3 and S4 of the SI, are both chiral and achiral \textit{azo} compounds. The base HChM, consisting of 65~wt\% COC and 35~wt\%  nematic E7, was chosen for all studies, because this cholesteric mixture is characterized by a broad BP temperature range during both heating (\textit{i.e.}, $\Delta{T}_{\text{BP}}\approx1^\circ\text{C}$) and cooling (\textit{i.e.}, $\Delta{T}_{\text{BP}}\approx4.2^\circ\text{C}$) at a relatively low concentration of COC compared to similar mixtures that possess comparable BP temperature ranges at a higher concentration of COC\cite{KasianGvozd2025}. 
The concentration (\textit{C}) of \textit{azo} compounds in the base HChM was within the range from 0 to 7.5~wt\%. A comparison of $\Delta{T}_{\text{BP}}$ change for mixtures containing either chiral or achiral \textit{azo} compounds was carried out at a dopant concentration of $\sim5~\text{wt\%}$ in the base HChM, because the maximum BP temperature range was observed at this concentration. To prepare uniform chiral mixtures, a Fisher Vortex Genie 2 mixer (Fisher Scientific, USA) was used with periodic heating the mixture to the Iso (vortex speed 4--5 over 30~min).

To measure the specific rotation $[\alpha]^{20^{\circ}\text{C}}_\text{D}$ of chiral \textit{azo} compounds, the MPC 300 polarimeter (Anton Paar GmbH, Germany) characterized by a wide range of measurements from 0 to $89.9^{\circ}$ and high precision $\pm 0.003^{\circ}$ was used. Measurements were carried out at the Na D-line ($\lambda_{\text{D}} = 589$~nm) at a sample temperature $21^\circ\text{C}$. The temperature stability of the samples (\ce{CHCl3} solution of chiral \textit{azo} compounds) was $\pm0.1^\circ\text{C}$.

To measure both the melting point ($T_{\text{m.p.}}$) of \textit{azo} compounds and phase transition temperatures of studied HChMs by recording the changes of textures, we used a thermostable heater based on a temperature regulator MikRa 603 (LLD `MikRa', Kyiv, Ukraine) equipped with a platinum resistance thermometer Pt1000 (PJSC `TERA', Chernihiv, Ukraine)\cite{KasianGvozd2025}. During the BP phase transition studies, the cooling rate was $\sim 0.5^{\circ}\text{C/min}$.

The liquid-crystalline textures during phase transitions of the studied mixtures were observed by means of the polarizing optical microscope (POM) BioLar PI (Warsaw, Poland) equipped with a  Nikon D80 digital camera.

We also made attempts to check the $\text{N}^*$ -- BP-I -- BP-II transition temperatures using differential scanning calorimetry (Mettler DSC 1, $\sim 15$~mg and scan rates 0.5--2~K$\cdot\text{min}^{-1}$), but the transitions could not be reliably detected under such conditions.

\begin{table*}
\begin{adjustwidth}{-1in}{0in}
  \caption{\ Characteristics of chiral \textit{azo} compounds as photosensitive dopants added to nematic host E7 and HChM}
  \end{adjustwidth}
  \centering
\small
  \medskip
  \label{tbl:Table1}
  \begin{tabular*}{\textwidth}{@{\extracolsep{\fill}}lllll}
    \hline
         \quad\,\,\, Chiral & \,Number of carbon & \,$[\alpha]^{20^{\circ}\text{C}}_\text{D}$, & HTP, &Handedness of helix\\
     \textit{azo} compound & atoms in alkyl chain & \,\,\,deg & $\mu\text{m}^{-1}$ & \qquad E7\, /\, HChM \\
    \hline
    \quad ChD-3793 & \qquad\qquad\, 0 & -76.88& -12.8 & left-handed \, /\, left-handed \\
    \quad ChD-3610 & \qquad\qquad 10 & -53,24 & -5.8 & left-handed \, /\, left-handed \\
    \quad ChD-3795 & \qquad\qquad\, 6 & -44,63 & -4 & left-handed \, /\, left-handed \\
    \quad ChD-3805 & \qquad\qquad 12 & -51,93 & -3.95 & left-handed \, /\, left-handed \\
    \quad ChD-3816 & \qquad\qquad\, 6 & -52,60 & -8 & left-handed \, /\, left-handed \\
    \quad ChD-3501 & \qquad\qquad 10 & -22,51 & -3.3 & left-handed \, /\, left-handed \\
    \hline
  \end{tabular*}
\end{table*}

\section{Results and discussions}
\subsection{The highly chiral mixture doped with \textit{azo} compounds: impact of the \textit{cis}-isomer on broadening the blue phase temperature range}
This section presents results for the base HChM (Fig.~3) doped with  \textit{azo} compounds, both chiral and achiral. The temperatures of phase transitions of the CLC mixtures with different concentrations of the chiral COC and nematic E7 have been studied in detail\cite{KasianGvozd2025}. The base HChM is characterized by the formation of BPs within a relatively wide temperature range of 46.4 to $42.2^\circ\text{C}$ upon cooling\cite{KasianGvozd2025}. Spectral characteristics of various BPs of the base HChM and phase transition textures are shown in Fig.~S2.1 and S2.2 (SI, Section~S2). Special attention is paid to the effects of UV irradiation on the broadening of the BP temperature range of these mixtures (caused by \textit{trans--cis} photo-isomerization). The influence of \textit{azo} compounds added to CLCs\cite{LiYe2025} and LC polymer composites\cite{MaKim2025} on thermal and optical properties has been reported earlier for the $\text{N}^*$ and the blue phases, respectively.

The list of various chiral and achiral \textit{azo} compounds added in the course of our experiments to the HChM (Fig.~3) as photosensitive dopants is given in Table~1 and Table~2, respectively. Table~1 shows the chemical abbreviations of \textit{trans}-isomers of chiral \textit{azo} compounds and their specific rotation in \ce{CHCl3} solvent, HTP values in nematic E7 and handedness of cholesteric helix as the primary characteristics of these ChDs.

\begin{table}[ht]
\begin{adjustwidth}{-1in}{0in}
  \caption{\ Achiral \textit{azo} compounds as photosensitive dopants added to the HChM}
  \end{adjustwidth}
  \centering
\small
  \medskip
  \label{tbl:Table2}
  \begin{tabular*}{0.66\textwidth}{@{\extracolsep{\fill}}lll}
    \hline
    \quad  \,\, Achiral & \,\, Number of carbon &Handedness of helix\\
     \,\textit{azo} compound & \,atoms in alkyl chain & \qquad\quad HChM \\
    \hline
    \,\,\, aChD-3490 & \qquad\qquad\, \,0 &  \qquad left-handed \\
    \,\,\, aChD-3496 & \qquad\qquad 10 & \qquad left-handed \\
    \,\,\, aChD-3497 & \qquad\qquad10 & \qquad left-handed \\
    \,\,\, aChD-3862 & \qquad\qquad12 & \qquad left-handed \\
    \,\,\, aChD-4195 & \qquad\qquad\, \,0 & \qquad left-handed \\
    \,\,\, aChD-4197 & \qquad\qquad\, \,6 & \qquad left-handed \\
    \hline
  \end{tabular*}
\end{table}

The chiral \textit{azo} compounds listed in Table~1 possess structural variations, including different alkyl chain lengths (the number of carbon atoms, \ce{C_n}) and various linking groups between molecular fragments. For example, the alkyl chains can be linked to the \textit{azo} moiety via ester (\ce{-COO-} as in ChD-3805, ChD-3816) or ether (\ce{-C-O-C-} as in ChD-3610, ChD-3501) bridges, or a combination thereof (\textit{i.e.}, as in ChD-3795). The chiral fragment can be linked to the \textit{azo}~moiety via either a direct bond (\textit{e.g.}, ChD-3793, ChD-3610, ChD-3795, ChD-3816, and ChD-3805) or a carboxymethylene bridge (\ce{-COO-CH2-}, \textit{i.e.}, as in ChD-3501). This consequently leads to changes in the flexibility of the chiral fragment, which influences the resulting properties.

Table~2 shows the structure and number of carbon atoms in alkyl chain of the studied achiral molecules. These achiral compounds differ in alkyl chain length and linkages between adjacent molecular fragments. For example, on one side, an alkyl chain can be linked via either an ester (\ce{-COO-}) or an ether (\ce{-C-O-C-}) group, while the other side of the \textit{azo} moiety is linked to another alkyl chain (as in aChD-3496) or a cyclohexyl group (as in aChD-4195 and aChD-4197).

Additionally, reversible \textit{cis--trans} isomerization can occur due to thermal annealing of the mixture at a certain temperature above the isotropic transition temperature (${T}_\text{Iso}$). In this study, thermally induced reversible \textit{cis--trans} isomerization was used.

\noindent The studied compounds (Tables~1 and 2), which contain an \textit{azo} moiety (\textit{i.e.}, \ce{-N=N-}), undergo \textit{trans--cis} photo-isomerization under UV irradiation ($\lambda_{\text{max}} = 365$~nm). This involves a transformation from a \textit{trans}-isomer, characterized by a \textit{``baseball bat-shaped''} (achiral) or \textit{``golf club-shaped''} (chiral) geometry, to a \textit{cis}-isomer, which exhibits a \textit{``boomerang-shaped''} (achiral) or \textit{``hockey stick-shaped''} (chiral) molecular structure. The reversible \textit{cis--trans} photo-isomerization by visible light illumination (tungsten lamp), as schematically shown in Fig.~4.

\begin{figure}[!ht]
\begin{adjustwidth}{-1in}{0in}
\centering
  \includegraphics[width=\textwidth]{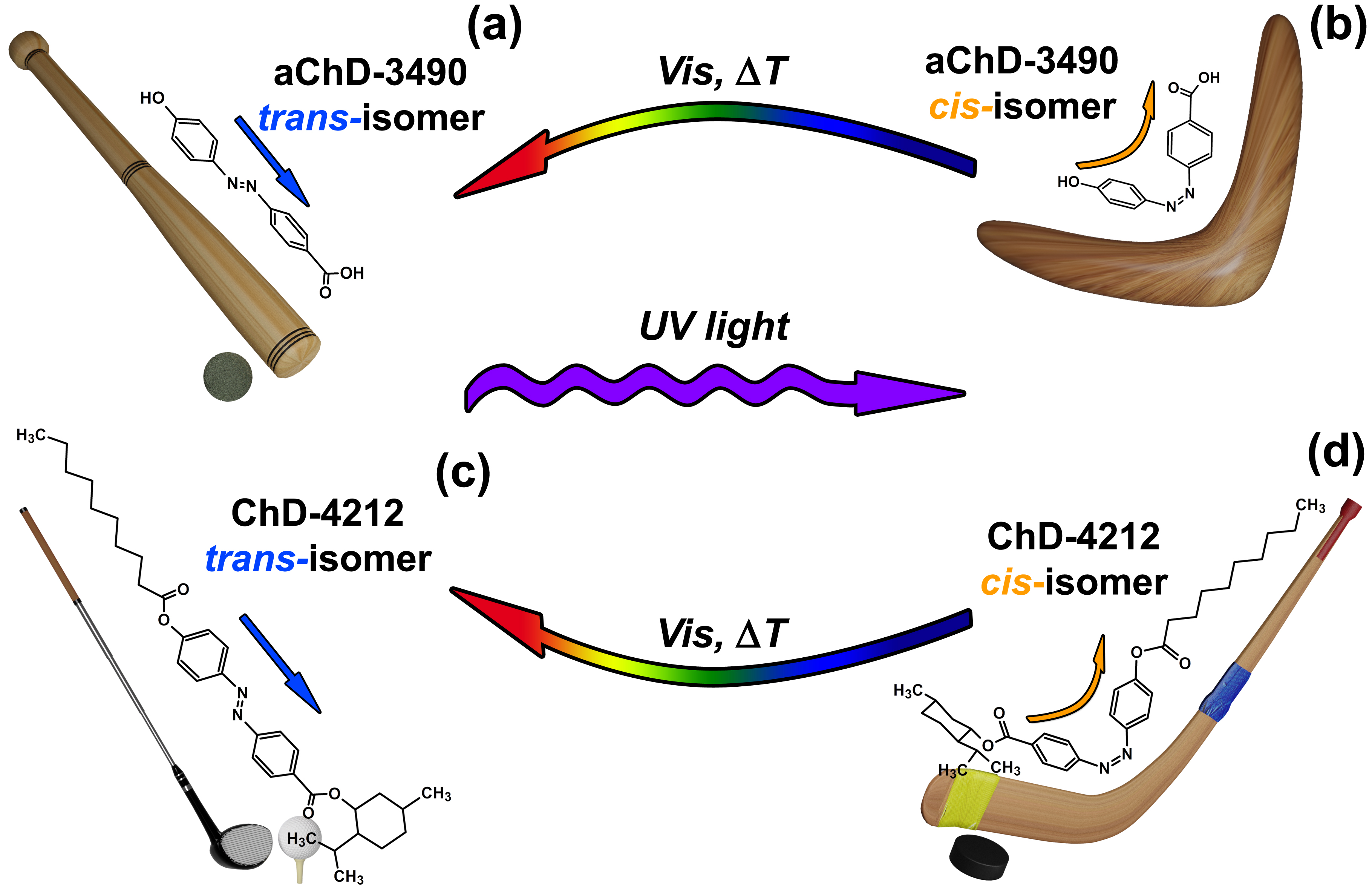}
  \caption{Schematic illustration of the photo-induced transformation of achiral (a, b) and chiral (c, d) dopants, using aChD-3490 and ChD-4212 as examples, respectively. The \textit{trans}-isomers are represented by: (a) a \textit{``baseball bat-shaped''} molecule for achiral compounds and (b) a \textit{``golf club-shaped''} molecule for chiral compounds. The \textit{cis}-isomers are represented by: (c) a \textit{``boomerang-shaped''} molecule for achiral compounds and (d) a \textit{``hockey stick-shaped''} molecule for chiral compounds. The reversible photo- and thermal isomerization of \textit{azo} molecule is indicated by rainbow arrows.}
  \label{fgr:Figure4}
  \end{adjustwidth}
\end{figure}

Fig.~5 shows the BP temperature range ($\Delta{T}_{\text{BP}}$) of various BPMs containing either chiral (Fig.~5(a)) or achiral (Fig.~5(b)) \textit{azo} compounds, in comparison with the base HChM (solid black squares). As shown in Fig.~5, the presence of any \textit{azo} compound in the HChM leads to a broadening of the BP temperature range, sometimes modest but consistently reproducible, for both chiral and achiral \textit{azo} compounds. 

\begin{figure*}[!ht]
\begin{adjustwidth}{-1in}{0in}
 \centering
 \includegraphics[width=\textwidth]{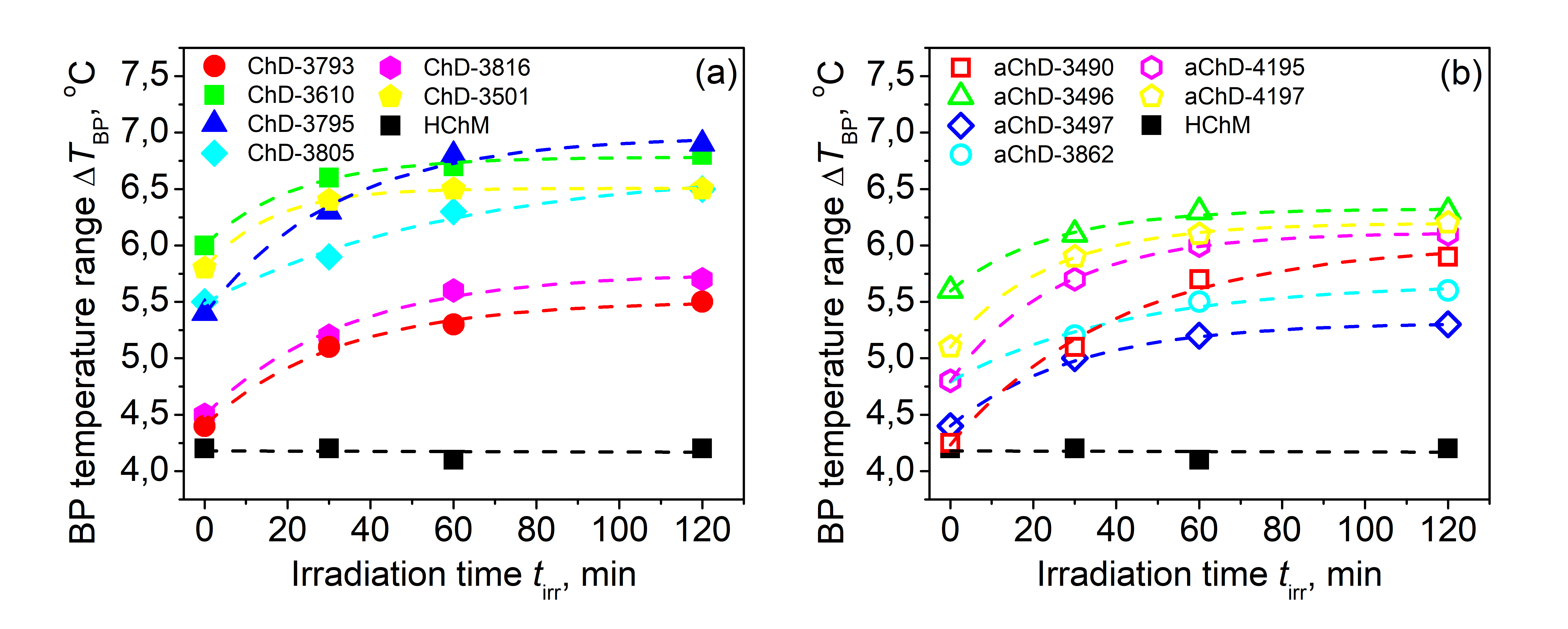}
 \caption{BP temperature range of the base HChM (solid black squares), doped with 5~wt\% of (a) chiral and (b) achiral \textit{azo} compounds as a function of irradiation time $\textit{t}_{\text{irr}}$.}
 \label{fgr:Figure5}
 \end{adjustwidth}
\end{figure*}

Fig.~5 also displays the BP temperature range ($\Delta{T}_{\text{BP}}$) for various BPMs containing chiral (Table~1) or achiral (Table~2) \textit{azo} compounds after UV irradiation. The prolonged UV irradiation of BPMs results in the transformation of maximum amounts of \textit{trans}-isomers to \textit{cis}-isomers, leading to the maximum broadening of the BP temperature range (\textit{i.e.}, saturation of the BP range is observed). The diagrams showing the BP phase transition temperatures of these mixtures are presented in detail in the SI (Section~S4).

Thermal annealing of the pre-irradiated BPMs at approximately 60--65$^{\circ}\text{C}$ (\textit{i.e.}, above the BP--Iso transition temperature) leads to  reverse \textit{cis-trans} isomerization of the \textit{azo} compounds. This process is accompanied by a decrease in $\Delta{T}_{\text{BP}}$ to its initial value, corresponding to the state of the BP range before UV irradiation ($t_{\text{irr}} = 0$~min). Detailed data on phase transition temperatures can be found in the SI (Section~S4).

A comparative assessment of the effects of specific compounds on the broadening of the BP temperature range is not straightforward, due to varying \textit{cis}-isomer contents in the photostationary state during UV irradiation of BPMs and their structural variations (Tables~1 and 2). Studies on the effects of chemical structure on $\Delta{T}_{\text{BP}}$ of the BP will be considered in subsequent sections. 

To estimate the degree of influence of the \textit{trans}--\textit{cis} photo-isomerization of various \textit{azo} compounds on the broadening of the BP temperature range, we introduce a parameter $\xi$ (referred to as the \textit{cis}-isomer efficiency) as follows:

\begin{equation}
  \xi = \frac{\Delta T^{t_{\text{irr}}=i}_{\text{BP}}}{\Delta T^{t_{\text{irr}}=0}_{\text{BP}}} 
\end{equation}

where $\Delta T^{t_{\text{irr}}=0}_{\text{BP}}$ and $\Delta T^{t_{\text{irr}}=i}_{\text{BP}}$ are the BP temperature ranges (including BP-I, BP-II and BP-III) before ($t_{\text{irr}} = 0$~min) and after ($t_{\text{irr}} = i$~min, where $0 < i \leq 30$) UV irradiation of the BPM, respectively. 

The temperature ranges ($\Delta{T}_{\text{BP}}$) of the studied BPMs can be determined from the phase diagrams presented in Tables~S4.1, S4.3 and S4.5 of the SI. Thus, the broadening of the BP temperature range in HChMs containing various \textit{azo} compounds upon UV irradiation for 30 min is a key characteristic of the \textit{azo} compound. The reason for choosing this time range of UV irradiation of samples (where the concentration of the formed \textit{cis}-isomer changes almost linearly) is described in detail in Section~S5 of the SI.

Fig.~6 shows the \textit{cis}-isomer efficiency $\xi$ of various \textit{azo} compounds with respect to the broadening of the BP temperature range of the BPMs containing chiral and achiral compounds, listed in Tables~1 and 2, respectively. The UV irradiation time was fixed at 30~min.

As shown in Figs.~5 and 6, certain \textit{azo} compounds clearly broaden the BP temperature range before UV irradiation (\textit{e.g.}, ChD-3610, solid green squares in Fig.~5(a)). In contrast, the \textit{cis}-isomer efficiency $\xi$ (after UV irradiation) is more pronounced for other \textit{azo} dopants (\textit{e.g.}, ChD-3795, solid blue triangles in Figs.~5(a) and 6(a)).

The broadening of the BP temperature range of the base HChM upon addition of \textit{azo} compounds (Fig.~4) can be explained by analogy with the stabilization of BPs by means of nanoparticles (NPs)\cite{Muševič2010,Dierking2012,Wang2012,Lavrič2013,Dierking2020,Oton2022} or polymers\cite{ColesPivnenko2005,Kikuchi2002,Kikuchi2021,PabloGonzalez2025,Gvozd2016}, which can be ascribed to their accumulation in the disclination lines of the lattice (Fig.~1(b) and (c)). In this case, \textit{azo} compounds can be considered as molecular-sized ``particles''. As in the case of NPs added to BPs\cite{Wang2012,Lavrič2013}, the BP temperature range $\Delta{T}_{\text{BP}}$ depends on the concentration $C$ of the introduced \textit{azo} compounds. Therefore, to verify this assumption  (\textit{i.e.}, that \textit{azo} compound molecules act as direct analogue of NPs), we analyzed the dependence of $\Delta{T}_{\text{BP}}$ on the concentration $C$ of \textit{azo} compound, using the chiral \textit{azo} compound ChD-3795 as an example.

\begin{figure}[ht]
\begin{adjustwidth}{-1in}{0in}
 \centering
 \includegraphics[width=\textwidth]{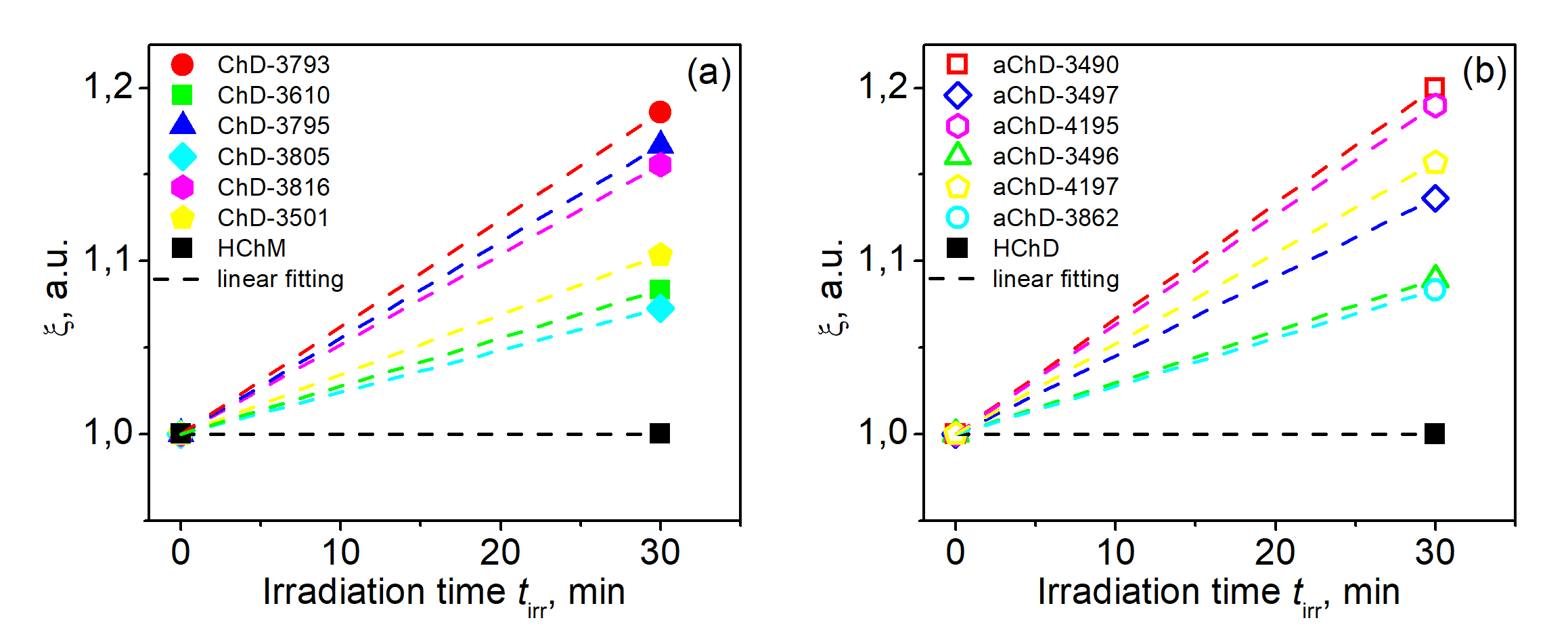}
 \caption{The effect of \textit{cis}-isomer efficiency $\textit{t}_{\text{irr}}$ on the broadening of the BP temperature range of the BPMs. The composition of the mixtures and symbols are the same as in Fig.~5. The irradiation of 20~$\mu$m thick LC cells was carried out by UV lamp with wavelength $\lambda_{\text{max}} = 365$~nm over 30~min.
}
 \label{fgr:Figure6}
 \end{adjustwidth}
\end{figure}

Fig.~7(a) shows the concentration dependence of the BP temperature range ($\Delta{T}_{\text{BP}}$) for the BPM containing the chiral \textit{azo} compound ChD-3795 (Table~1), at various UV irradiation times ($t_{\text{irr}}$). Maximum broadening of $\Delta{T}_{\text{BP}}$ is observed at about 5~wt\% of \textit{azo} compound ChD-3795. At concentrations below and above $\sim 5$~wt\% the broadening of the BP temperature range of the BPM is less pronounced. Similar dependencies were also observed at various UV irradiation times. The addition of the left-handed chiral dopant S-811 to the base HChM is characterized by a similar concentration dependence for the broadening of the BP temperature range as shown in Fig.~S6.1 of the SI.

\begin{figure}[ht]
\begin{adjustwidth}{-1in}{0in}
\centering
  \includegraphics[width=\textwidth]{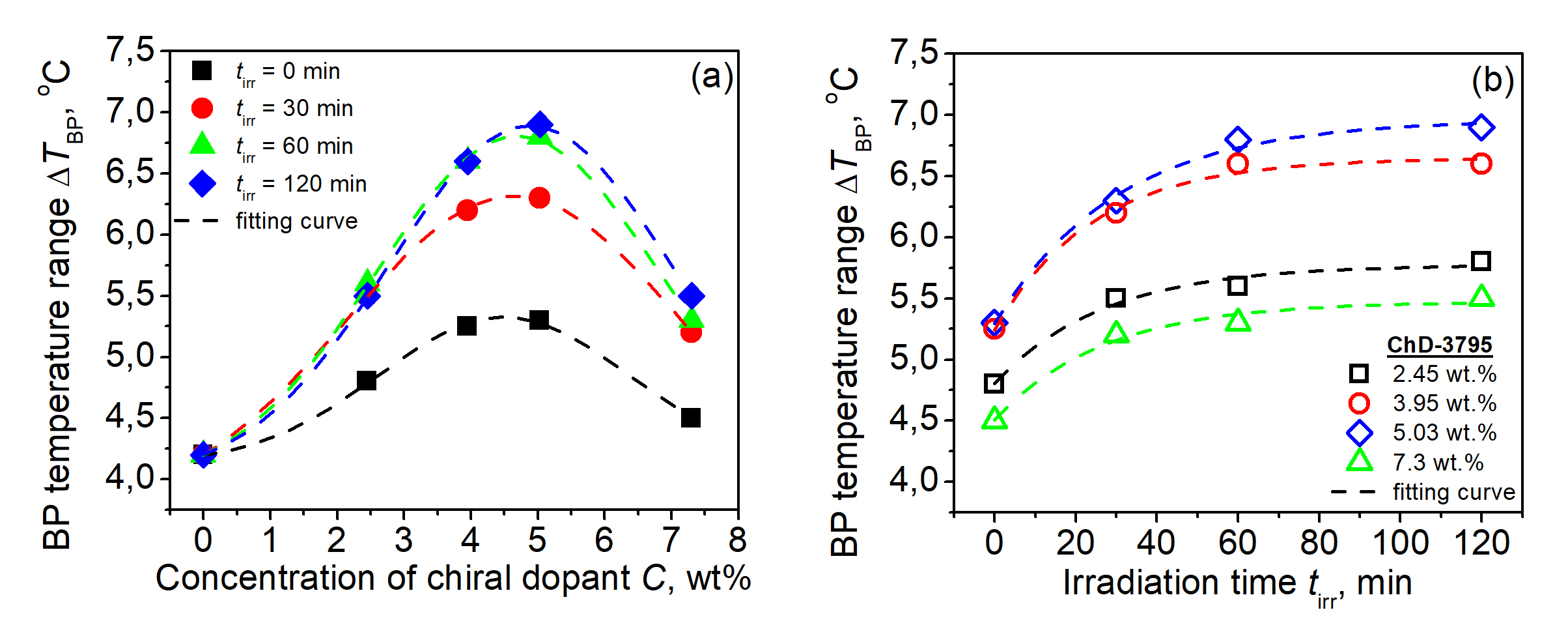}
  \caption{Dependence of the BP temperature range $\Delta{T}_{\text{BP}}$ of the BPM containing the base HChM and chiral \textit{azo} compound ChD-3795 on the concentration of \textit{azo} compound ChD-3795 (solid symbols, a) and UV irradiation time $t_{\text{irr}}$ (open symbols, b). The concentration of chiral \textit{azo} compound ChD-3795 in the HChM is: 1) 2.45~wt\% (open black squares); 2) 3.95~wt\% (open red circles); 3) 5.03~wt\% (open blue diamonds); 4) 7.3~wt\% (open green triangles).}
  \label{fgr:Figure7}
  \end{adjustwidth}
\end{figure}

To explain the results shown in Fig.~7(a), a \textit{``cinema hall''} model is proposed, providing a simplified view of the observed concentration dependence of the BP temperature range (see Section~6 in the SI). This suggests that at low concentrations of ChD-3795 in the HChM all molecules of \textit{azo} compound, analogously to NPs in BPs\cite{Wang2012,KasianGvozd2025}, tend to accumulate at disclination lines of BPs lattice (Fig.~S6.2(b) in the SI). At the \textit{``threshold''} concentration of ChD-3795 close to 5~wt\% the dopants very tightly occupy the space at disclination lines of BPs lattice (Fig.~S6.2(c) in the SI). At a concentration of ChD-3795 above 5~wt\%, the decrease in the BP temperature range ($\Delta{T}_{\text{BP}}$) is observed (Fig.~7(a)). Such dependence can be explained by the fact that \textit{azo} molecules ChD-3795 occupy not just the disclination lines of BP lattice, but also penetrate into the free volume in BPM (Fig.~S6.2(d) in the SI). The guest \textit{azo} molecules ChD-3795 occupying the free volume cause destabilization of the BPs, \textit{i.e.}, a decrease in $\Delta{T}_{\text{BP}}$ (Fig.~7(a)).

Fig.~7(b) shows the influence of UV irradiation on the broadening of the BP temperature range for BPMs with different concentrations of ChD-3795. The BP temperature range of the BPMs expands with increasing UV irradiation time ($t_{\text{irr}}$); however, $\Delta{T}_{\text{BP}}$ values also depend on the concentration of the chiral \textit{azo} compound ChD-3795 added to the base HChM (Fig.~3).

{\sloppy \subsection{Influence of the structure of \textit{azo} compounds assembled from ``LEGO'' molecular units on the blue phase temperature range}}

In this section, the influence of various \textit{azo} compound geometries introduced to the base HChM mixture on the broadening of the BP temperature range is analyzed. In our experiments, various fragments of \textit{azo} compounds are assembled via a ``LEGO'' modular approach (Fig.~2). This section consists of several subsections, where the \textit{azo} molecule aChD-3490~(Table~2)  is considered as the base ``LEGO'' molecular unit for all studied \textit{azo} compounds. The achiral \textit{azo} compound aChD-3490 was modified through the systematic variation of its structural fragments, \textit{e.g.}, by changing the number of carbon atoms (\ce{C_n}) in the alkyl chain or adding a chiral fragment (Fig.~2). In the studied \textit{azo} dopants, the linkages between the \textit{azo} moiety and alkyl chains, as well as between the \textit{azo} moiety and chiral fragment (\textit{l}-menthol) were realized through ether (\ce{-C-O-C-}) and ester (\ce{-COO-}) bridges.

\subsubsection{Broadening of blue phase temperature range by the base \textit{azo} molecule as a key element of ``LEGO'' molecular unit.~~}

The chemical structure of the base \textit{azo} molecule aChD-3490 used for the synthesis of chiral and achiral \textit{azo} compounds is shown, in Fig.~S1.1 of the SI. The influence of \textit{trans}--\textit{cis} photo-isomerization of these \textit{azo} dopants on the BP temperature range ($\Delta{T}_{\text{BP}}$) is examined.
The \textit{azo} compound aChD-3490 exhibits a more linear \textit{trans}-isomer structure in comparison to that of the recently studied \textit{azo} molecule 2-(4-hydroxyphenylazo)benzoic acid (HABA)\cite{MaKim2025}. This difference in linearity is due to the different positioning of the carboxylic group relative to the \textit{azo} moiety; namely, in aChD-3490, the carboxylic group is in the \textit{para}-position, whereas in HABA, it occupies the \textit{ortho}-position.

The addition of 5~wt\% aChD-3490 to the base HChM ($\Delta{T}_{\text{BP}}\approx4.2^\circ\text{C}$) results in the formation of a BPM characterized by a slightly broader BP temperature range (to $\approx4.4^\circ\text{C}$ for $t_{\text{irr}} = 0$~min), as shown in Fig.~5(b) (open red squares). Prolonged UV irradiation leads to the accumulation of the aChD-3490 \textit{cis}-isomer within the BPM, which, in turn, results in a significant broadening of $\Delta{T}_{\text{BP}}$ (reaching the value $\approx5.9^\circ\text{C}$ at $t_{\text{irr}} = 120$~min).

\begin{figure*}[ht]
\begin{adjustwidth}{-1in}{0in}
\centering
  \includegraphics[height=7cm]{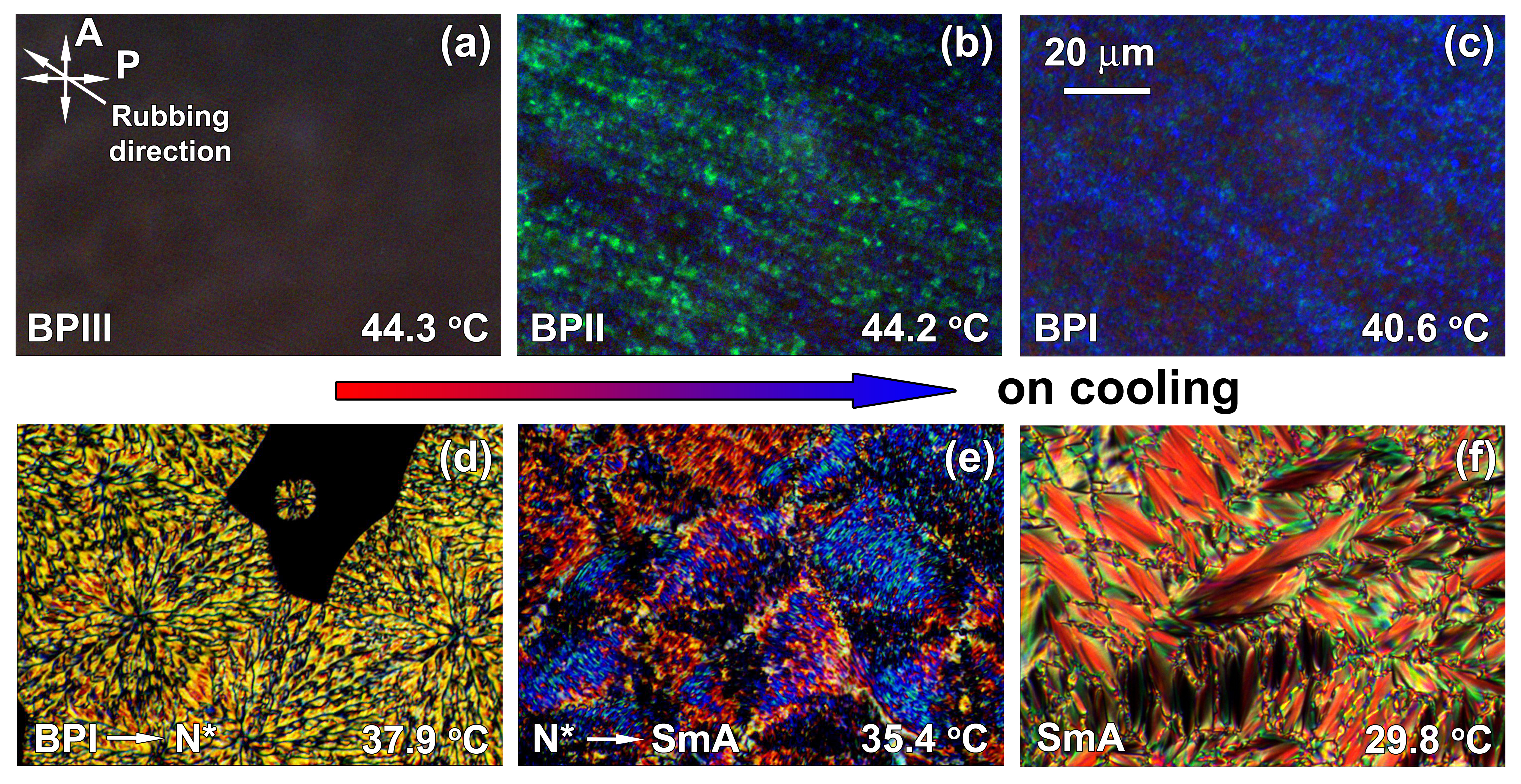}
  \caption{Sequential changes of textures of the BPM containing the base HChM (\textit{i.e.}, 65~wt\%~COC and 35~wt\%~nematic E7) doped with 5~wt\%~aChD-3940 irradiated for 120~min by UV light before the recorded cooling process: (a) BP-III (\textit{i.e.}, ``fog'' phase characterized by amorphous structure) at $44.3^\circ\text{C}$; (b) BP-II at $44.2^\circ\text{C}$; (c) BP-I at $40.6\,^\circ\text{C}$; (d) BP-I -- $\text{N}^*$ transition at $37.9^\circ\text{C}$; (e) $\text{N}^*$ -- SmA transition at $35.4^\circ\text{C}$, and (f) SmA phase at $29.8^\circ\text{C}$. LC cell was placed between crossed polarizers of POM. The angle between the rubbing direction and crossed polarizers was about 45~degrees. Thickness of LC cell was 19.8~$\mu$m.}
  \label{fgr:Figure8}
  \end{adjustwidth}
\end{figure*}
    
Fig.~8 shows the sequence phase transition textures during the cooling of the BPM, consisting of 95~wt\% HChM and 5~wt\% aChD-3490, after 120 min of UV irradiation. The microphotographs illustrate all inherent BPM phases, including BPs, $\text{N}^*$ and SmA. As compared to the phases observed for the base HChM\cite{KasianGvozd2025}, it can be concluded that the addition of \textit{azo} compound aChD-3490 does not modify the general phase behavior.

\subsubsection{Influence of the chiral fragment of \textit{azo} compounds on the photoinduced broadening of the blue phase temperature range.~~}

This subsection focuses on the effects of a chiral fragment attached to the base \textit{azo} molecule aChD-3490~(Fig.~9(a)). In this work, the synthesis of all studied chiral \textit{azo} compounds was carried out using only \textit{l}-menthol as the source of chirality. As mentioned above, the base \textit{azo} compound aChD-3490 possesses a relatively linear, \textit{``short-length''} molecular structure (a so-called \textit{``baseball bat-shaped''} molecule). Upon the addition of the chiral fragment, the molecule becomes distinctly nonlinear (a so-called \textit{``golf club-shaped''} molecule). To compare the influence of reducing molecular linearity (\textit{i.e.}, decreasing anisometry) of the \textit{azo} molecule ChD-3793 on $\Delta{T}_{\text{BP}}$, here we also examine a modification of the base \textit{azo} compound aChD-3490 featuring a similar achiral fragment (\textit{e.g.}, the cyclohexyl group in aChD-4195). These modified \textit{azo} compounds are considered \textit{``short-length''} molecules.

\begin{figure*}[ht]
\begin{adjustwidth}{-1in}{0in}
 \centering
 \includegraphics[width=\textwidth]{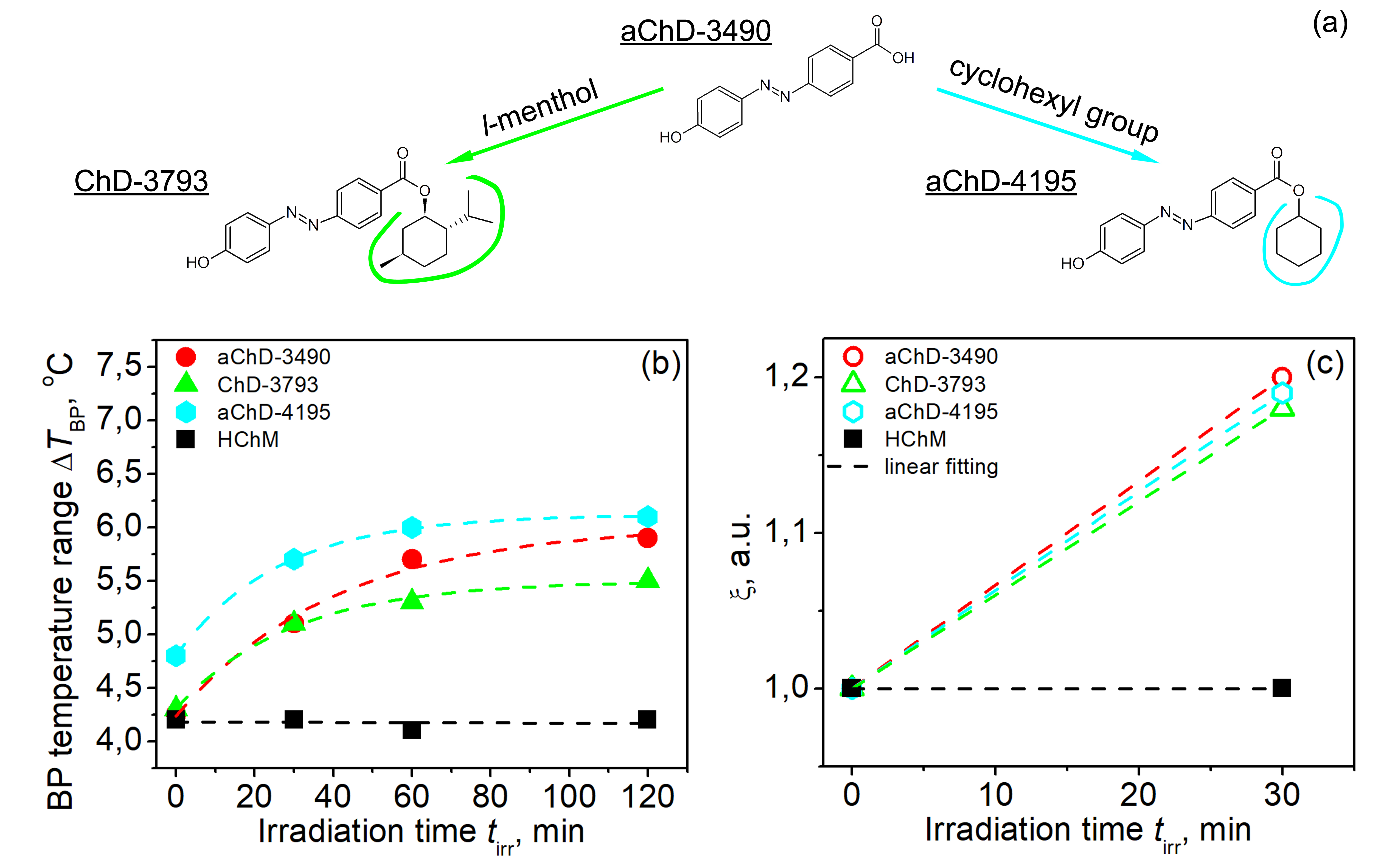}
 \caption{(a) Chemical structures of \textit{azo} compounds derived from the base \textit{azo} compound aChD-3490 by incorporating a chiral fragment (\textit{l}-menthol, as in ChD-3793) or an achiral fragment (cyclohexyl group as in aChD-4195). Dependence of (b) the BP temperature range and (c) \textit{cis}-isomers efficiency $\xi$ in broadening the BP for \textit{azo} compounds of \textit{``short-length''} structure on UV irradiation time ($t_{\text{irr}}$) of an HChM doped with 5~wt\% of the base achiral aChD-3490~(red circles) or its modifications: chiral ChD-3793~(green triangles) and achiral aChD-4195~(cyan hexagons).}
 \label{fgr:Figure9}
 \end{adjustwidth}
\end{figure*}

Fig.~9(b) shows the BP temperature ranges ($\Delta{T}_{\text{BP}}$) of the BPMs before ($t_{\text{irr}} = 0$~min) and after UV irradiation for various $t_{\text{irr}}$. The introduction of any of the \textit{azo} compounds, such as aChD-3490~(solid red circles), ChD-3793 (solid green triangles) and aChD-4195~(solid cyan hexagons), leads to a broadening of the BP temperature range compared to the base HChM~(solid black squares). The modification of aChD-3490 with either a chiral fragment (ChD-3793) or a cyclohexyl group (aChD-4195) leads to the broadening of the BP temperature range. However, the value of $\Delta{T}_{\text{BP}}$ for chiral \textit{azo} compound ChD-3793~(solid green triangles) is lower ($\Delta{T}_{\text{BP}}\approx4.3^\circ\text{C}$) than in the case of \textit{azo} compound with a cyclohexyl group, aChD-4195~(solid cyan hexagons) with $\Delta{T}_{\text{BP}}\approx5.2^\circ\text{C}$ (Fig.~9(b)). During UV irradiation of \textit{azo}-containing BPMs, the increase in \textit{cis}-isomer concentration leads to an expansion of $\Delta{T}_{\text{BP}}$ compared to the light-insensitive base HChM (solid black squares).

Fig.~9(c) shows the UV irradiation time dependence of the \textit{cis}-isomer efficiency $\xi$ in the broadening of the BP temperature range for BPMs containing \textit{``short-length''} \textit{azo} molecules. The BPM containing the linear molecule aChD-3490~(open red circles) exhibits the maximum value of \textit{cis}-isomer efficiency, $\xi\approx 1.2$. This value is higher than those observed for the less linear \textit{azo} structures: achiral aChD-4195~(open cyan hexagons) and chiral ChD-3793 (open green triangles), characterized by $\xi$ values of approximately 1.19 and 1.18, respectively. During UV irradiation, the formed \textit{cis}-isomers of \textit{azo} compounds with either a bulky chiral fragment (open green triangles) or a flat cyclohexyl group (open cyan hexagons) exhibit a more bent spatial conformation compared to the \textit{trans}-isomers. These \textit{cis}-isomers are less anisometric, forming \textit{``hockey stick-shaped''} and \textit{``boomerang-shaped''} molecules, respectively. This reduces the rate of \textit{trans}--\textit{cis} isomerization compared to the linear \textit{azo} compound aChD-3490~(open red circles) due to resistance from the LC environment (\textit{i.e.}, the HChM).
The presence of either a chiral fragment or an achiral cyclohexyl group was found to increase the BP temperature range ($\Delta{T}_{\text{BP}}$). To make these conclusions more convincing, three other \textit{azo} compounds, namely aChD-3497, ChD-3610 and aChD-3496 (Fig.~10(a)), were considered. 

\begin{figure*}[ht]
\begin{adjustwidth}{-1in}{0in}
 \centering
 \includegraphics[width=\textwidth]{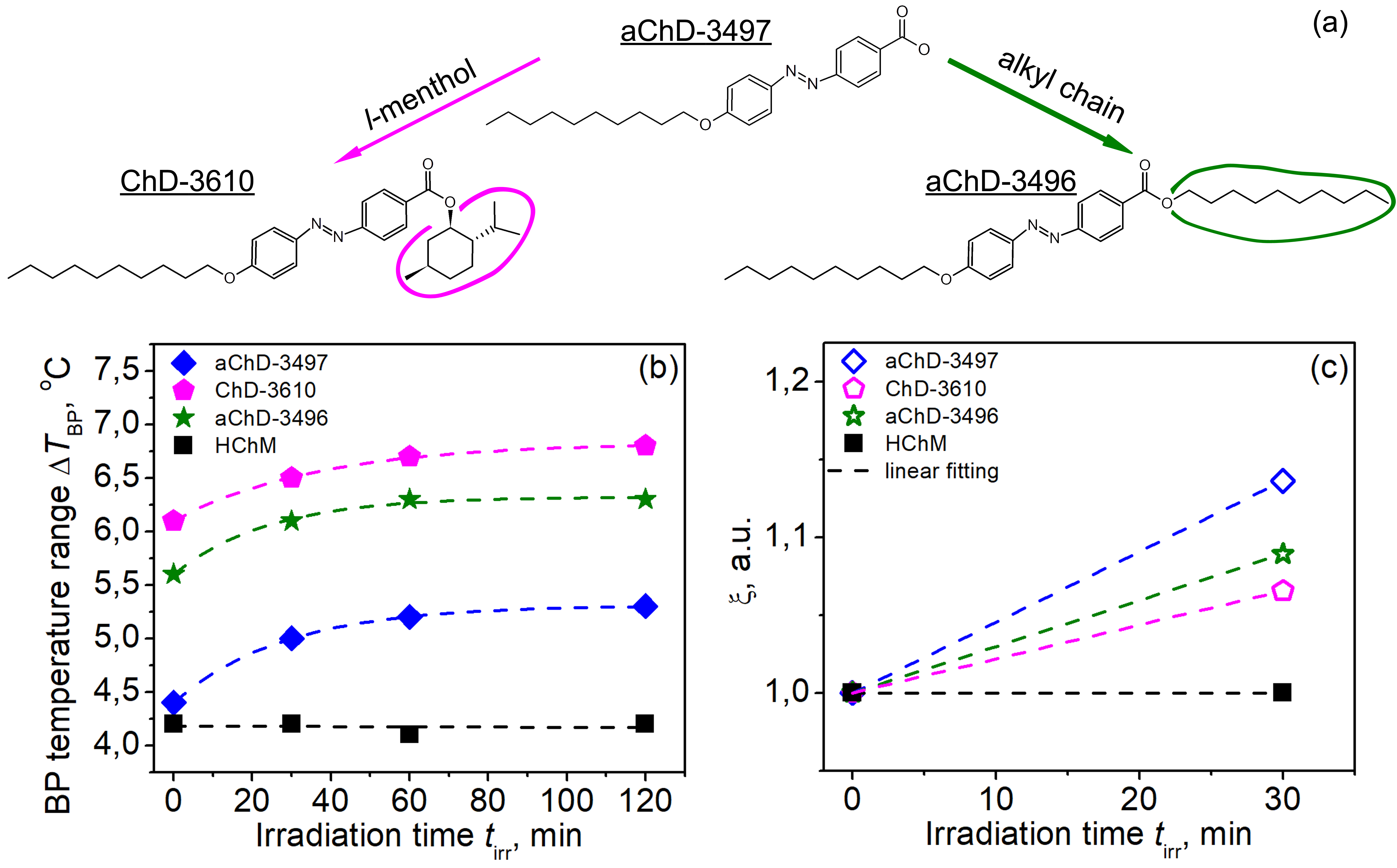}
 \caption{(a) Chemical structures of \textit{azo} compounds obtained by introducing a chiral fragment (\textit{l}-menthol, as in ChD-3610) or an alkyl chain (\ce{C_10}, as in aChD-3496) to the base compound aChD-3497. Dependence of the BP temperature range (b) and \textit{cis}-isomer efficiency $\xi$ in broadening the BP temperature range of the BPMs (c) on UV irradiation time ($t_{\text{irr}}$) for base HChM doped with 5~wt\% of achiral aChD-3497 (blue diamonds) or its two modifications, \textit{i.e.}, chiral ChD-3610 (magenta pentagons) and achiral aChD-3496 (olive stars).}
 \label{fgr:Figure10}
 \end{adjustwidth}
\end{figure*}

Fig.~10(a) shows the modified \textit{azo} compounds, which have the same alkyl chain length (10 carbon atoms, \ce{C_10}). They differ by the presence of either a chiral fragment as in ChD-3610 or an alkyl chain (\ce{C_10}) linked to the opposite side of the \textit{azo} moiety as in aChD-3496. These molecules are derivatives of the base molecule aChD-3497 obtained by incorporating either the additional alkyl chain or the chiral fragment. All these \textit{azo} compounds are considered \textit{``long-length''} molecules.

Figs.~10(b) and (c) show the UV irradiation time ($t_{\text{irr}}$) dependence of both the BP temperature range $\Delta{T}_{\text{BP}}$ and the \textit{cis}-isomer efficiency $\xi$ in broadening the temperature range of the BPMs. A comparison of $\Delta{T}_{\text{BP}}$ for all three BPMs before UV irradiation ($\textit{t}_{\text{irr}}$ = 0~min) shows that the presence of either a chiral fragment in ChD-3610 (solid magenta pentagons) or an additional alkyl chain linked to the opposite side of the \textit{azo} moiety in aChD-3496 (solid olive stars) leads to a more significant broadening of the BP temperature range compared to the base \textit{azo} compound aChD-3497 (solid blue diamonds). This is in good agreement with the results for the \textit{``short-length''} molecules shown in Fig.~9(b). It should also be emphasized that the presence of the chiral fragment in ChD-3610 leads to a further broadening of the BP temperature range ($\Delta{T}_{\text{BP}}$) compared to aChD-3496, which incorporates an additional alkyl chain (Fig.~10(b)). We hypothesize that the  broadening of the BP temperature range in ChD-3610 results from the deviation from the linearity of the molecule (\textit{i.e.}, a \textit{``golf club-shaped''} structure) induced by the chiral fragment, which distinguishes it from the linear structures of aChD-3497 and aChD-3496. The \textit{``golf club-shaped''} molecule is analogous to the \textit{cis}-isomer, which induces an increase in $\Delta{T}_{\text{BP}}$ as opposed to the \textit{trans}-isomer (Fig.~10(b)). However, after a specific UV irradiation time $t_{\text{irr}}$ (\textit{e.g.}, over 30 min), the \textit{cis}-isomer efficiency $\xi$  for the achiral \textit{azo} compound aChD-3497 (open blue diamonds) is sufficiently higher (\textit{i.e.}, $\xi \approx1.14$) than the values of 1.08 and 1.06 observed for achiral aChD-3496 (open olive stars) and chiral ChD-3610 (open magenta pentagons), respectively (Fig.~10(c)).

The inverse relationship between the broadening of the temperature range ($\Delta{T}_{\text{BP}}$) and the \textit{cis}-isomer efficiency ($\xi$) of this broadening can be explained by the different molecular shapes of the dopants. It can be assumed that during \textit{trans}--\textit{cis} photo-isomerization, the variations in \textit{cis}-isomer efficiency ($\xi$) can be explained by the interaction between the \textit{azo} compound incorporated into the BP lattice and the lattice itself. 
For the \textit{``baseball bat-shaped''} molecule aChD-3497, the \textit{cis}-isomer efficiency $\xi$ is higher, because the absence of a chiral fragment allows for easier incorporation into the BP lattice, owing to the lower resistance of the liquid crystalline medium during \textit{trans}--\textit{cis} photo-isomerization. 
Fig.~10(c) shows that for the \textit{cis}-isomer of ChD-3610, the rigid linkage between the chiral fragment and the \textit{azo} moiety results in a strong resistance from the liquid crystalline medium, causing a smaller broadening of the BP temperature range ($\Delta{T}_{\text{BP}}$).

\begin{figure*}[ht]
\begin{adjustwidth}{-1in}{0in}
 \centering
 \includegraphics[width=\textwidth]{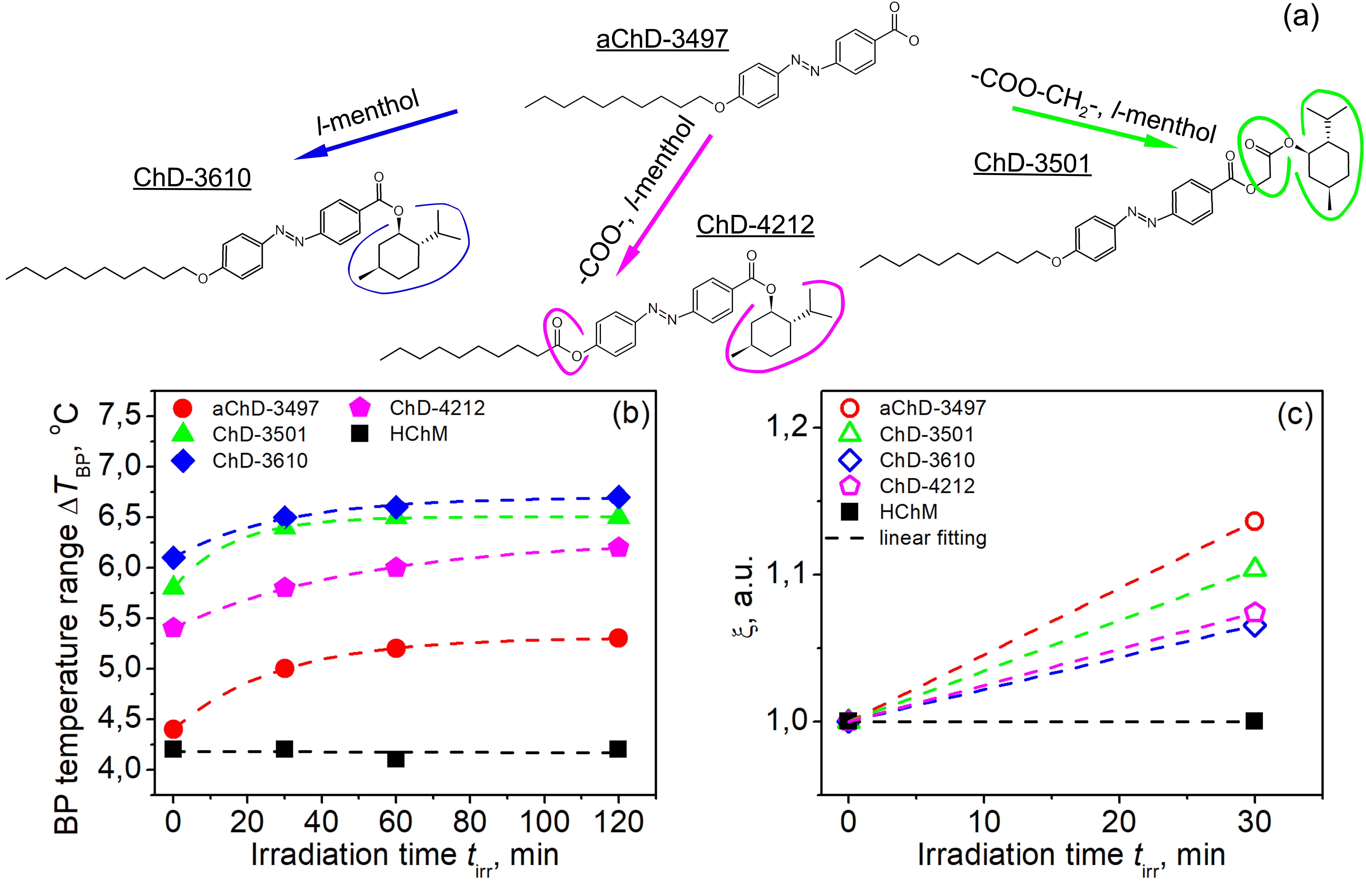}
 \caption{(a) Chemical structures of \textit{azo} compounds obtained by modification of aChD-3497 via: 1) adding a chiral fragment (\textit{l}-menthol as in ChD-3610); 2) using ester group (\ce{-COO-}) to attach the alkyl chain in ChD-4212; and 3) using a carboxymethylene bridge (\ce{-COO-CH2-}) to attach the chiral fragment in ChD-3501. UV irradiation time ($t_{\text{irr}}$) dependence of (b) the BP temperature range and (c) the \textit{cis}-isomer efficiency $\xi$ for the base HChM doped with 5~wt\% of either the achiral aChD-3497 (red circles) or its modifications: chiral ChD-3610 (blue diamonds), ChD-4212 (magenta pentagons) and ChD-3501 (green triangles).}
 \label{fgr:Figure11}
 \end{adjustwidth}
\end{figure*}

\subsubsection{Influence of various linking groups in \textit{azo} compounds on the photoinduced broadening of the blue phase temperature range.~~} 

This subsection examines the impact of various bridges---such as ester (\ce{-COO-}), ether (\ce{-C-O-C-}) and methylene (\ce{-CH2-}) groups---used as modular ``LEGO'' molecular units to link the \textit{azo} moiety (\textit{i.e.}, the base \textit{azo} compound aChD-3490) with the chiral fragment (\textit{l}-menthol) and the alkyl chains. We hypothesize that the use of various linking bridges can result in different degrees of flexibility between adjacent molecular fragments.

For this purpose, we consider various modifications of the base achiral \textit{azo} compound aChD-3497 (\textit{``long-length''} structure), such as the derivatives ChD-3501, ChD-3610 and ChD-4212 (Fig.~11(a)). These compounds are characterized by an alkyl chain length of 10 carbon atoms (\ce{C_10}). However, these \textit{azo} compounds differ in their fragment linkage: an ether bridge (\ce{-C-O-C-}) is employed for aChD-3497, ChD-3501 and ChD-3610, while ChD-4212 incorporates an ester bridge (\ce{-COO-}). The linkage between the chiral fragment and \textit{azo} moiety in ChD-3501 is achieved through a carboxymethylene bridge (\ce{-COO-CH2-}).

Fig.~11(b) shows the dependence of $\Delta{T}_{\text{BP}}$ on UV irradiation time ($t_{\text{irr}}$) for the BPMs containing the base HChM doped with 5~wt\% of various \textit{azo} compounds. As established for the \textit{``short-length''} \textit{azo} compounds aChD-3490 and ChD-3793 (Fig.~5(a)), the studied BPMs containing chiral \textit{azo} derivatives---such as ChD-3501 (solid green triangles), ChD-3610 (solid blue diamonds) and ChD-4212 (solid magenta pentagons)---exhibit a broader BP temperature range $\Delta{T}_{\text{BP}}$ compared to the BPM containing the achiral \textit{azo} compound aChD-3497 (solid red circles). Notably, this broadening occurs despite the use of structural variations in bridging groups between the chiral fragment and \textit{azo} moiety within the molecule.   

Fig.~11(b) also shows that before UV irradiation ($t_{\text{irr}} = 0$~min), the maximum broadening of the BP temperature range occurs for the BPM containing ChD-3610 (solid blue diamonds). ChD-3610 is characterized by a rigid linkage via two ether bridges (\ce{-C-O-C-}) that join both the alkyl chain and the chiral fragment to the \textit{azo} moiety (Fig.~11(a)). ChD-3501 (solid green triangles) is derived from ChD-3610 by incorporating a carboxymethylene bridge (\ce{-COO-CH2-}), which modifies the rigidity of the chiral fragment within the \textit{azo} compound. This leads to a reduction in the BP temperature range ($\Delta{T}_{\text{BP}}$) compared to the BPM based on ChD-3610 (solid blue diamonds). ChD-4212 (solid magenta pentagons) is derived from ChD-3610 by introducing an ester bridge (\ce{-COO-}), which modifies the flexibility between the alkyl chain and the \textit{azo} moiety (Fig.~11 (a)). The substitution of an ether with an ester linkage, as in ChD-4212, leads to a reduction in the BP temperature range ($\Delta{T}_{\text{BP}}$) compared to ChD-3610 and ChD-3501, both of which contain an ether bridge between the alkyl chain and the \textit{azo} moiety (Fig.~11(b)). 
\begin{sloppypar}
UV irradiation of the \textit{azo}-containing BPMs (Fig.~11(a)) induces \textit{trans}--\textit{cis} photo-isomerization, leading to a further broadening of the BP temperature range (Fig.~11(b)). With prolonged UV exposure of the BPMs ($t_{\text{irr}} > 60$~min), the concentration of the \textit{cis}-isomer in these liquid crystalline medium reaches saturation, and the rate of $\Delta{T}_{\text{BP}}$ broadening slows down.
Fig.~11(c) shows that the \textit{cis}-isomer efficiency $\xi$ in broadening the BP temperature range is higher ($\xi\approx~1.1$) for the BPM containing the most flexible derivative (ChD-3501), compared to the less flexible ChD-4212 ($\xi\approx~1.07$) or the rigid ChD-3610 ($\xi\approx1.06$). The incorporation of the base \textit{azo} compound aChD-3497---characterized by a \textit{``rod-like''} (so-called \textit{``baseball bat-shaped''}) molecular structure---into the BPM results in a higher value of the \textit{cis}-isomer efficiency $\xi$ ($\approx1.14$). Taking into account the molecular structures of the \textit{azo} compounds (Fig.~11(a)), the \textit{cis}-isomer efficiency values $\xi$ are in good agreement with the explanation proposed in Subsection~3.2.2.
\end{sloppypar}

\subsubsection{Effect of alkyl chain length on the broadening of the blue phase temperature range in homologous \textit{azo} compounds.~~}

\begin{table*}
\begin{adjustwidth}{-1in}{0in}
  \caption{\ Characteristics of the base compound ChD-3793 and a homologous series of chiral \textit{azo} compounds as photosensitive dopants added to the nematic host E7 and HChM}
  \end{adjustwidth}
  \centering
\small
\medskip
  \label{tbl:Table3}
  \begin{tabular*}{\textwidth}{@{\extracolsep{\fill}}lllll}
    \hline
       \quad \, Chiral &  \, Number of carbon & $[\alpha]^{20^{\circ}\text{C}}_\text{D}$, & HTP, &\quad Handedness of helix\\
    \textit{azo} compound &atoms in alkyl chain & deg & $\mu\text{m}^{-1}$ & \qquad \quad \, E7\, /\, HChM \\
    \hline
    \quad ChD-3793 & \quad \qquad\,\,\,\, 0 & -76.88 & -12.8 & left-handed \, /\, left-handed \\
   \quad ChD-4185 & \quad \qquad\,\,\,\, 2 & -63.17 & -8.5 & left-handed \, /\, left-handed \\
   \quad ChD-4187 & \quad \qquad\,\,\,\, 4 & -55.19 & -8.1 & left-handed \, /\, left-handed \\
   \quad ChD-3816 & \quad \qquad\,\,\,\, 6 & -52.60 & -8 & left-handed \, /\, left-handed \\
   \quad ChD-4211 & \quad \qquad\,\,\,\, 8 & -43.16 & -6.63 & left-handed \, /\, left-handed \\
   \quad ChD-4212 & \quad \qquad\,\,\,10 & -47.6 & -5.81 & left-handed \, /\, left-handed \\
   \quad ChD-3805 & \quad \qquad\,\,\,12 & -51.93 & -3.95 & left-handed \, /\, left-handed \\
    \hline
  \end{tabular*}
\end{table*}

This subsection examines the influence of alkyl chain length within a homologous series of the chiral \textit{azo} compounds on broadening of the BP temperature range ($\Delta{T}_{\text{BP}}$) for BPMs containing 5~wt\% of the dopants listed in Table~3. These compounds were synthesized by modifying of the base molecule ChD-3793 (Table~1), which incorporates a terminal hydroxyl group (\ce{-OH}). Table~3 shows the chiral \textit{azo} compounds, which are characterized by a rigid linkage between the \textit{azo} moiety and chiral fragment and a flexible ester bridge (\ce{-COO-}) between the alkyl chain and \textit{azo} moiety. Notably, the influence of the alkyl chain lengths (\ce{C_n}) in nematic LCs on the broadening of the BP temperature range ($\Delta{T}_{\text{BP}}$) has been reported in detail\cite{MillerGlesson1993}.

  Table~3 summarizes the key characteristics of the base compound ChD-3793 and its derivatives, which form a homologous series with varying alkyl chain lengths (\ce{C_n}). These \textit{azo} compounds exhibit levorotatory activity (\textit{i.e.}, they rotate the plane of linearly polarized light counterclockwise), as evidenced by both the negative signs ($-$) in Table~3 and the red $y$-axis in Fig.~12. Their specific rotation $[\alpha]^{20^{\circ}\text{C}}_\text{D}$ values (in \ce{CHCl3}) vary non-monotonically as the alkyl chain length increases (Fig.~12, solid red squares). For \textit{azo} compounds with the alkyl chain lengths from \ce{C_2} to \ce{C_8}, a clear decrease in specific rotation values is observed, whereas the increase occurs for derivatives with \ce{C_10} and \ce{C_12}  (Fig.~12). The maximum specific rotation ($\sim -76.9^\circ$) is observed for the base \textit{azo} compound ChD-3793, which incorporates a terminal hydroxyl group (\ce{-OH}) and lacks an alkyl chain (\ce{C_0}) (Table~3). 
  
\begin{figure}[h]
\begin{adjustwidth}{-1in}{0in}
\centering
  \includegraphics[width=0.7\textwidth]{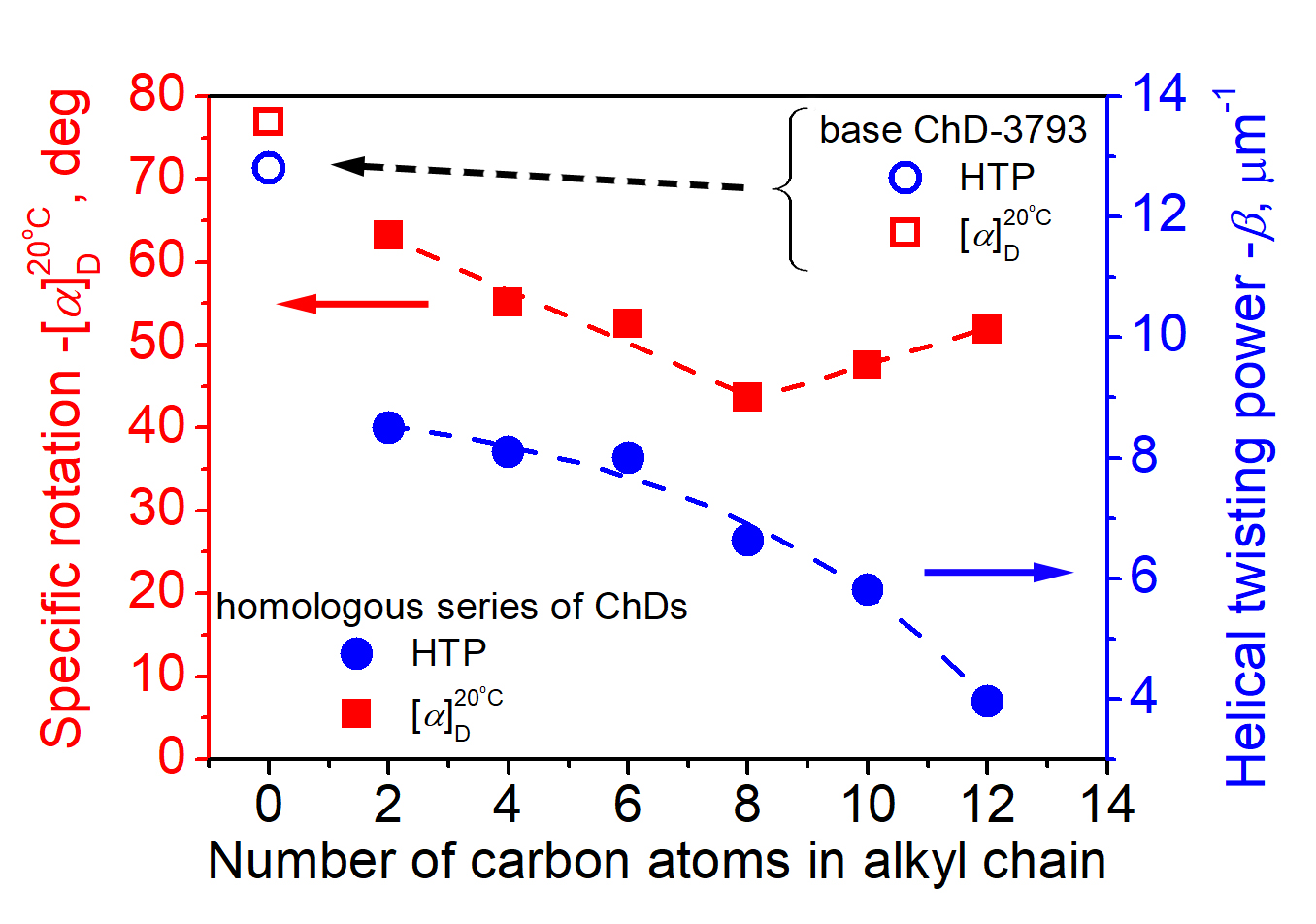}
  \caption{Dependence of helical twisting power $\beta$ in nematic E7 (solid blue circles) and specific rotation $[\alpha]^{20\,^{\circ}\text{C}}_\text{D}$ in \ce{CHCl3} (solid red squares) of the homologous series of chiral \textit{azo} compounds on the number of carbon atoms in alkyl chains. The HTP (open blue circle) and $[\alpha]^{20\,^{\circ}\text{C}}_\text{D}$ (open red square) of the base chiral \textit{azo} compound ChD-3793.}
  \label{fgr:Figure12}
  \end{adjustwidth}
\end{figure}
  
  The left-handed helical structure of CLCs (indicated by the negative signs ($-$) in Table~3), induced by adding \textit{azo} dopants to the nematic host E7, was characterized by the methods previously described in detail\cite{GvozdTeren2002,GvozdLis2007,GvozdLis2008,Gerber1980}. The base chiral \textit{azo} compound ChD-3793, which lacks an alkyl chain, exhibits a higher HTP ($\sim -12.8~\mu\mathrm{m}^{-1}$) than its derivatives with various alkyl chain lengths (Table~3). It should be mentioned that the HTP gradually decreases with increasing alkyl chain length.

  Fig.~12 shows the dependence of the HTP values on the alkyl chain length (\ce{C_n}) for the chiral \textit{azo} compounds in nematic E7. Considering the HTP values of both the base ChD-3793 (Tables~1 and 3) and its homologous series of chiral \textit{azo} dopants (Table~3), it can be concluded that for this class of \textit{``golf club-shaped''} \textit{azo} compounds, the \textit{``short-chain''} molecular structures (up to 6 carbon atoms) exhibit higher HTP values ($\sim -8~\mu\mathrm{m}^{-1}$. The HTP values of the \textit{``short-chain''} \textit{azo} compounds are comparable to those of the well-known chiral dopant S-811\cite{LCbrochure1994}. 

   Fig.~13(a) shows the broadening of the BP temperature range ($\Delta{T}_{\text{BP}}$) as a function of UV irradiation time ($t_{\text{irr}}$). The addition of approximately 5~wt\% of any chiral \textit{azo} compound listed in Table~3 to the base HChM broadens the BP temperature range. A certain correlation is observed between $\Delta{T}_{\text{BP}}$ and the alkyl chain length of the dopants.

\begin{figure}[ht]
\begin{adjustwidth}{-1in}{0in}
\centering
  \includegraphics[width=\textwidth]{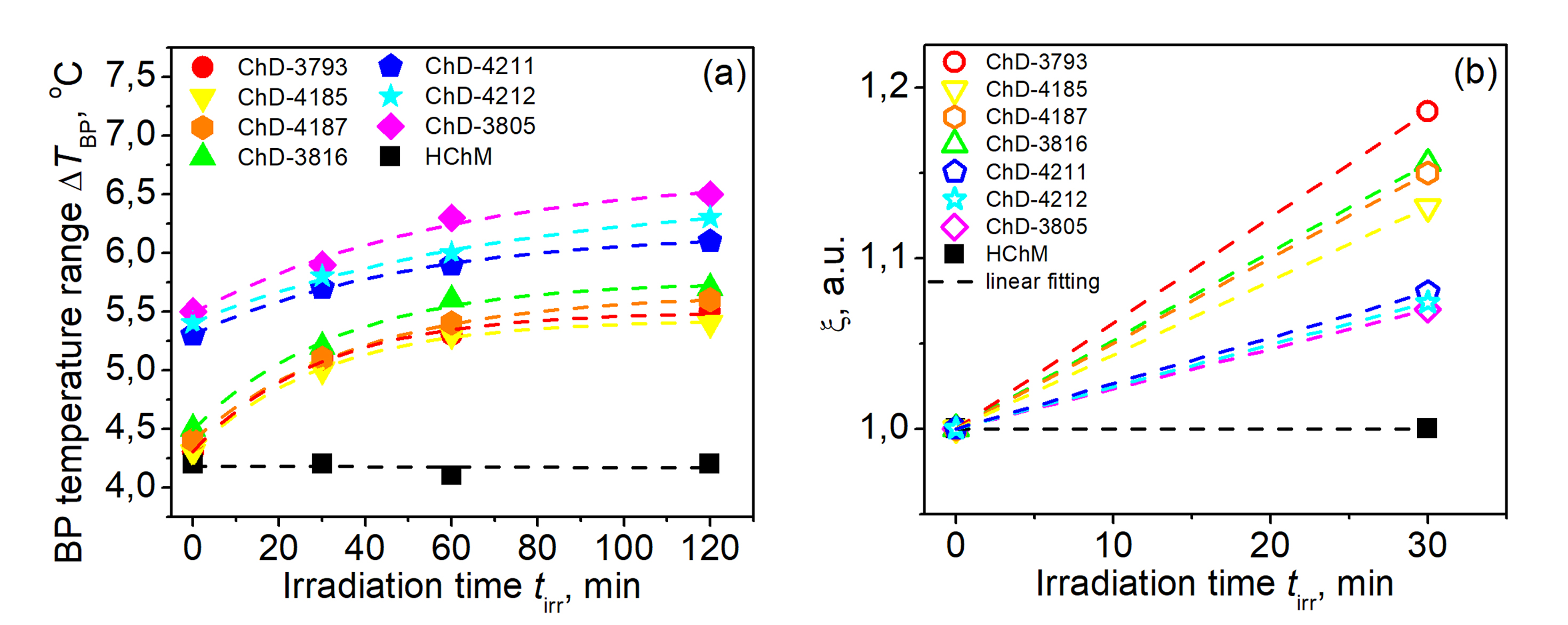}
  \caption{(a) Dependence of BP temperature ranges $\Delta{T}_{\text{BP}}$ on UV irradiation time ($t_{\text{irr}}$) for the base HChM (solid black squares) and chiral \textit{azo}-containing HChMs with different alkyl chain lengths (number of carbon atoms \ce{C_n}): \ce{C_0} – ChD-3793 (solid red circles); \ce{C_2} – ChD-4185 (solid yellow inverted triangles); \ce{C_4} – ChD-4187 (solid orange hexagons); \ce{C_6} – ChD-3816 (solid green up-pointing triangles); \ce{C_8} - ChD-4211 (solid blue pentagons); \ce{C_10} - ChD-4212 (solid cyan stars); and \ce{C_12} – ChD-3805 (solid magenta diamonds). (b) The \textit{cis}-isomer efficiency, $\xi$, for the base HChM (solid black squares) and mixtures doped with 5~wt\% of a homologous series of chiral \textit{azo} compounds (open color symbols) after 30~min of irradiation.}
  \label{fgr:Figure13}
  \end{adjustwidth}
\end{figure} 
  
  Compared to the base HChM (which contains no chiral or achiral \textit{azo} compounds), a slight increase in the broadening of the BP temperature range is observed for the BPMs containing \textit{azo} compounds with \textit{``short-chain''} alkyl fragments, such as ChD-4185 (solid yellow inverted triangles), ChD-4187 (solid orange hexagons) and ChD-3816 (solid green up-pointing triangles), characterized by 2, 4 and~6 carbon atoms in the alkyl chain, respectively. 
 The maximum $\Delta{T}_{\text{BP}}$ is observed for chiral dopants with \textit{``long-chain''} alkyl fragments, \textit{e.g.}, ChD-4211 (solid blue pentagons) with \ce{C_8}, ChD-4212 (solid cyan stars) and ChD-3805 (solid magenta diamonds) with \ce{C_10} and \ce{C_12}, respectively.

Fig.~13(b) shows that upon UV irradiation, BPMs containing \textit{azo} compounds with \textit{``long-chain''} alkyl fragments exhibit a slightly smaller increase in the temperature range compared to those with \textit{``short-chain''} alkyl fragments. The increase in the concentration of \textit{cis}-isomer, which possesses a \textit{``hockey stick-shaped''} molecular structure, leads to the broadening of the temperature range ($\Delta{T}_{\text{BP}}$). It is observed that the \textit{cis}-isomer efficiency $\xi$ in broadening the BP temperature range depends on the size of the \textit{azo} compound, specifically the alkyl chain length (\ce{C_n}). Compared to \textit{``short-chain''} molecules, the \textit{``long-chain''} analogous may experience higher steric hindrance from the liquid crystalline medium during \textit{trans--cis} photo-isomerization. This causes a decrease in the \textit{cis}-isomer efficiency, $\xi$, for \textit{azo} compounds with \textit{``long-chain''} alkyl fragments, such as ChD-4211, ChD-4212 and ChD-3805.

  \section*{Conclusions}
In this study, the photosensitive blue phase (BP) mixtures based on cholesteryl oleyl carbonate (COC) and nematic E7 doped with either chiral or achiral \textit{azo} compounds were investigated. It was demonstrated that the photosensitive \textit{azo} dopants broaden the BP temperature range. This broadening effect depends on both the impact of UV irradiation and the chemical structure of the \textit{azo} compounds, which can be assembled from modular "LEGO" molecular units. Furthermore, it was found that the \textit{cis}-isomers are more effective in broadening the BP temperature range compared to the \textit{trans}-isomers.

\bibliography{library} 
\bibliographystyle{unsrtnat}




\clearpage
\begin{center}
   \fontsize{17pt}{20pt}\selectfont \textbf{Supporting Information} 
\end{center}

\setcounter{section}{0}
\setcounter{figure}{0}
\counterwithin{figure}{section}
\numberwithin{table}{section}
\renewcommand{\thesection}{S\arabic{section}}
\renewcommand{\thetable}{\thesection.\arabic{table}}


\begin{center}
    \linespread{1.2}\selectfont 
    {\large \textbf{Broadening the temperature range of blue phases using \\ \textit{azo} compounds of various molecular geometries assembled from modular ``LEGO'' molecular units}}
\end{center}

\begin{center}
{Igor Gvozdovskyy\textsuperscript{1,}\footnote[1]{igvozd@gmail.com}, Vitalii Chornous\textsuperscript{2,}\footnote[2]{chornous.vitalij@bmsu.edu.ua}, Halyna Bogatyryova\textsuperscript{1}, Oleksandr Samoilov\textsuperscript{3}, Longin Lisetski\textsuperscript{3,}\footnote[3]{lcsciencefox@gmail.com}, Serhiy Ryabukhin\textsuperscript{4,5}, Yurii Dmyriv\textsuperscript{5,6} and Mykhaylo Vovk\textsuperscript{7}}
\end{center}

\section{Synthesis and chemical characterization of \textit{azo} compounds}

In this section, we describe the syntheses of various chiral and achiral \textit{azo} compounds. The initial materials and chemical reagents were obtained from Enamine Ltd (Kyiv, Ukraine). Syntheses were carried out using the “p.a.” grade solvents. 
NMR spectra were recorded on the Varian VXR-400(500) instrument (400~MHz for 1H and 125.7, 150.8~MHz for $^{13}$C~NMR) in DMSO-$d_6$ and \ce{CDCl3} solutions, using TMS as an internal standard. Chemical shifts ($\delta$) and \textit{J} values of the spectra are given in ppm and Hz, respectively. Structures of spectral lines are abbreviated as follows: s~(singlet), d~(doublet), t~(triplet), td~(triplet of doublet) and m~(multiplet).
LC-MS spectra of compounds were recorded by means of an Agilent 1100 Series high-performance liquid chromatograph (Hewlett-Packard, California, USA) equipped with a diode array detector with an Agilent LC/MSD SL mass selective detector. Melting points of synthesized compounds were determined using the polarizing optical microscope (POM) BioLar PI (Warsaw, Poland) equipped with a thermostable heating stage based on a MikRa 603 temperature regulator (LLD ``MikRa'', Kyiv, Ukraine) and a Pt1000 platinum resistance thermometer (PJSC ``TERA'', Chernihiv, Ukraine).
The synthesis of the first group of \textit{azo} compounds was performed according to the following scheme:

\begin{figure}[!ht]
\begin{adjustwidth}{-1in}{0in}
\centering
  \includegraphics[width=12cm]{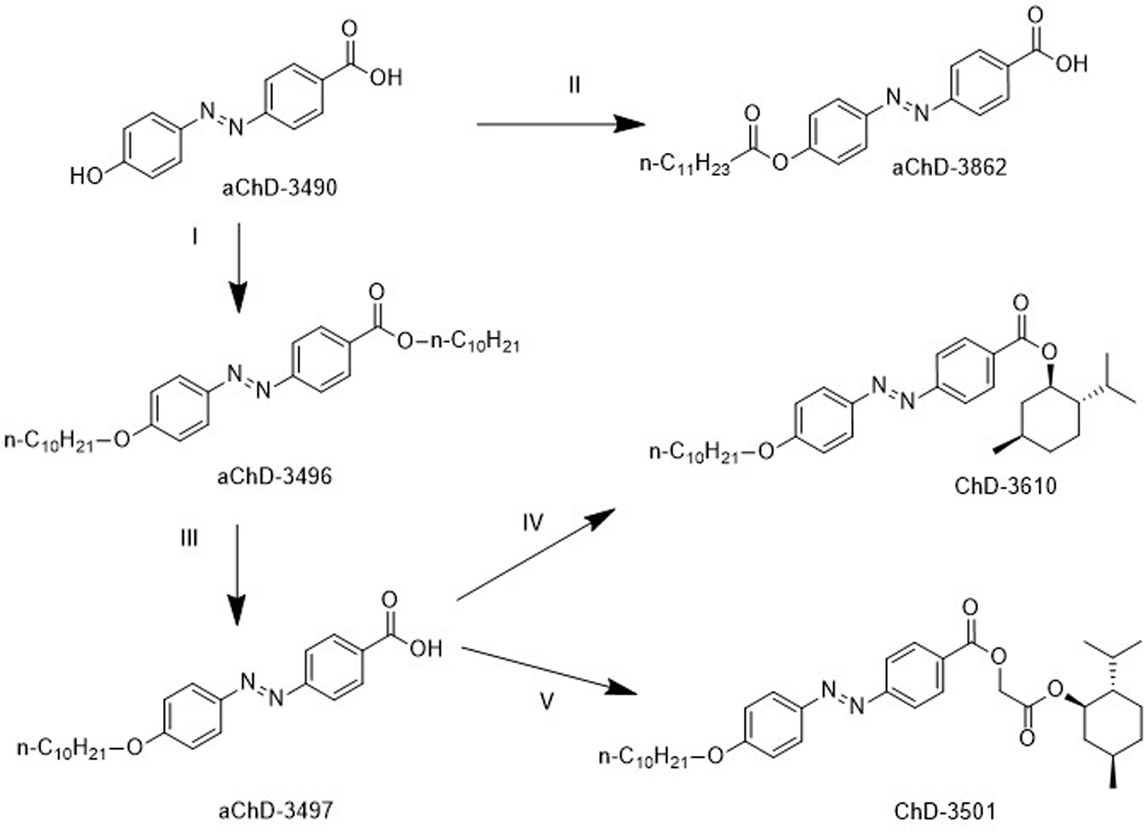}
  \caption{Synthesis of the first group of \textit{azo} compounds (aChD-3862, ChD-3610 and ChD-3501). \textbf{I}: \ce{n-C10H21Br}, \ce{K2CO3}, DMF, 6~h, reflux; \textbf{II}: \ce{n-C11H23COCl}, \ce{Et3N}, \ce{CH2Cl2}, 3~h, reflux; \textbf{III}: a) EtOH, KOH, 10~h, reflux, b) $\text{H}^+$; \textbf{IV}: a) \ce{SOCl2}, \ce{CH2Cl2}, 1~h reflux; b) $\textit{l}$-menthol, \ce{Et3N}, 3~h reflux; \textbf{V}: \ce{ClCH2COO(l-Menthyl)}, \ce{K2CO3}, DMF, 12~h r.t.}
  \label{fgr:Figure1S1}
  \end{adjustwidth}
\end{figure} 

Notably, the synthesis and characterization of compounds ChD-3490, ChD-3496, ChD-3497 and ChD-3501 were previously reported in Ref \cite{ChornousGvozd2022}.
\bigskip

\noindent \textbf{\seqsplit{(1R,2S,5R)-2-isopropyl-5-methylcyclohexyl 4-((E)-(4-(decyloxy)phenyl)diazenyl)benzoate} (ChD-3610 for short).}\smallskip

3.82~g (0.01~mol) of ChD-3497, 100~ml of dry toluene and 1 drop of DMF were added to a 500~ml reactor equipped with a magnetic stirrer. 1.8~g (0.015~mol) of \ce{SOCl2} were added dropwise to the reaction mixture with stirring and external cooling. Then the reaction mixture was heated under reflux within 2~hours to complete gas evolution and after this the solvent was evaporated under reduced pressure. The residue was dissolved in 100~ml of acetonitrile, 1.56~g (0.01~mol) of \textit{l}-menthol was added, and to 1.333~g (0.013~mol) of triethylamine was added dropwise to the resulting mixture with stirring and external cooling. The mixture was heated under reflux within 2~hours, and the solvent was evaporated under reduced pressure and washed with water. The precipitate was crystallized from ethylacetate.\smallskip

M.p. $75.7^\circ\text{C}$. \textbf{\ce{^{1}H}~NMR} (400~MHz, Chloroform-d) $\delta$ 8.21 (d, \textit{J} = 8.4~Hz, 2H), 8.01 (d, \textit{J} = 8.7~Hz, 2H), 7.96 (d, \textit{J} = 8.4~Hz, 2H), 7.26 (d, \textit{J} = 8.4~Hz, 2H), 4.96 (td, \textit{J} = 10.8, 4.2~Hz, 1H), 4.01 (t, \textit{J} = 7.5~Hz, 2H), 2.21--2.14 (m, 1H), 2.04-1.96 (m, 1H), 1.85-1.72 (m, 4H), 1.65--1.56 (m, 4H), 1.50-1.26 (m, 13H), 1.20--1.10 (m, 2H), 1.04-0.84 (m, 9H), 0.82 (d, \textit{J} = 6.9~Hz, 3H). $^{13}$C~NMR (101~MHz, Chloroform-$d$) $\delta$ 171.90, 154.94, 153.32, 150.08, 132.57, 130.57(2C), 124.38(2C), 122.61(2C), 122.34(2C), 75.23, 67.9, 47.28, 40.97, 34.43, 34.31, 31.87, 31.47, 29.42, 29.27 (2C), 29.11, 26.57, 24.89, 23.67, 22.68, 22.06, 20.78, 16.57, 14.13. MS, m/z (\%): 521(100) [M+1]. 
\bigskip

\noindent \textbf{\seqsplit{(E)-4-(4-(dodecanoyloxy)phenyl)diazenyl)benzoic acid} (aChD-3862 for short).}\smallskip

To the mixture 2.4~g (0.01~mol) of ChD-3490 and 1~g (0.01~mol) \ce{Et3N} in 100~ml of dry \ce{Ch2Cl2}, 2.2~g (0.01~mol) \ce{n-C9H19COCl} were added dropwise. After the reaction, this mixture was heated under reflux for 3~hours. Further the solvent was evaporated under reduced pressure and washed with water. The compound aChD-3862 was crystallized from acetic acid. The small orange-white solid crystals of aChD-3862 with yield of 65~\% were obtained.\medskip

M.p. $248.7^\circ\text{C}$. \textbf{\ce{^{1}H}~NMR} (400~MHz, DMSO-$d_6$): $\delta$ 12.11 (s, 1H), 8.21(d, \textit{J} = 8.4~Hz, 2H), 8.02 (d, \textit{J} = 8.7~Hz, 2H), 7.97 (d, \textit{J} = 8.4~Hz, 2H), 7.27 (d, \textit{J} = 8.8~Hz, 2H), 2.56 (t, \textit{J} = 7.5~Hz, 2H), 1.36--1.30 (m, 2H), 1.26--1.15 (m, 16H), 0.89 (d, \textit{J} = 6.9~Hz, 3H).  $^{13}$C~NMR (151~MHz, DMSO-$d_6$): $\delta$ 172.4, 167.3, 157.7, 157.1, 152.4, 130.9, 130.2, 130.0, 128.5, 126.3, 124.2, 123.9, 122.6, 122.4, 33.5, 29.7, 28.4, 28.1, 27.6, 27.2, 26.8, 25.6, 24.3, 22.4, 14.7. LC-MS, m/z (\%): 425(100) [M+1].\bigskip

The synthesis of the second group of \textit{azo} compounds was performed according to the following scheme:

\begin{figure}[!ht]
\begin{adjustwidth}{-1in}{0in}
\centering
  \includegraphics[width=12cm]{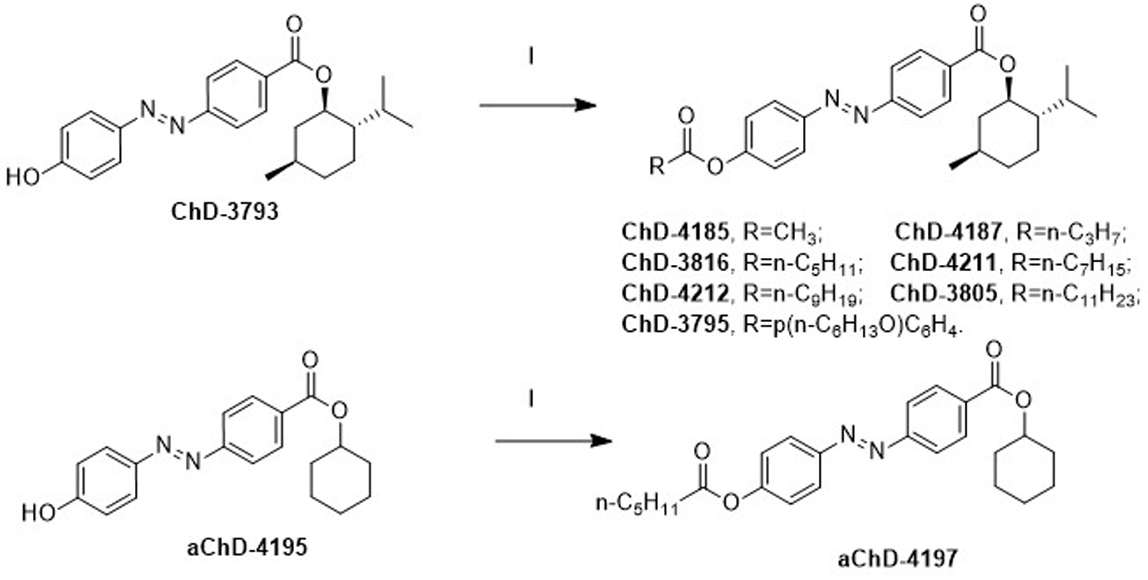}
  \caption{Synthesis of the second group of \textit{azo} compounds (ChD-4185, ChD-4187, ChD-3816, ChD-4211, ChD-4212, ChD-3805, ChD-3795 and aChD-4197). \textbf{I}: RCOCl, ChD-3673 or aChD-4195, \ce{Et3N}, acetonitrile, 3~h reflux.}
  \label{fgr:Figure1S2}
  \end{adjustwidth}
\end{figure} 

The synthesis, reaction conditions and characterization of compound ChD-3793 and ChD-3816 were previously reported in \cite{ChornousGvozd2023}. All other compounds were synthesized according to the method reported for the synthesis of ChD-3816 \cite{ChornousGvozd2023}.

To the solution of (0.03~mol) of compound ChD-3673 or aChD-4195 and 3~g (0.03~mol) of triethylamine in 50~ml of acetonitrile, 0.03~mol of alkanoyl chloride was added. The reaction mixture was heated under reflux within 2~hours; the solvent was evaporated under reduced pressure. The red precipitate was filtered off, dried and purified by silica gel column chromatography. Hexane was used as eluent. The product was obtained as yellow plates.\bigskip

\noindent \textbf{\seqsplit{(1R,2S,5R)-2-isopropyl-5-methylcyclohexyl-(E)-4-((4-((4-(hexyloxy)benzoyl)oxy)phenyl)diazenyl)benzoate} (ChD-3795 for short).}
\smallskip

M.p. $144.3^\circ\text{C}$. \textbf{\ce{^{1}H}~NMR} (400~MHz, Chloroform-$d$) $\delta$ 8.20 (d, \textit{J} = 8.4 Hz, 2H), 8.00--7.94 (m, 4H), 7.90--7.84 (m, 4H), 7.27 (d, \textit{J} = 8.4 Hz, 2H), 4.98 (td, \textit{J} = 10.8, 4.2~Hz, 1H), 4.04 (t, \textit{J} = 7.5~Hz, 2H), 2.20--2.14 (m, 1H), 2.06--1.94 (m, 1H), 1.85--1.70 (m, 2H), 1.65--1.55 (m, 2H), 1.51--1.24 (m, 11H), 1.22-1.10 (m, 2H), 1.02--0.87 (m, 9H), 0.79 (d, \textit{J} = 6.9~Hz, 3H). $^{13}$C~NMR (101~MHz, Chloroform-$d$) $\delta$ 171.91, 165.50, 154.95, 153.34, 150.06(2), 132.58(2), 130.55(2C), 124.38(2C), 122.60(2C), 122.32(2C), 75.24, 68.3, 47.26, 40.97, 34.43, 34.33, 31.65, 31.48, 29.07, 28.92, 26.56(2C), 24.84(2C), 23.68, 22.61, 22.05, 20.75, 16.56, 14.07. MS, m/z (\%): 585(100) [M+1].\bigskip

\noindent\textbf{\seqsplit{(1R,2S,5R)-2-isopropyl-5-methylcyclohexyl-(E)-4-((4-(dodecanoyloxy)phenyl)diazenyl)benzoate} (ChD-3805 for short).}
\smallskip 

M.p. $58.6^\circ\text{C}$. \textbf{\ce{^{1}H}~NMR} (400~MHz, Chloroform-$d$) $\delta$ 8.21 (d, \textit{J} = 8.4~Hz, 2H), 8.02 (d, \textit{J} = 8.7~Hz, 2H), 7.96 (d, \textit{J} = 8.4~Hz, 2H), 7.28 (d, \textit{J} = 8.4~Hz, 2H), 4.98 (td, \textit{J} = 10.8, 4.2~Hz, 1H), 2.62 (t, \textit{J} = 7.5~Hz, 2H), 2.22--2.14 (m, 1H), 2.04--1.98 (m, 1H), 1.86--1.72 (m, 2H), 1.65--1.54 (m, 6H), 1.52-1.24 (m, 15H), 1.20--1.10 (m, 2H), 1.02--0.86 (m, 9H), 0.81 (d, \textit{J} = 6.9~Hz, 3H). $^{13}$C~NMR (101~MHz, Chloroform-$d$) $\delta$ 171.92, 165.54, 154.90, 153.32, 150.06, 132.53, 130.57(2C), 124.38(2C), 122.61(2C), 122.32(2C), 75.22, 47.28, 40.95, 34.46, 34.31, 31.85, 31.47, 29.44, 29.27 (2C), 29.11, 26.55(2C), 24.87(2C), 23.67, 22.68, 22.04, 20.78, 16.55, 14.12. MS, m/z (\%): 563(100) [M+1].\bigskip

\noindent\textbf{\seqsplit{(1R,2S,5R)-2-isopropyl-5-methylcyclohexyl 4-((E)-(4-acetoxyphenyl)diazenyl)benzoate} (ChD-4185 for short)}.\smallskip

M.p. $89.7^\circ\text{C}$. \textbf{\ce{^{1}H}~NMR} (400~MHz, Chloroform-$d$) $\delta$ 8.20 (d, \textit{J} = 8.4~Hz, 2H), 8.01 (d, \textit{J} = 8.7~Hz, 2H), 7.95 (d, \textit{J} = 8.4~Hz, 2H), 7.29 (d, \textit{J} = 8.8~Hz, 2H), 4.98 (td, \textit{J} = 10.8, 4.2~Hz, 1H), 2.36 (s, 3H), 2.20--2.14 (m, 1H), 2.04--1.96 (m, 1H), 1.80--1.72 (m, 2H), 1.65--1.55 (m, 2H), 1.22--1.10 (m, 2H), 1.00--0.92 (m, 7H), 0.83 (d, \textit{J} = 6.9~Hz, 3H). $^{13}$C~NMR (101~MHz, Chloroform-$d$) $\delta$ 169.04, 165.52, 154.92, 153.16, 150.15, 132.60, 130.58(2C), 124.41(2C), 122.62(2C), 122.34(2C), 75.25, 47.27, 40.97, 34.31, 31.47, 26.56, 23.66, 22.06, 21.20, 20.79, 16.56. LC-MS, m/z (\%): 423(100) [M+1].\bigskip

\noindent\textbf{\seqsplit{(1R,2S,5R)-2-isopropyl-5-methylcyclohexyl (E)-4-((4-(butyryloxy)phenyl)diazenyl)-benzoate} (ChD-4187 for short)}.\smallskip

M.p. $77.6^\circ\text{C}$. \textbf{\ce{^{1}H}~NMR} (400~MHz, Chloroform-$d$) $\delta$ 8.17 (d, J = 8.4~Hz, 2H), 7.97 (d, \textit{J} = 8.7~Hz, 2H), 7.92 (d, \textit{J} = 8.4~Hz, 2H), 7.28 (d, \textit{J} = 8.8~Hz, 2H), 4.95 (td, \textit{J} = 10.8, 4.2~Hz, 1H), 2.58 (t, \textit{J} = 7.5~Hz, 2H), 2.15--2.12 (m, 1H), 1.97--1.95 (m, 1H),  1.84-1.71 (m, 4H), 1.59--1.54 (m, 2H), 1.15--1.10 (m, 2H), 1.06--1.04 (m, 3H), 0.94--0.90 (m, 7H), 0.79 (d, \textit{J} = 6.9~Hz, 3H). $^{13}$C~NMR (126~MHz, Chloroform-$d$) $\delta$ 171.68, 165.51, 154.98, 153.32, 150.12, 132.61, 130.57(2C), 124.37(2C), 122.60(2C), 122.34(2C), 75.26, 47.31, 40.96, 36.27, 34.34, 31.48, 26.60, 23.70, 22.04, 20.77, 18.41, 16.59, 13.63. MS, m/z (\%): 451(100) [M+1].\bigskip

\noindent\textbf{\seqsplit{Cyclohexyl (E)-4-((4-(hexanoyloxy)phenyl)diazenyl)benzoate} (aChD-4197 for short)}.\smallskip

M.p. $66.4^\circ\text{C}$. \textbf{\ce{^{1}H}~NMR} (400~MHz, Chloroform-$d$) $\delta$ 8.17 (d, \textit{J} = 8.4~Hz, 2H), 7.97 (d, \textit{J} = 8.7~Hz, 2H), 7.91 (d, \textit{J} = 8.4~Hz, 2H), 7.24 (d, \textit{J} = 8.4~Hz, 2H), 5.04 (tt, \textit{J} = 8.3, 3.5~Hz, 1H), 2.57 (t, \textit{J} = 7.5~Hz, 2H), 1.97--1.94 (m, 2H), 1.80--1.72 (m, 4H), 1.66--1.52 (m, 3H), 1.51--1.30 (m, 7H), 0.92 (t, \textit{J} = 6.9~Hz, 3H). $^{13}$C~NMR (126~MHz, Chloroform-$d$) $\delta$ 171.85, 165.38, 154.96, 153.34, 150.12, 132.78, 130.54(2C), 124.36(2C), 122.57(2C), 122.32(2C), 73.46, 34.40, 31.66(2C), 31.26, 25.48, 24.56, 23.69(2C), 22.31, 13.89. MS, m/z (\%): 423(100) [M+1].\bigskip

\noindent\textbf{\seqsplit{(1R,2S,5R)-2-isopropyl-5-methylcyclohexyl (E)-4-((4-(octanoyloxy)phenyl)diazenyl)-benzoate} (ChD-4211 for short)}. \smallskip

M.p. $63.4^\circ\text{C}$. \textbf{\ce{^{1}H}~NMR} (400~MHz, Chloroform-$d$) $\delta$ 8.20 (d, \textit{J} = 8.4~Hz, 2H), 8.00 (d, \textit{J} = 8.7~Hz, 2H), 7.95 (d, \textit{J} = 8.4~Hz, 2H), 7.27 (d, \textit{J} = 8.4~Hz, 2H), 4.98 (td, \textit{J} = 10.8, 4.2~Hz, 1H), 2.60 (t, \textit{J} = 7.5~Hz, 2H), 2.20--2.14 (m, 1H), 2.06--1.94 (m, 1H), 1.85--1.70 (m, 2H), 1.65--1.55 (m, 2H), 1.51--1.24 (m, 11H), 1.22--1.10 (m, 2H), 1.02--0.87 (m, 9H), 0.79 (d, \textit{J} = 6.9~Hz, 3H). $^{13}$C~NMR (101~MHz, Chloroform-$d$) $\delta$ 171.91, 165.52, 154.95, 153.32, 150.09, 132.58, 130.57(2C), 124.38(2C), 122.60(2C), 122.34(2C), 75.24, 47.28, 40.97, 34.43, 34.31, 31.65, 31.47, 29.06, 28.92, 26.56, 24.88, 23.67, 22.61, 22.05, 20.77, 16.56, 14.07. MS, m/z (\%): 507(100) [M+1].\bigskip

\noindent\textbf{\seqsplit{(1R,2S,5R)-2-isopropyl-5-methylcyclohexyl (E)-4-((4-(decanoyloxy)phenyl)diazenyl)-benzoate} (ChD-4212 for short)}.\smallskip

M.p. $60.8^\circ\text{C}$. \textbf{\ce{^{1}H}~NMR} (400~MHz, Chloroform-$d$) $\delta$ 8.20 (d, \textit{J} = 8.4~Hz, 2H), 8.00 (d, \textit{J} = 8.7~Hz, 2H), 7.95 (d, \textit{J} = 8.4~Hz, 2H), 7.28 (d, \textit{J} = 8.4~Hz, 2H), 4.99 (td, \textit{J} = 10.8, 4.2~Hz, 1H), 2.60 (t, \textit{J} = 7.5~Hz, 2H), 2.21--2.14 (m, 1H), 2.04--1.96 (m, 1H), 1.85--1.70 (m, 2H), 1.65--1.55 (m, 4H), 1.50--1.24 (m, 13H), 1.20-1.10 (m, 2H), 1.04--0.86 (m, 9H), 0.80 (d, \textit{J} = 6.9~Hz, 3H). $^{13}$C~NMR (101~MHz, Chloroform-$d$) $\delta$ 171.90, 165.51, 154.94, 153.32, 150.08, 132.57, 130.57(2C), 124.38(2C), 122.61(2C), 122.34(2C), 75.23, 47.28, 40.97, 34.43, 34.31, 31.87, 31.47, 29.42, 29.27 (2C), 29.11, 26.57, 24.89, 23.67, 22.68, 22.06, 20.78, 16.57, 14.13. MS, m/z (\%): 535(100) [M+1].\bigskip

The melting points of various chiral \textit{azo} compounds (\textit{i.e.}, ChD-4185, ChD-4187, ChD-3816 \cite{ChornousGvozd2023}, ChD-4211, ChD4112 and ChD-3805) with different numbers of carbon atoms in their alkyl chains are shown in Fig.~S1.3.

\begin{figure}[!ht]
\begin{adjustwidth}{-1in}{0in}
\centering
  \includegraphics[width=9cm]{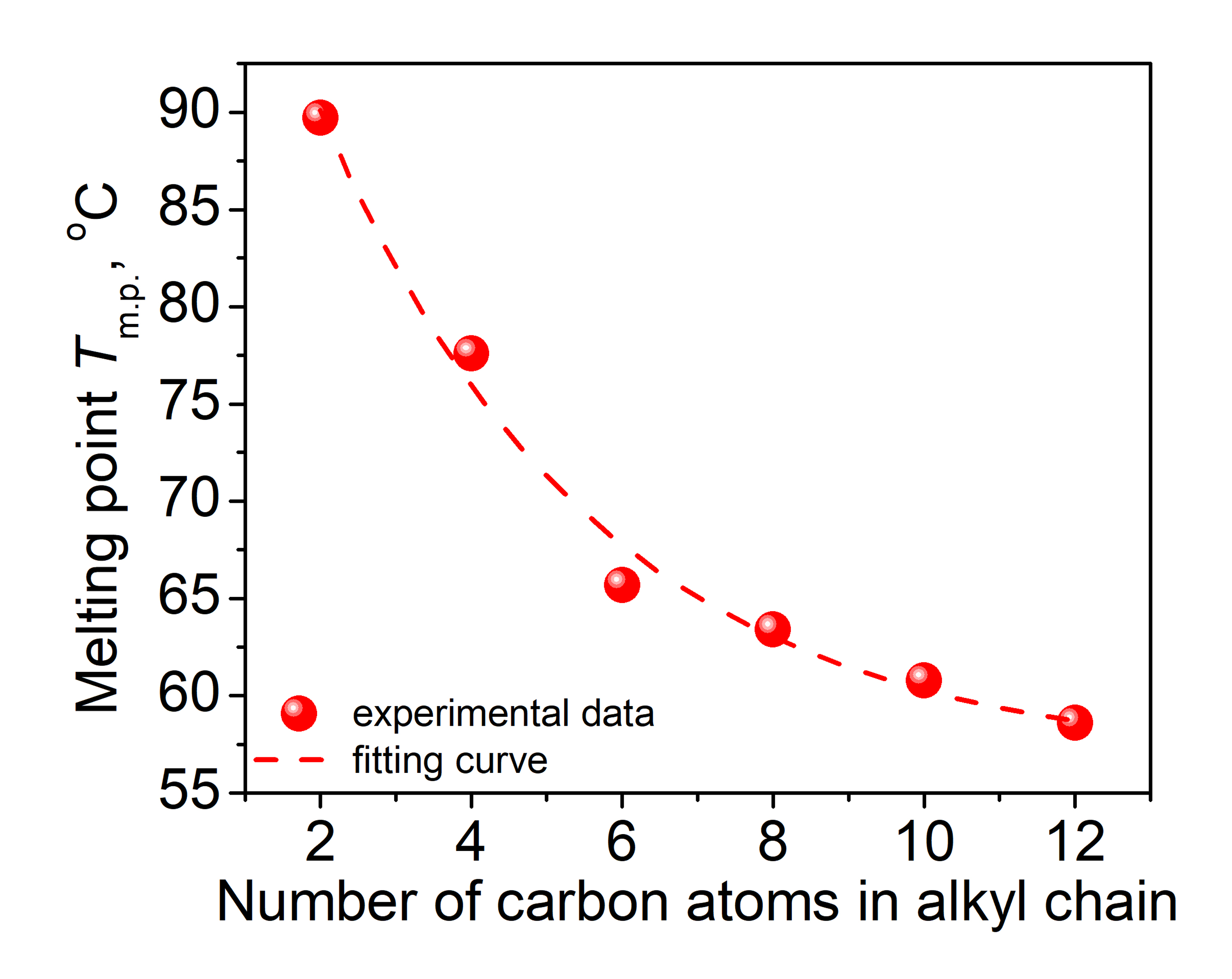}
  \caption{Melting points of various chiral \textit{azo} compounds (\textit{i.e.}, ChD-4185, ChD-4187, ChD-3816\cite{ChornousGvozd2023}, ChD-4211, ChD-4112 and ChD-3805) as function of the number of carbon atoms in the alkyl chain \ce{C_n}.}
  \label{fgr:Figure1S3}
  \end{adjustwidth}
\end{figure} 
\newpage

\section{Characterization of the base highly chiral mixture: phase transitions and spectral properties of the blue phases}

The base highly chiral mixture (HChM) was prepared by mixing 65~wt\% cholesteryl oleyl carbonate (COC) and 35~wt\% nematic E7. Typical spectra of the HChM in the isotropic (Iso), blue phases (BP-II and BP-I) and cholesteric phase ($\text{N}^*$) are shown in Fig.~S2.1.

\begin{figure}[!ht]
\begin{adjustwidth}{-1in}{0in}
\centering
  \includegraphics[width=9cm]{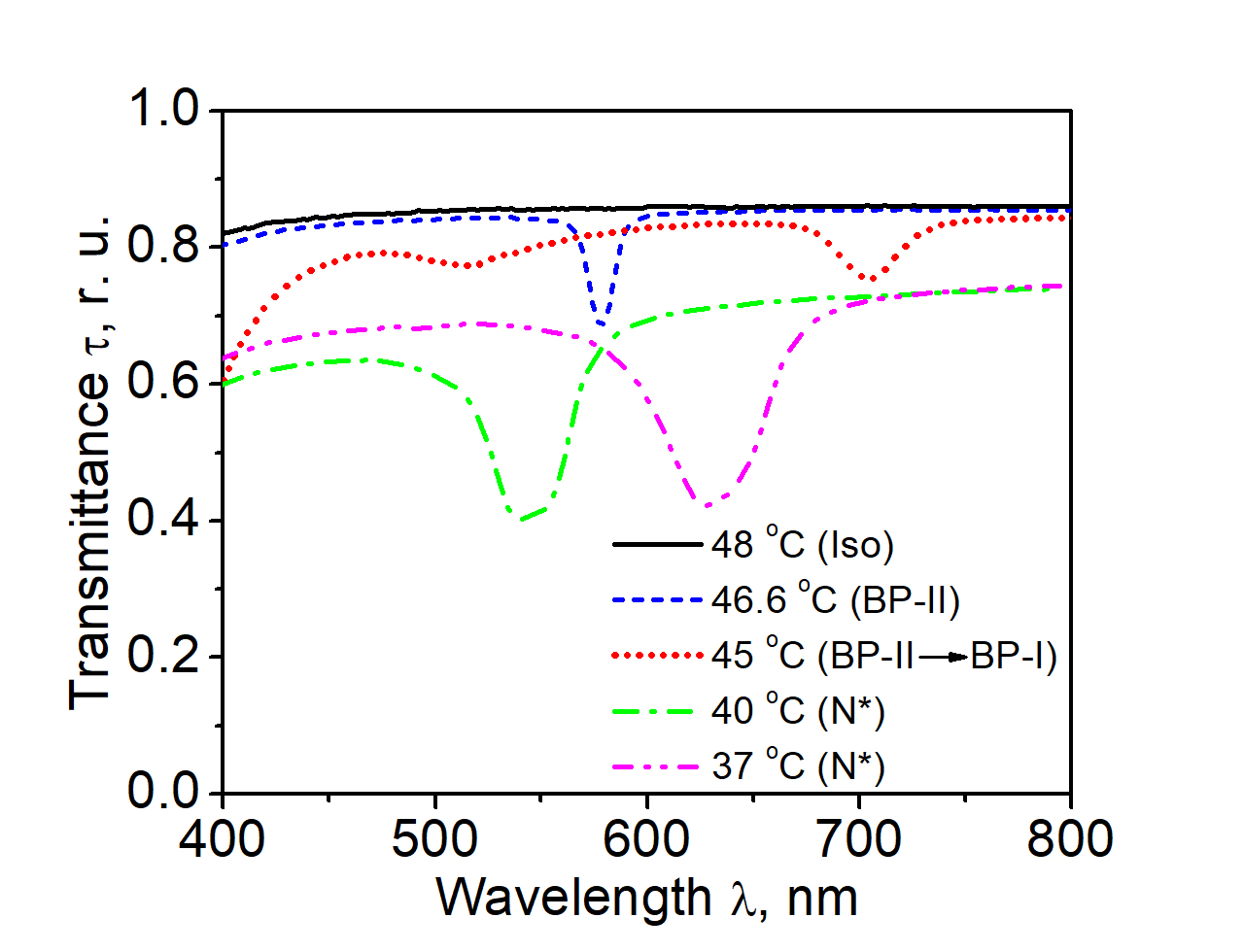}
  \caption{\textbf{Transmission spectra of the base HChM (65~wt\% COC and 35~wt\% E7) at various temperatures: (1) isotropic phase (Iso) at $48^\circ\text{C}$ (black); (2) BP-II at $46.6^\circ\text{C}$ (dashed blue) and (3) transition of BP-II to BP-I (dotted red), (4) $\text{N}^*$ at $40^\circ\text{C}$ (dash-dotted green) and (5) $\text{N}^*$ at $37^\circ\text{C}$ (dash-dot-dotted magenta).}}
  \label{fgr:Figure2S1}
  \end{adjustwidth}
\end{figure} 

Fig.~S2.2 shows microphotographs of the sequential textures of phase transitions of the base HChM (65 wt\% COC and 35 wt\% E7) during both heating and cooling processes. HChM is characterized by various phases, \textit{i.e.}, smectic A (SmA), cholesteric ($\text{N}^*$), blue phases (BP-I and BP-II) and isotropic phase (Iso).

\begin{figure}[!ht]
\begin{adjustwidth}{-1in}{0in}
\centering
  \includegraphics[width=15cm]{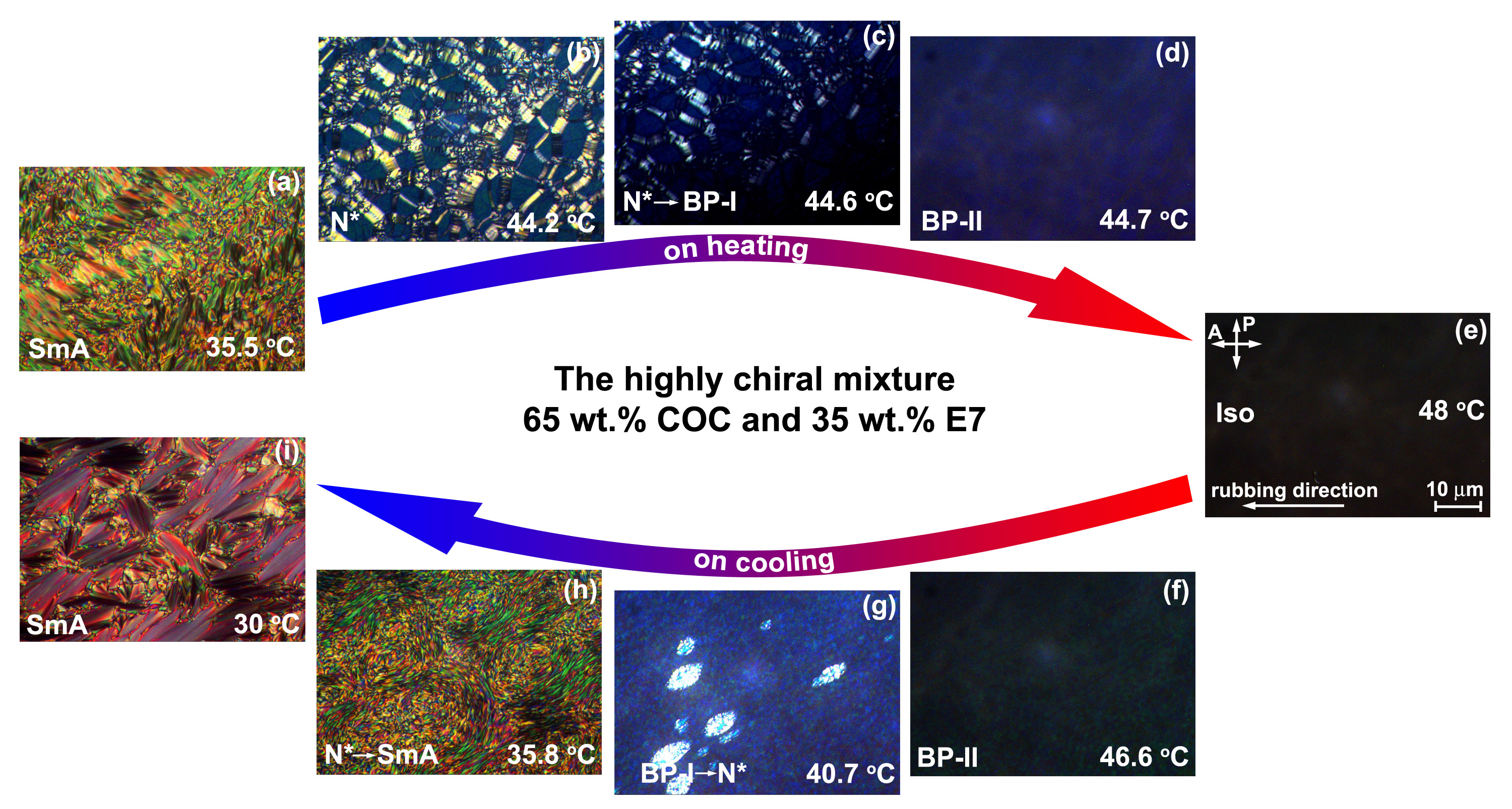}
  \caption{\textbf{Sequential POM textures of the base HChM (65~wt\% COC and 35~wt\% E7) during the heating (a)--(e) and cooling (e)--(i) cycles.}}
  \label{fgr:Figure2S2}
  \end{adjustwidth}
\end{figure} 

\section{Determination of the helical twisting power (HTP) of chiral \textit{azo} compounds}

The helical twisting power of chiral \textit{azo} compounds synthesized in Section~S1 and recently described in Refs. \cite{ChornousGvozd2022,ChornousGvozd2023} was determined as follows:

\begin{equation}
\text{HTP} = \frac{1}{P \cdot C}
\end{equation}

where \textit{P} and \textit{C} are the pitch of cholesteric helix and concentration of chiral \textit{azo} compound in the nematic LC (\textit{e.g.}, the nematic mixture E7), respectively.

First of all, the measurement of \textit{P} for a weakly chiral mixture (\textit{i.e.}, when the concentration of ChD in the nematic host E7 is within the range of 0 to 5~wt\%) was carried out by means of the Grandjean-Cano method as described elsewhere in detail\cite{ChornousGvozd2022}. 

To measure the cholesteric pitch, we used a wedge-shaped cell assembled with glass substrates coated with PI2555, featuring both a thin ($d_0$) and a thick ($d$) end (Fig.~S3.1). The thickness d of the thick end was changed in the range from 18 to 30~$\mu$m. The length \textit{P} of cholesteric helix is related to the number $N_C$ of Grandjean-Cano zones as follows\cite{ChornousGvozd2022}:

\begin{equation}
P = 2 \cdot \frac{d - d_0}{N_{\text{C}}}
\end{equation}

\begin{figure}[!ht]
\begin{adjustwidth}{-1in}{0in}
\centering
  \includegraphics[width=16cm]{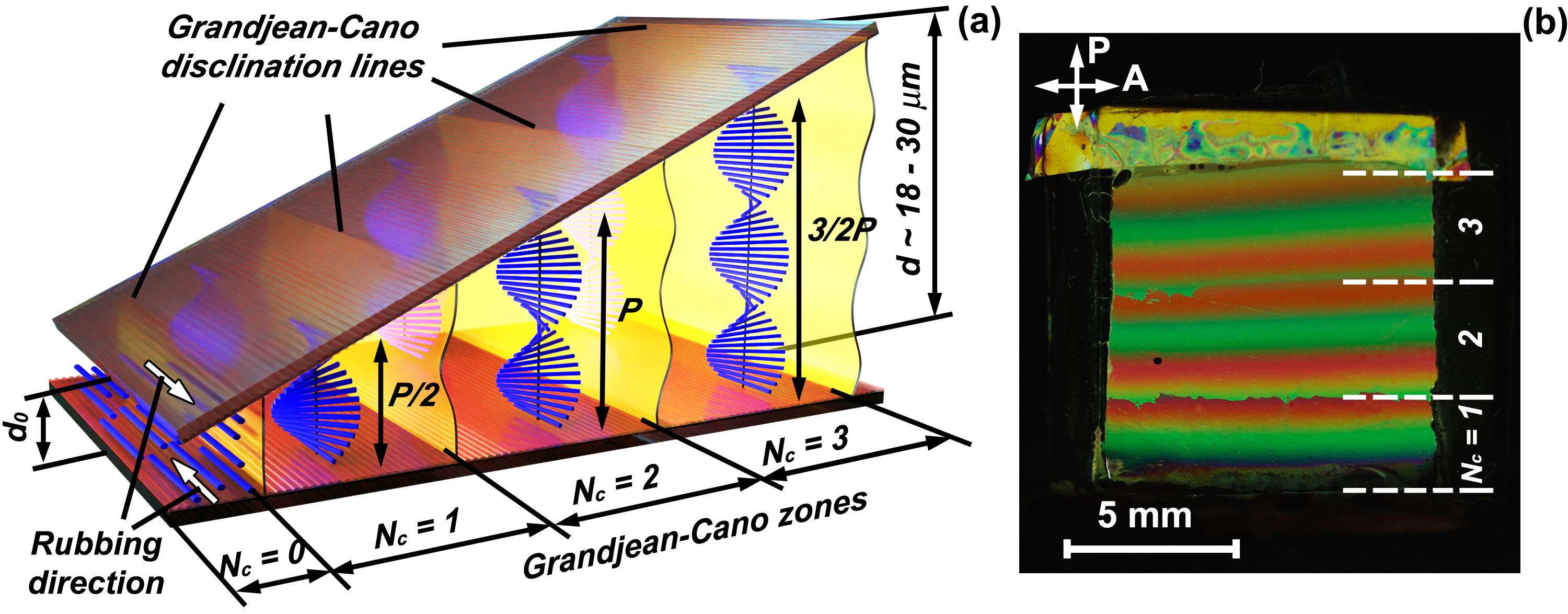}
  \caption{\textbf{(a) Schematic illustration of wedge-like LC cell and (b) the photograph of wedge-like cell filled with cholesteric mixture (\textit{e.g.}, 1.95~wt\% ChD-4187 and 98.05~wt\% E7) between crossed polarizers.}}
  \label{fgr:Figure3S1}
  \end{adjustwidth}
\end{figure} 

The HTP of various chiral \textit{azo} compounds in the nematic host E7 was determined from the slope of the linear dependence of inverse pitch ($1/P$) on dopant concentration $C$ (Fig.~S3.2). In all cases, the linear fits were constrained to pass through the origin (Fig.~S3.2(b)).\cite{ChornousGvozd2022}

\newpage

\begin{figure}[!ht]
\begin{adjustwidth}{-1in}{0in}
\centering
  \includegraphics[width=16cm]{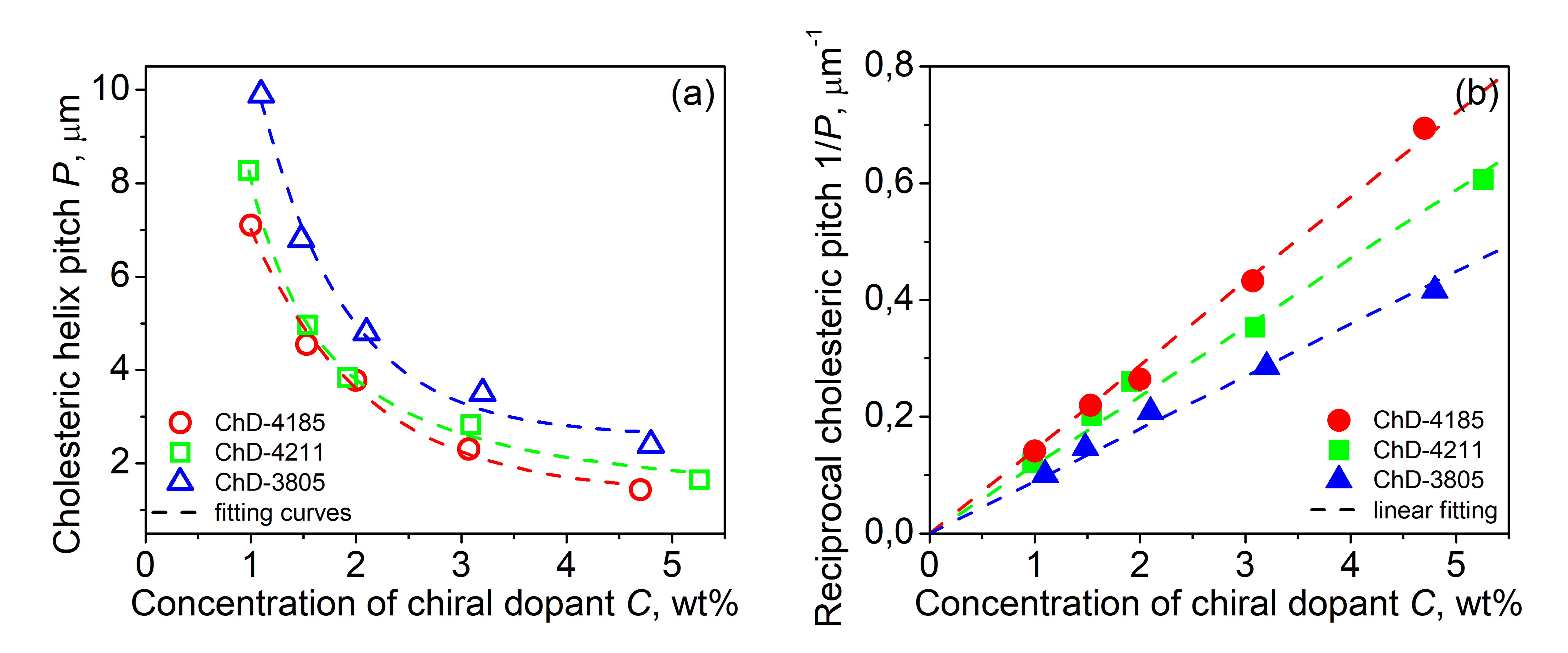}
  \caption{\textbf{Dependence of (a) cholesteric helix pitch and (b) reciprocal cholesteric pitch on concentration of chiral dopants taking ChD-4185 (circles symbols), ChD-4211 (squares symbols) and ChD-3805 (triangles symbols) as examples.}}
  \label{fgr:Figure3S2}
  \end{adjustwidth}
\end{figure} 

The HTP values of various chiral \textit{azo} compounds differed by their structures are listed in Table~S3.1.

\begin{table}[!ht]
\begin{adjustwidth}{-0.3in}{0in}
   \caption{Chiral \textit{azo} compounds as photosensitive dopants added to nematic host E7}
   \medskip
  \label{tbl:Table3S1}
\begin{tabular}{|>{\centering\arraybackslash}m{5cm}|>{\centering\arraybackslash}m{6cm}|>{\centering\arraybackslash}m{2cm}|}
  \hline
  \textbf{\textit{Azo} compound} & \textbf{Chemical formula of chiral molecules in \textit{trans}-isomer} & \shortstack{\textbf{HTP,}\\ \textbf{$\mu$m$^{-1}$}} \\
  \hline
  ChD-3793 & \includegraphics[width=2.8cm, valign=c]{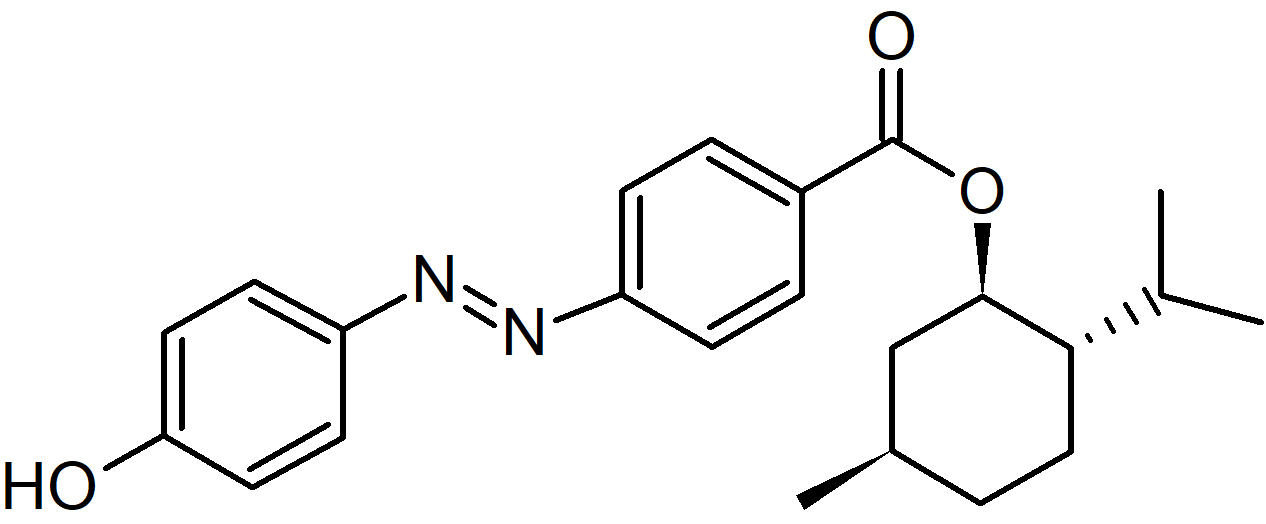} & $-12.8$ \\
  \hline
  ChD-3610 & \includegraphics[width=4.6cm, valign=c]{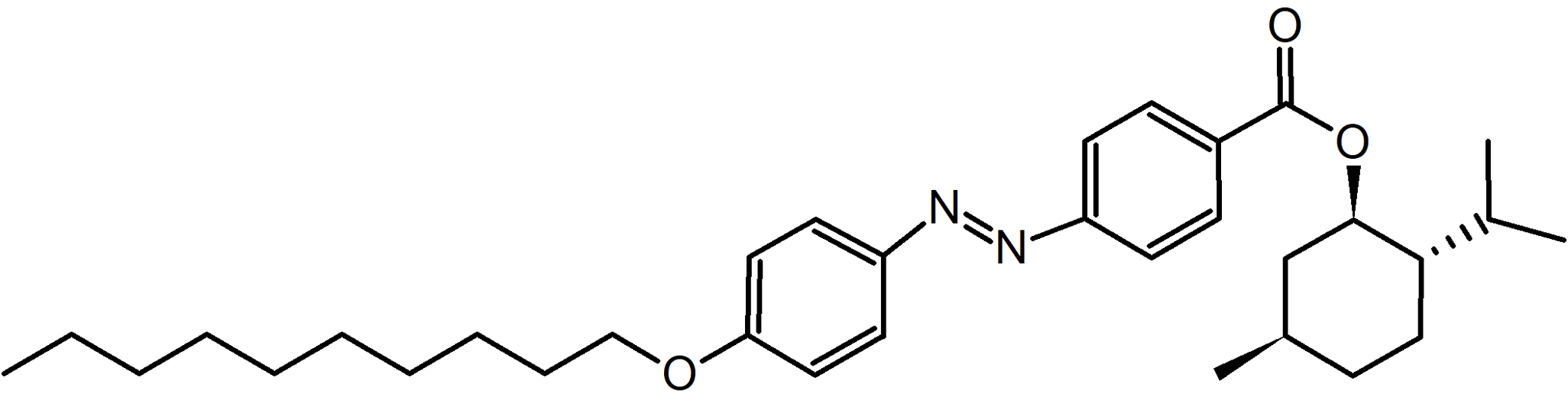} & $-5.8$ \\
  \hline
  ChD-3795 & \includegraphics[width=5cm, valign=c]{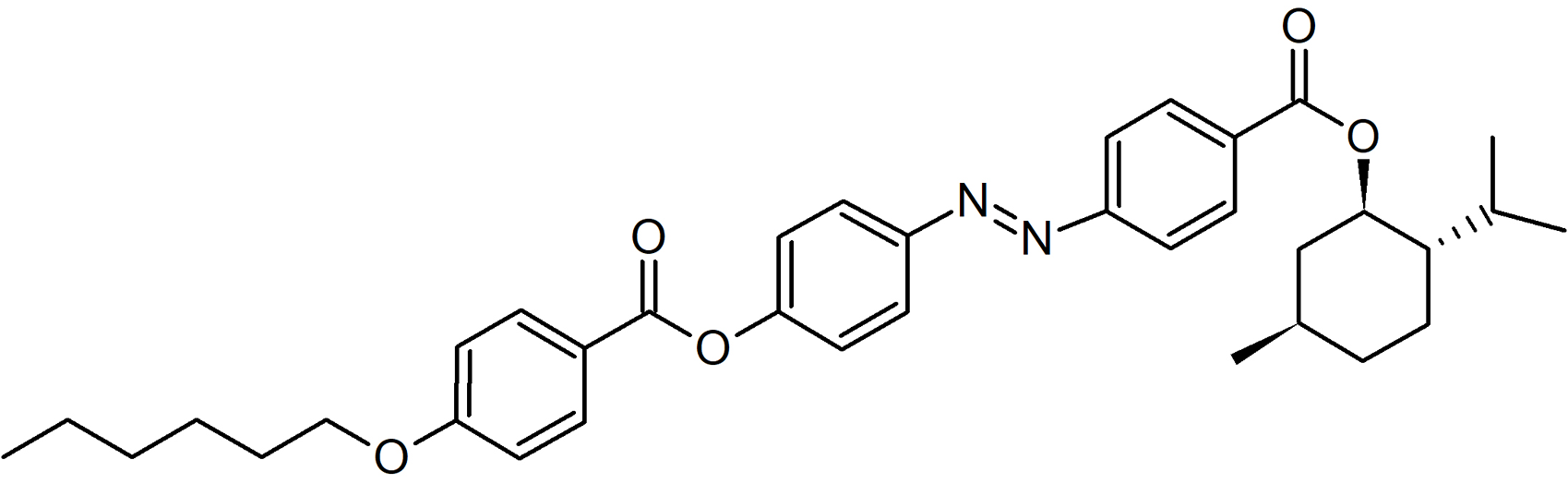} & $-4$ \\
  \hline
  ChD-3805 & \includegraphics[width=4.9cm, valign=c]{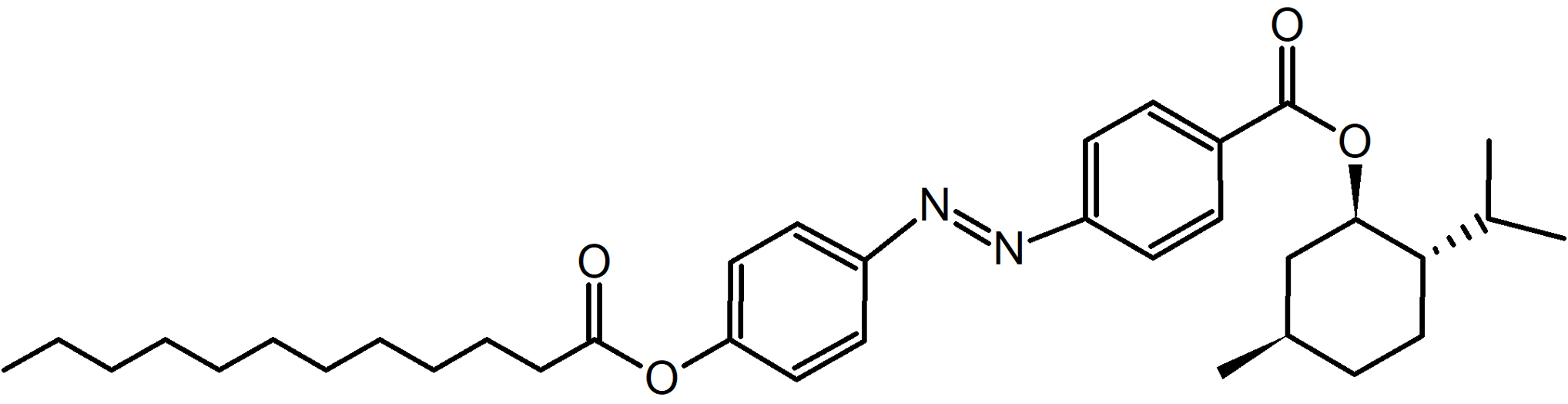} & $-3.95$ \\
  \hline
  ChD-3816 & \includegraphics[width=4cm, valign=c]{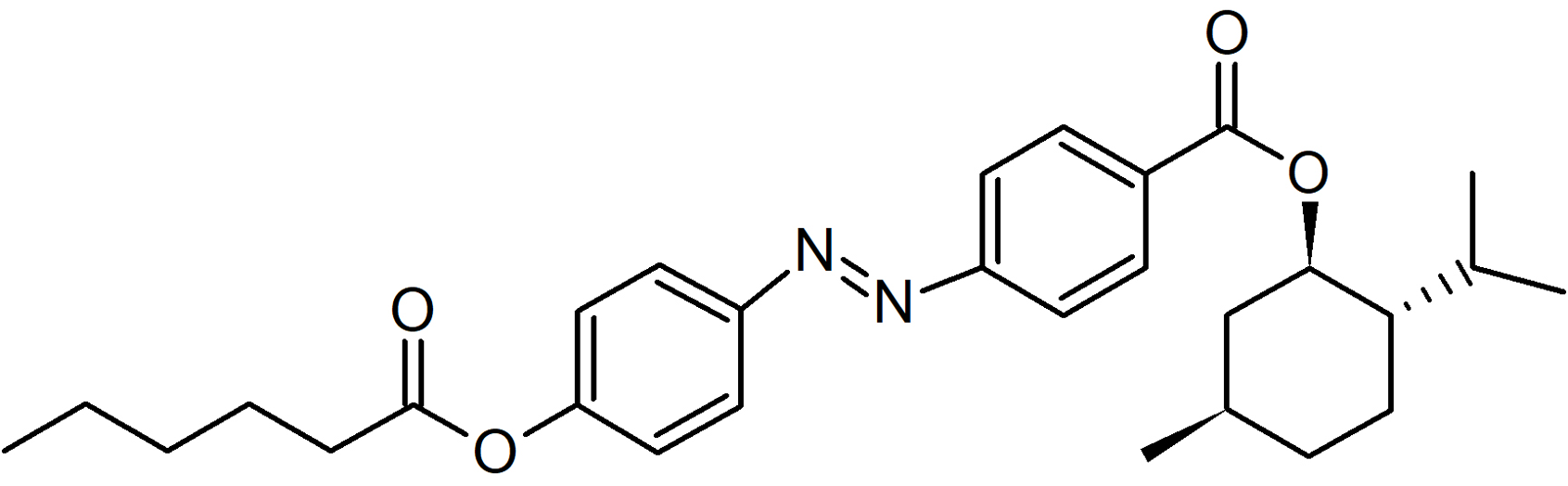} & $-8$ \\
  \hline
  ChD-3501 & \includegraphics[width=5cm, valign=c]{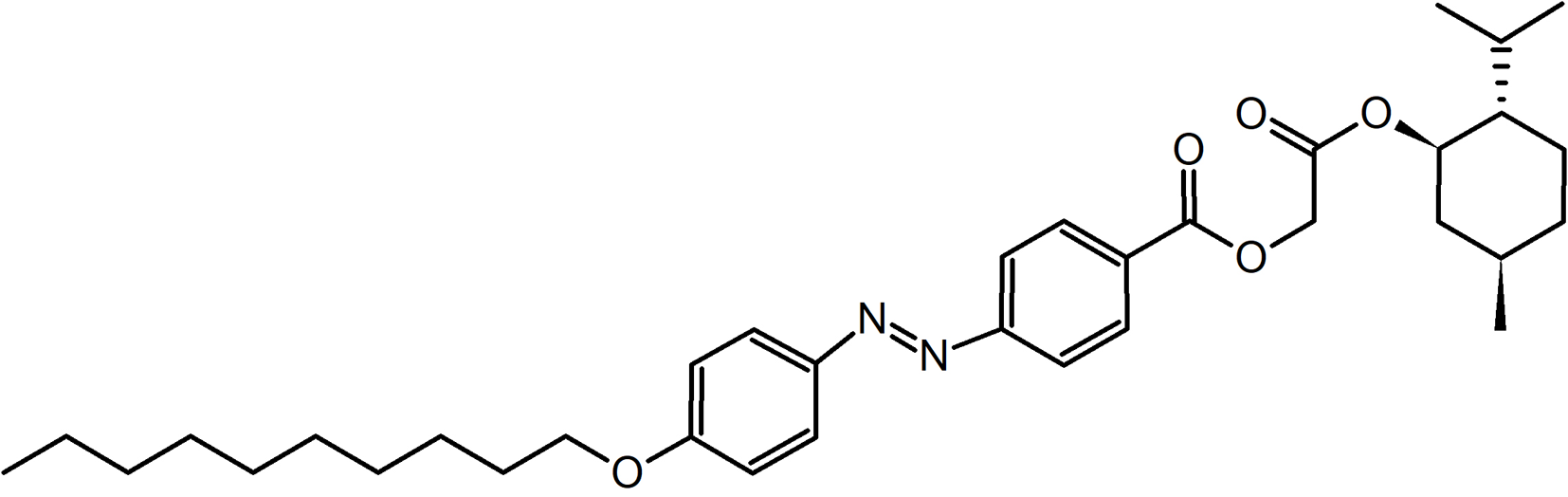} & $-3.3$ \\
  \hline
   \end{tabular}
   \end{adjustwidth}
    \end{table}

\begin{table}[!ht]
\begin{adjustwidth}{-0.3in}{-0in}
   \caption{\ Homologous series of chiral \textit{azo} compounds added to the nematic host E7}
   \medskip
  \label{tbl:Table3S2}
\begin{tabular}{|>{\centering\arraybackslash}m{2.5cm}|>{\centering\arraybackslash}m{6cm}|>{\centering\arraybackslash}m{2cm}|>{\centering\arraybackslash}m{2cm}|}
  \hline
  \textbf{\textit{Azo} compound} & \textbf{Chemical formula of chiral molecules in \textit{trans}-isomer} & \textbf{Number of carbon atoms in alkyl chain} & \shortstack{\textbf{HTP,}\\ \textbf{$\mu$m$^{-1}$}} \\
  \hline
  ChD-4185 & \includegraphics[width=3.1cm, valign=c]{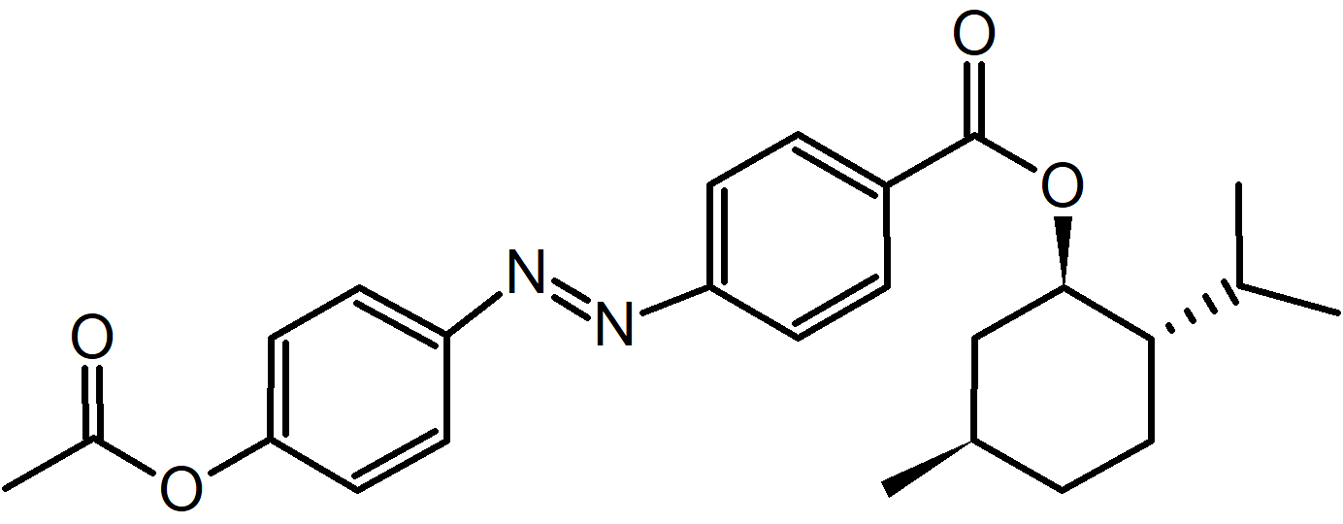} & $2$ & $-8.5$ \\
  \hline
  ChD-4187 & \includegraphics[width=3.4cm, valign=c]{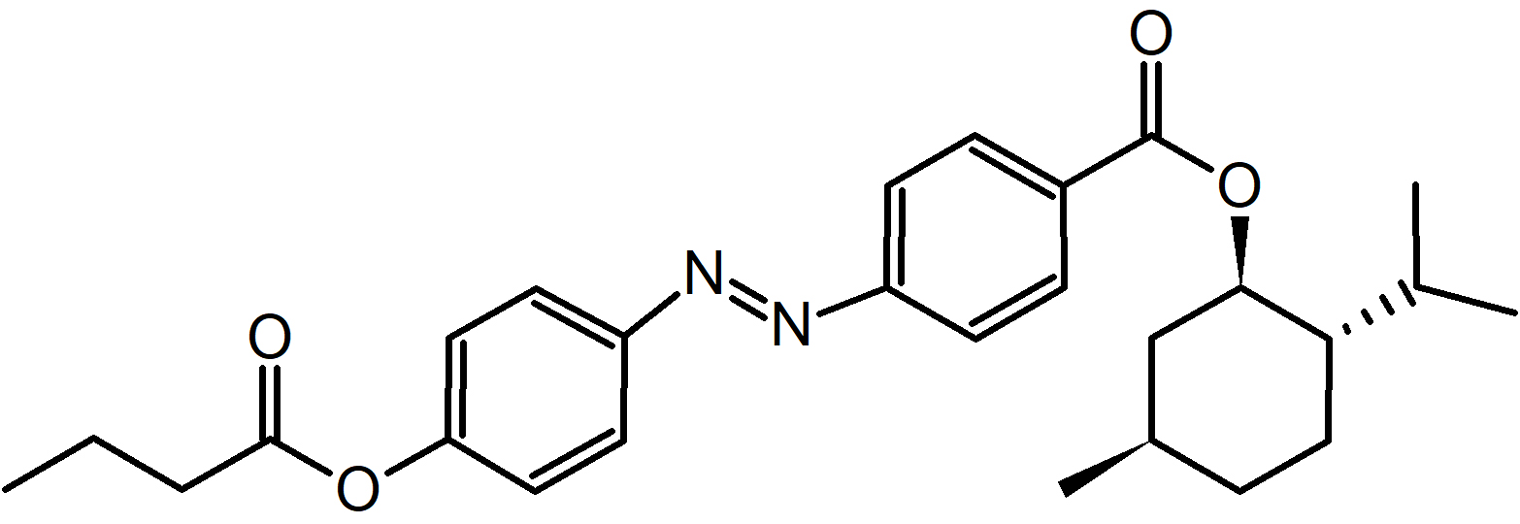} & $4$ & $-8.1$ \\
  \hline
  ChD-3816 & \includegraphics[width=3.8cm, valign=c]{ChD-3816.png} & $6$ & $-8$ \\
  \hline
  ChD-4211 & \includegraphics[width=4.5cm, valign=c]{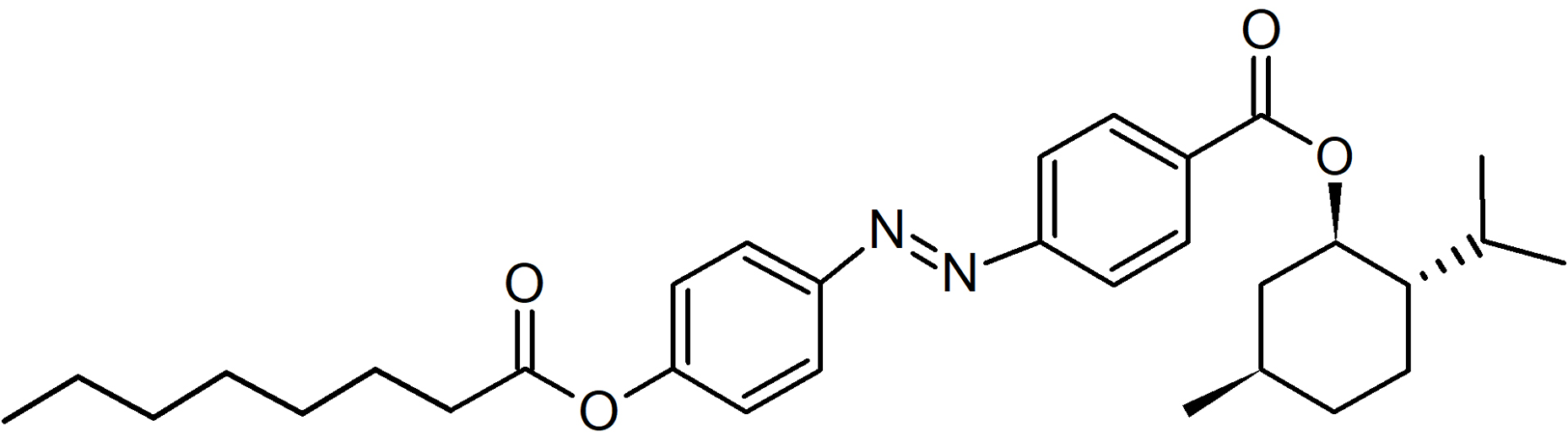} & $8$ & $-6.63$ \\
  \hline
  ChD-4212 & \includegraphics[width=4.9cm, valign=c]{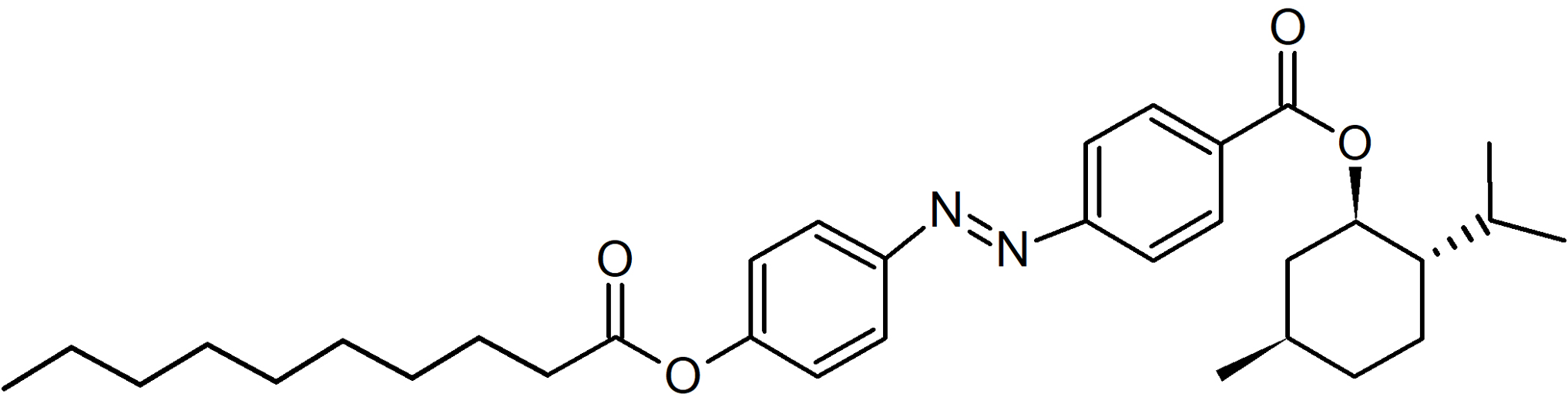} & $10$ & $-5.81$ \\
  \hline
  ChD-3805 & \includegraphics[width=4.9cm, valign=c]{ChD-3805.png} & $12$ & $-3.95$ \\
  \hline
\end{tabular}
\end{adjustwidth}
\end{table}

The HTP values of a homologous series of chiral \textit{azo} compounds (with varying alkyl chain lengths) are listed in Table~S3.2. 

The dependence of the HTP of various chiral \textit{azo} compounds on the number of carbon atoms in the alkyl chain is shown in Fig.~S3.3. An increase in the alkyl chain length leads to a decrease in HTP.

\begin{figure}[!ht]
\begin{adjustwidth}{-1in}{0in}
\centering
  \includegraphics[width=9cm]{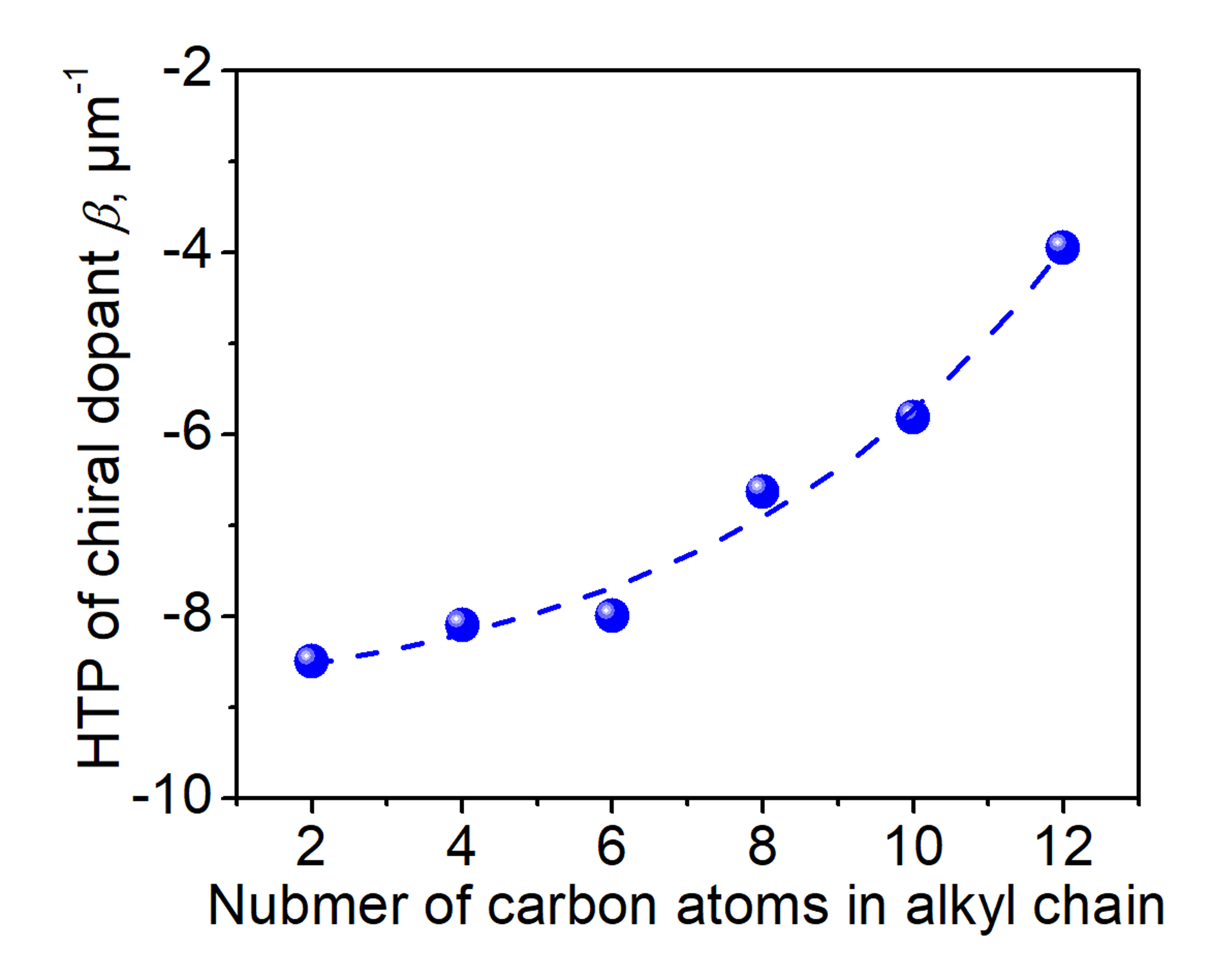}
  \caption{\textbf{Dependence of HTP of the chiral \textit{azo} compound in nematic E7 on the number of carbon atoms in alkyl chain.}}
  \label{fgr:Figure3S3}
  \end{adjustwidth}
\end{figure} 

\begin{figure}[!ht]
\begin{adjustwidth}{-1in}{0in}
\centering
  \includegraphics[width=11cm]{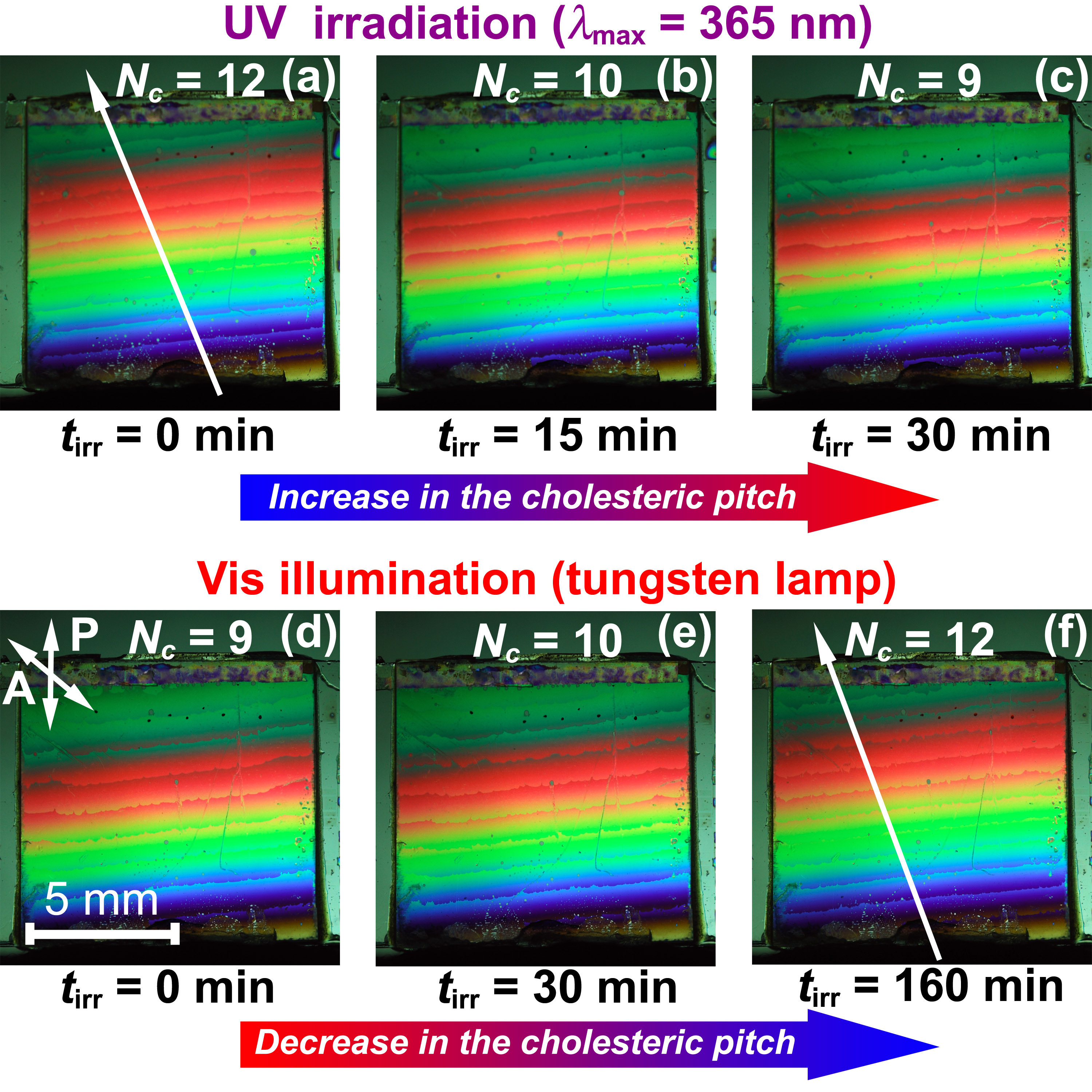}
  \caption{\textbf{Sequential microphotographs of a wedge-like LC cell filled with 3~wt\% ChD-4187 and 97~wt\% E7 during (a)--(c) UV irradiation and (d)--(f) Vis illumination for various times $t_{\text{irr}}$.}}
  \label{fgr:Figure3S4}
  \end{adjustwidth}
\end{figure}

\newpage

Sequential microphotographs of a wedge-like LC cell, filled with 3~wt\% ChD-4187 and 97~wt\% E7, obtained during UV irradiation ($\lambda_{\text{max}} = 365$~nm) for various irradiation times $t_\text{irr}$ are shown in Figs.~S3.4(a)--(c). One can see that the number of the Grandjean-Cano stripes is decreasing. The reversible change in the number of the stripes was observed under Vis illumination the wedge-like cell, which was carried out by tungsten lamp (Figs.~S3.4(d)--(f)).

\begin{figure}[!ht]
\begin{adjustwidth}{-1in}{0in}
\centering
 \includegraphics[width=16cm]{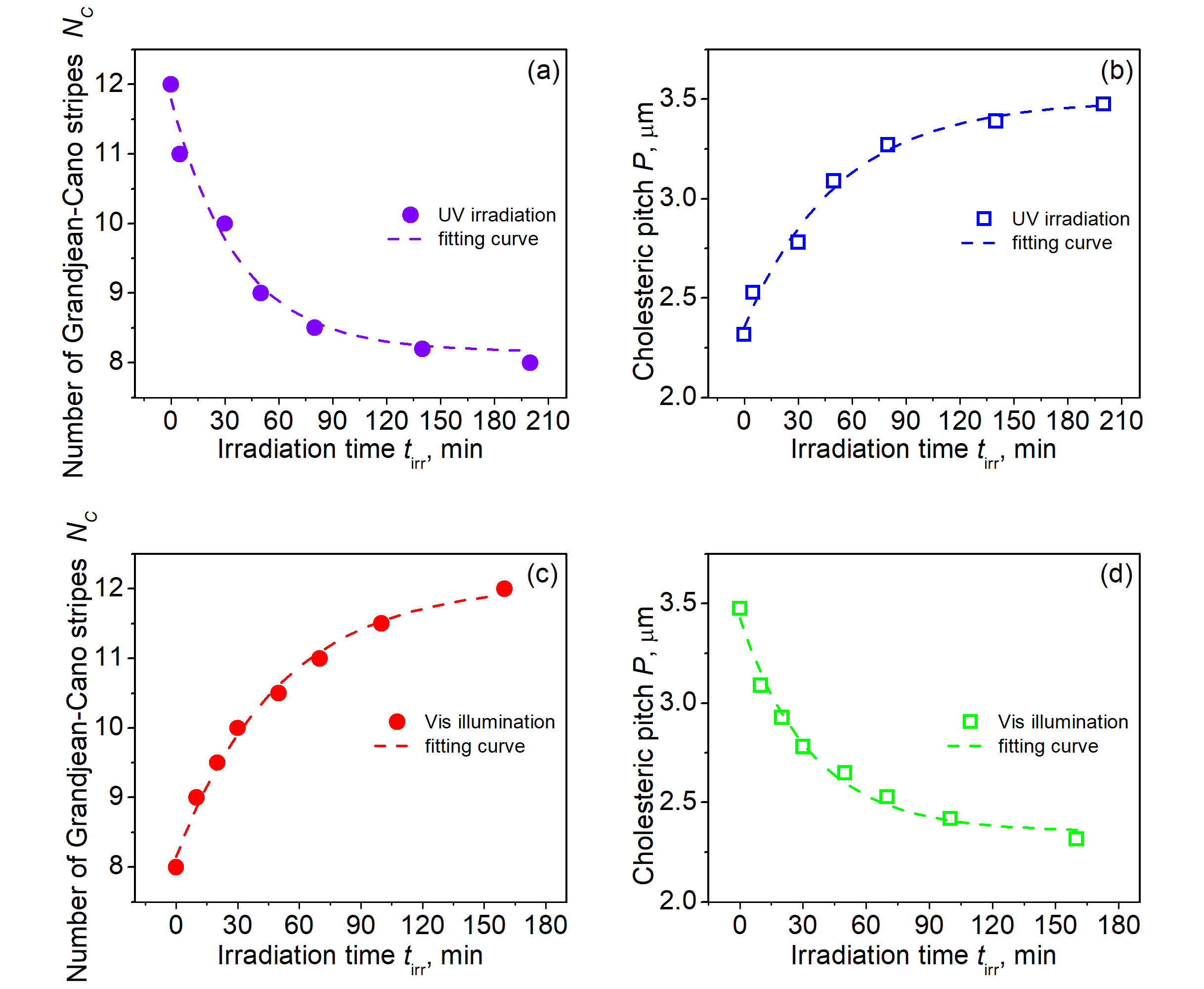}
  \caption{\textbf{Dependencies of both the number $N_C$ of Grandjean-Cano stripes (a), (c) and the cholesteric pitch \textit{P} (b), (d) on irradiation time $t_{\text{irr}}$. The irradiation was carried out by a UV lamp with $\lambda_{\text{max}} = 365$~nm (a), (b) and a tungsten lamp (c), (d).}}
 \label{fgr:Figure3S5}
  \end{adjustwidth}
\end{figure} 

Under UV irradiation of the wedge-like LC cell, the increasing concentration of the cis-isomer of ChD-4187 leads to a decrease in the number of Grandjean-Cano stripes $N_C$ (Fig.~S3.5(a)), \textit{i.e.}, the cholesteric pitch \textit{P} increases (Fig.~S3.5(b)). In contrast, Vis illumination (using a tungsten lamp) of the LC cell results in a reversible recovery of the number of stripes (Fig.~S3.5(c)), and accordingly, the cholesteric pitch \textit{P} (Fig.~S3.5(d)).

\newpage





\bigskip

\section{Phase transitions during heating and cooling of the base highly chiral mixture (HChM) doped with chiral and achiral \textit{azo} compounds}

In this section, all results are summarized in Tables~S4.1--S4.6.
\bigskip

Table~S4.1 presents the phase transition temperatures of the base HChM doped with 5~wt\% of a chiral \textit{azo} compound during heating and cooling processes. Table~S4.1 also summarizes the effects of \textit{trans}--\textit{cis} photo-isomerization during UV irradiation (as a function of irradiation time, $t_{\text{irr}}$) and the subsequent reversible \textit{cis}--\textit{trans} thermal isomerization after annealing at $80^\circ\text{C}$ for 30~min.
\bigskip

Table~S4.2 presents the photoinduced phase transitions of the base HChM (65 wt\% COC and 35 wt\% E7) doped with 5 wt\% of chiral \textit{azo} compound. The data cover both heating and cooling processes, showing the influence of UV irradiation for various irradiation times $t_{\text{irr}}$.
\bigskip

Table~S4.3 presents the phase transition temperatures of the HChM host doped with 5~wt\% of achiral \textit{azo}-compounds during heating and cooling processes. Data are presented for (i) UV-irradiated mixtures (\textit{trans}--\textit{cis} photoisomerization) and (ii) after reverse \textit{cis}--\textit{trans} thermal isomerization ($80^\circ\text{C}$ for 30~min).
\bigskip

Table~S4.4 summarizes the phase transition temperatures of the base HChM doped with 5~wt\% of an achiral \textit{azo} compound during both heating and cooling processes, illustrating the effects of UV irradiation for various irradiation times ($t_{\text{irr}}$).
\bigskip

Table~S4.5 summarizes the phase transition temperatures of the base HChM doped with 5~wt\% of a homologous series of \textit{azo} compounds possessing various alkyl chain lengths (\ce{C_n}) during both heating and cooling processes. It also details the effects of \textit{trans}--\textit{cis} photo-isomerization under UV irradiation ($t_{\text{irr}}$) and the reverse (thermal) \textit{cis}--\textit{trans} isomerization ($80^\circ\text{C}$ for 30~min).
\bigskip

Table~S4.6 shows the photo-induced phase transition temperatures (upon UV exposure for various times $t_{\text{irr}}$) of the base HChM (65~wt\% COC and 35~wt\% E7) doped with 5~wt\% of a homologous series of chiral \textit{azo} compounds possessing various alkyl chain lengths (\ce{C_n}) during both heating and cooling processes.

\clearpage

\clearpage

\begin{sidewaystable}[!p]
\centering
\vspace*{-1.5cm}
    \caption{Phase transition temperatures of the base HChM doped with 5 wt\% of chiral \textit{azo} compounds}
    \medskip
    \label{tab:Table4S1}
    \small
    \renewcommand{\arraystretch}{1.8} 
    \newcolumntype{C}{>{\centering\arraybackslash}X}
    
    \begin{tabularx}{\linewidth}{|>{\hsize=0.6\hsize}C|
    >{\hsize=1.4\hsize}C|c|C|C|C|C|C|}
       \hline 
        \textbf{\textit{Azo} compound} & \textbf{Chemical formula (\textit{trans}-isomer)} & \textbf{Process} & \multicolumn{5}{c|}{\textbf{Phase transitions ($^\circ$C)}} \\ 
        \cline{4-8} 
        & & & \cellcolor{green!60} \textbf{Before UV} & \cellcolor{magenta!20} \textbf{30~min} & \cellcolor{magenta!50} \textbf{60~min} & \cellcolor{magenta!80} \textbf{120~min} & \cellcolor{green!60}\textbf{Recovery$^a$} \\ 
        \midrule
            
        \multirow{2}{*}{ChD-3793} & 
        \multirow{2}{*}{\includegraphics[width=2.2cm, valign=c]{ChD-3793.png}} & \cellcolor{red!20}
        \text{Heating} & SmA 37.9 $\text{N}^*$ 41 BPs 42.5 Iso & SmA 37.2 $\text{N}^*$ 41 BPs 42.3 Iso & SmA 37 $\text{N}^*$ 40.8 BPs 42.2 Iso & SmA 36.8 $\text{N}^*$ 40.4 BPs 42 Iso & SmA 37.9 $\text{N}^*$ 41 BPs 42.5 Iso \\
        \cline{3-8}
        & & \cellcolor{cyan!20} \text{Cooling} & Iso 42.3 BPs 37.9 $\text{N}^*$ 36 SmA & Iso 41.9 BPs 36.8 $\text{N}^*$ 35.6 SmA & Iso 41.8 BPs 36.5 $\text{N}^*$ 35.2 SmA & Iso 41.6 BPs 36.1 $\text{N}^*$ 35 SmA & Iso 42.3 BPs 37.9 $\text{N}^*$ 36 SmA \\
        \hline
        \multirow{2}{*}{ChD-3610} & 
        \multirow{2}{*}{\includegraphics[width=3.6cm, valign=c]{ChD-3610.png}} & \cellcolor{red!20}
        \text{Heating} & SmA 35.6 $\text{N}^*$ 40.9 BPs 42.1 Iso & SmA 31.3 $\text{N}^*$ 35.6 BPs 38.2 Iso & SmA 30.9$\text{N}^*$ 35.2 BPs 38.1 Iso & SmA 29.4 $\text{N}^*$ 34.9 BPs 37.9 Iso & SmA 35.6 $\text{N}^*$ 40.9 BPs 42.1 Iso \\
        \cline{3-8}
        & & \cellcolor{cyan!20} \text{Cooling} & Iso 41.9 BPs 35.9$\text{N}^*$ 34 SmA & Iso 38.1 BPs 31.5 $\text{N}^*$ 30.3 SmA & Iso 38.1 BPs 31.4 $\text{N}^*$ 30 SmA & Iso 38.1 BPs 31.3 $\text{N}^*$ 29.8 SmA & Iso 41.9 BPs 35.9 $\text{N}^*$ 34 SmA \\
        \hline
       \multirow{2}{*}{ChD-3795} & 
        \multirow{2}{*}{\includegraphics[width=3.6cm, valign=c]{ChD-3795.png}} & \cellcolor{red!20}
        \text{Heating} & SmA 41.7 $\text{N}^*$ 48.4 BPs 49.2 Iso & SmA 36.9 $\text{N}^*$ 45 BPs 46.8 Iso & SmA 36.5 $\text{N}^*$ 44.8 BPs 46.7 Iso & SmA 36.5 $\text{N}^*$ 44.6 BPs 46.5 Iso & SmA 41.7 $\text{N}^*$ 48.4 BPs 49.2 Iso \\
        \cline{3-8}
        & & \cellcolor{cyan!20} \text{Cooling} & Iso 49.1 BPs 43.7 $\text{N}^*$ 40.5 SmA & Iso 46.7 BPs 40.4 $\text{N}^*$ 37.7 SmA & Iso 46.6 BPs 39.8 $\text{N}^*$ 37.5 SmA & Iso 46.4 BPs 39.5 $\text{N}^*$ 37.4 SmA & Iso 49.1 BPs 43.7 $\text{N}^*$ 40.5 SmA \\
        \hline
        \multirow{2}{*}{ChD-3805} & 
        \multirow{2}{*}{\includegraphics[width=3.6cm, valign=c]{ChD-3805.png}} & \cellcolor{red!20}
        \text{Heating} & SmA 36.3 $\text{N}^*$ 41.6 BPs 42.7 Iso & SmA 33.5 $\text{N}^*$ 39.3 BPs 41 Iso & SmA 33 $\text{N}^*$ 38.8 BPs 40.5 Iso & SmA 32.4 $\text{N}^*$ 38.7 BPs 40.5 Iso & SmA 36.3 $\text{N}^*$ 41.6 BPs 42.7 Iso \\
        \cline{3-8}
        & & \cellcolor{cyan!20} \text{Cooling} & Iso 42.6 BPs 37.1 $\text{N}^*$ 35.7 SmA & Iso 42.1 BPs 36.2 $\text{N}^*$ 33.5 SmA & Iso 41.7 BPs 35.4 $\text{N}^*$ 32.9 SmA & Iso 41.4 BPs 34.9 $\text{N}^*$ 32.1 SmA & Iso 42.6 BPs 37.1 $\text{N}^*$ 35.7 SmA \\
        \hline
        \multirow{2}{*}{ChD-3816} & 
        \multirow{2}{*}{\includegraphics[width=3cm, valign=c]{ChD-3816.png}} & \cellcolor{red!20}
        \text{Heating} & SmA 36.4 $\text{N}^*$ 44.4 BPs 44.7 Iso & SmA 36.1 $\text{N}^*$ 44.3 BPs 44.5 Iso & SmA 35.7 $\text{N}^*$ 44.2 BPs 44.5 Iso & SmA 35.5 $\text{N}^*$ 44 BPs 44.5 Iso & SmA 36.4 $\text{N}^*$ 44.4 BPs 44.7 Iso \\
        \cline{3-8}
        & & \cellcolor{cyan!20} \text{Cooling} & Iso 44.5 BPs 40 $\text{N}^*$ 35.8 SmA & Iso 43.8 BPs 38.6 $\text{N}^*$ 34.3 SmA & Iso 43.4 BPs 37.8 $\text{N}^*$ 33.7 SmA & Iso 43.1 BPs 37.4 $\text{N}^*$ 33.4 SmA & Iso 44.5 BPs 40 $\text{N}^*$ 35.8 SmA \\
        \hline
        \multirow{2}{*}{ChD-3501} & 
        \multirow{2}{*}{\includegraphics[width=3.6cm, valign=c]{ChD-3501.png}} & \cellcolor{red!20}
        \text{Heating} & SmA 35.5 N* 41.9 BPs 42.5 Iso & SmA 30.2 N* 36.5 BPs 39.1 Iso & SmA 29.5 N* 36.1 BPs 38.5 Iso & SmA 29.1 N* 35 BPs 37.5 Iso & SmA 35.5 N* 41.9 BPs 42.5 Iso \\
        \cline{3-8}
        & & \cellcolor{cyan!20} \text{Cooling} & Iso 42.4 BPs 36.6 $\text{N}^*$ 35.3 SmA & Iso 39 BPs 32.6 $\text{N}^*$ 29.8 SmA & Iso 38.5 BPs 32 $\text{N}^*$ 28.2 SmA & Iso 37.3 BPs 30.8 $\text{N}^*$ 28 SmA & Iso 42.4 BPs 36.6 $\text{N}^*$ 35.3 SmA \\
        \hline
    \end{tabularx}
\vspace{5pt}
    \flushleft \footnotesize $^a$ Reversible \textit{cis}--\textit{trans} isomerization at $80^\circ\text{C}$ for 30~min.
\end{sidewaystable}

\newpage
\clearpage

\begin{table}[!ht]
\centering
    \caption{Photo-induced phase transitions of the base HChM doped with 5~wt\% of a chiral \textit{azo} compound}
    \label{tab:Table4S2}
   \medskip
   \noindent\centerline{%
   \setlength{\tabcolsep}{2.5pt}
   \renewcommand{\arraystretch}{1.5}
\begin{tabularx}{0.99\textwidth}{|
    >{\centering\arraybackslash}m{1.8cm}|
    >{\centering\arraybackslash}m{4cm}|
    >{\centering\arraybackslash}X|
    >{\centering\arraybackslash}X|}
        \hline
        \vspace{4mm}\textbf{\shortstack{\textit{Azo}\\compound}} & 
        \vspace{3mm}\textbf{Chemical formula (\textit{trans}-isomer)} & 
        \multicolumn{2}{c|}{\textbf{Photo-induced phase diagrams}} 
        \\ 
        \cline{3-4}
        & & \cellcolor{red!20}\textbf{Upon heating} & \cellcolor{cyan!20}\textbf{Upon cooling} \\ 
        \hline
        ChD-3793 & 
        \includegraphics[width=2.5cm, valign=c]{ChD-3793.png} &
        \includegraphics[width=3.5cm, valign=c]{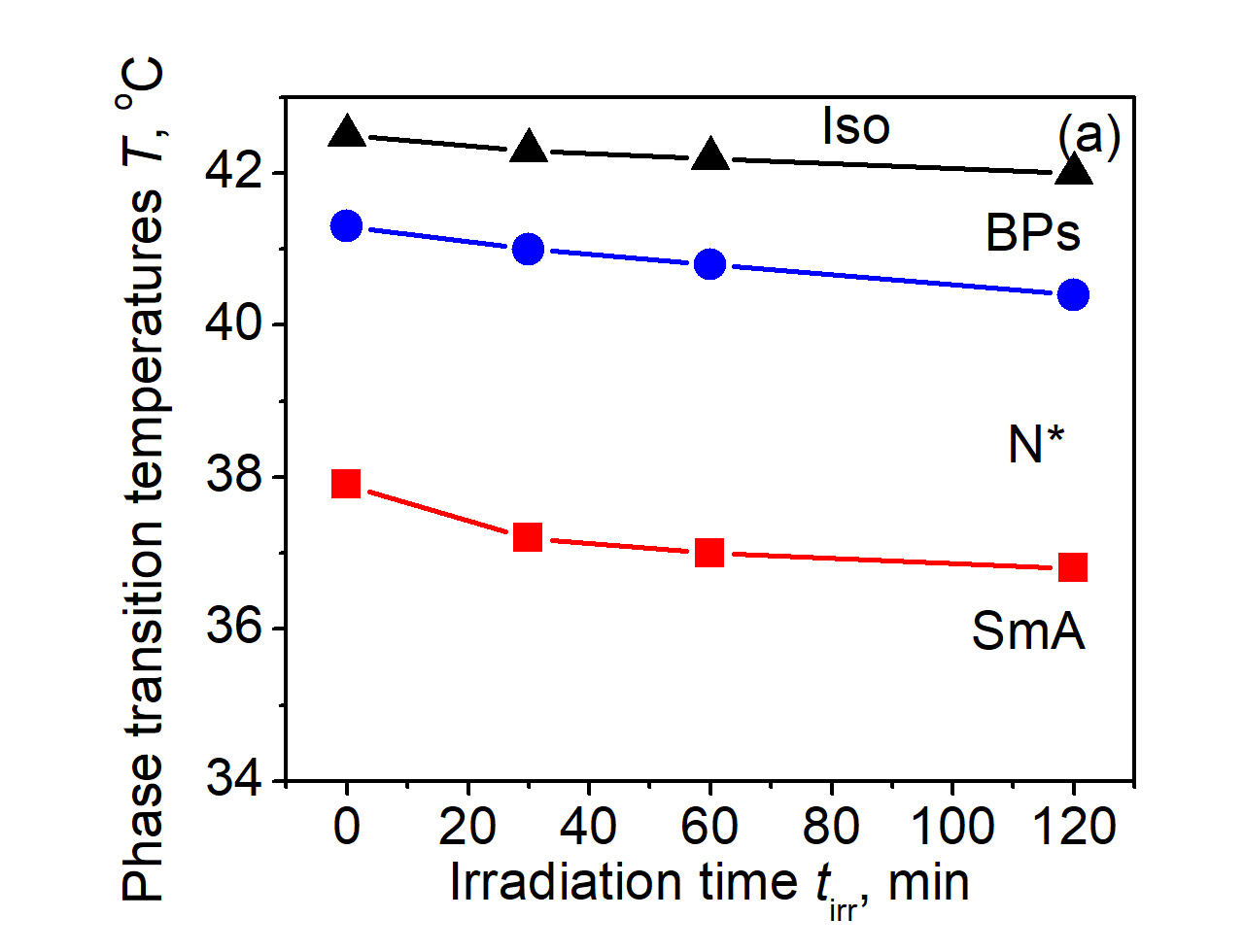} &\includegraphics[width=3.5cm, valign=c]{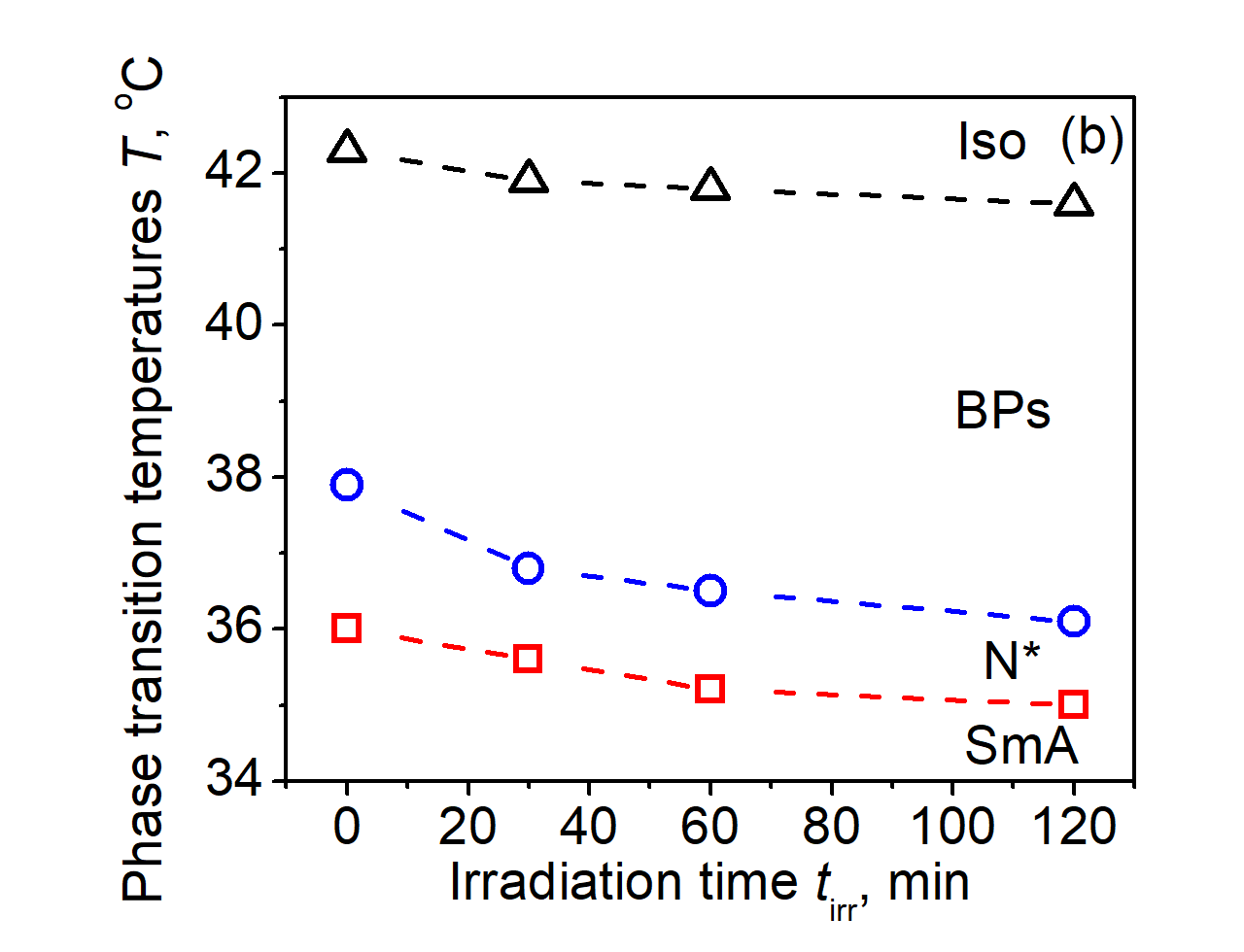} \\
       \hline
        ChD-3610 & 
      \includegraphics[width=4cm, valign=c]{ChD-3610.png} &
        \includegraphics[width=3.5cm, valign=c]{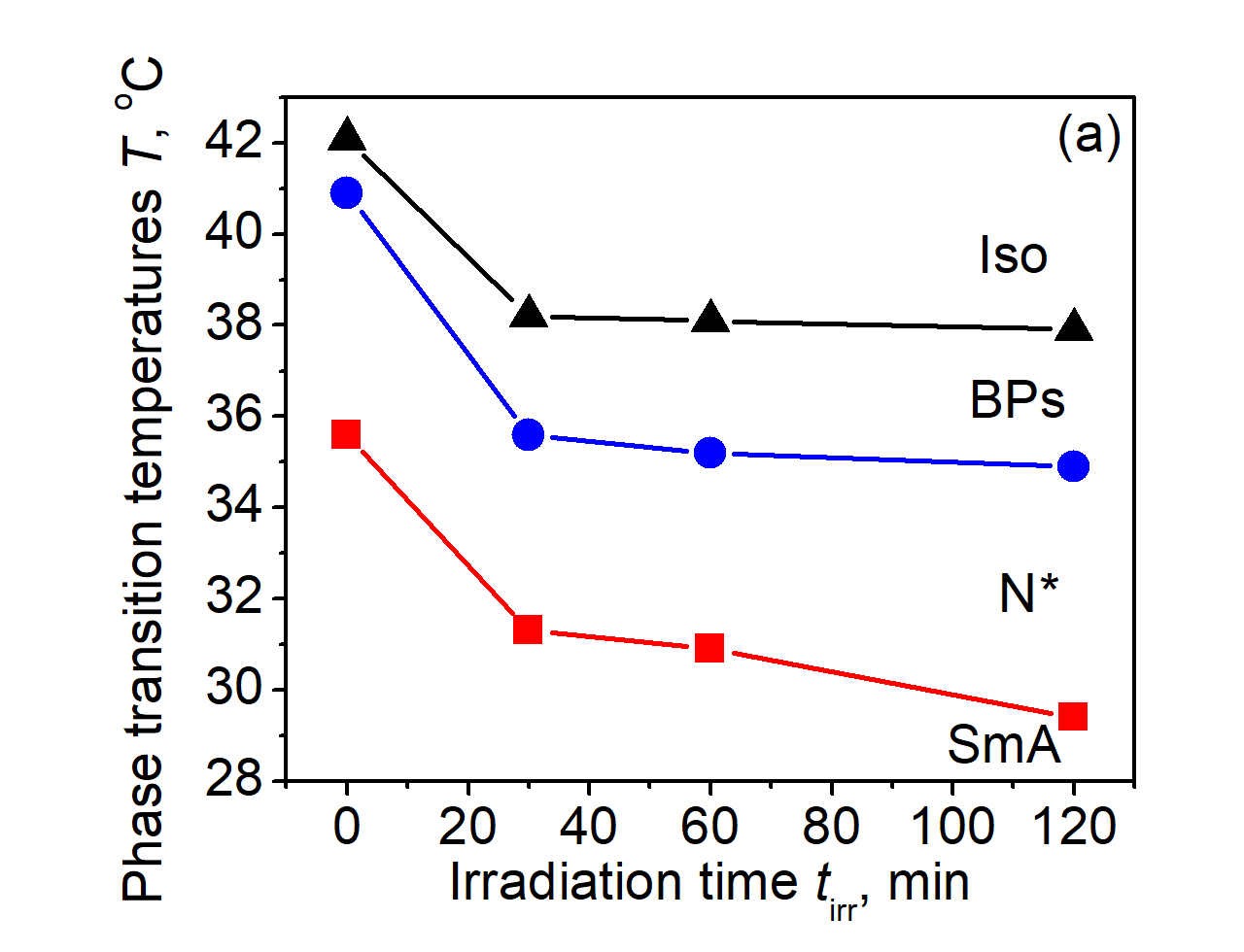} &\includegraphics[width=3.5cm, valign=c]{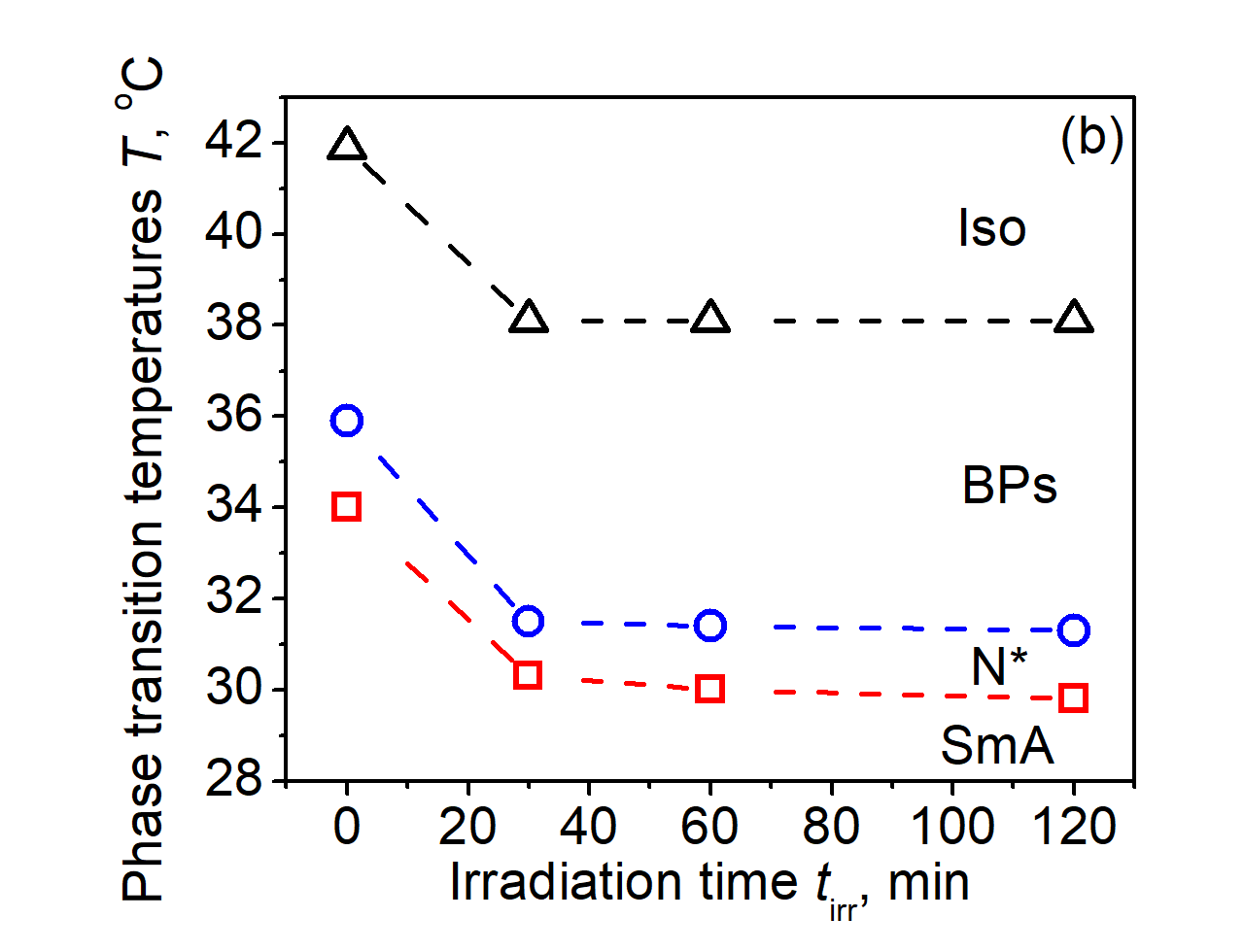} \\
        \hline
        ChD-3795 & 
       \includegraphics[width=4cm, valign=c]{ChD-3795.png} &
        \includegraphics[width=3.5cm, valign=c]{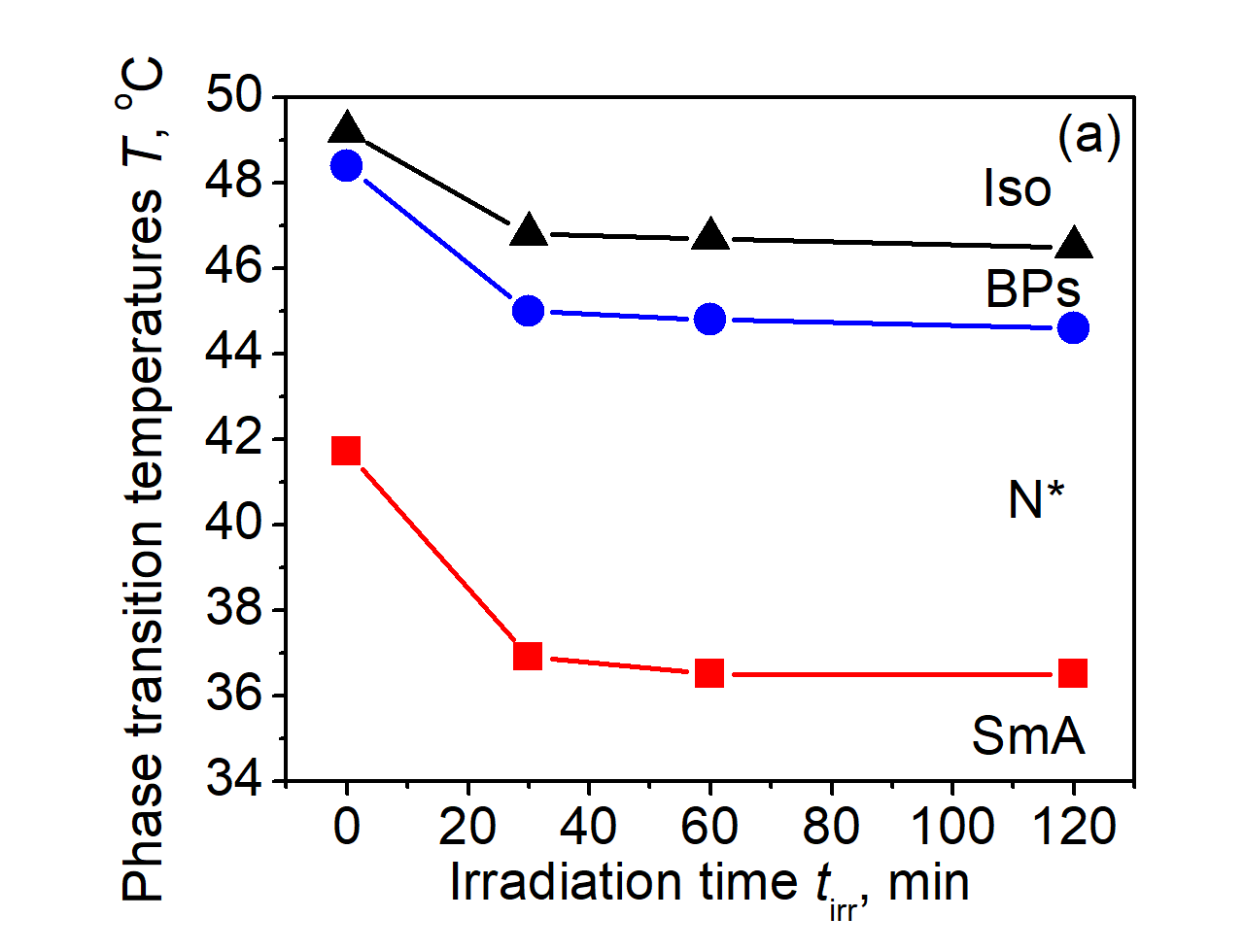} &\includegraphics[width=3.5cm, valign=c]{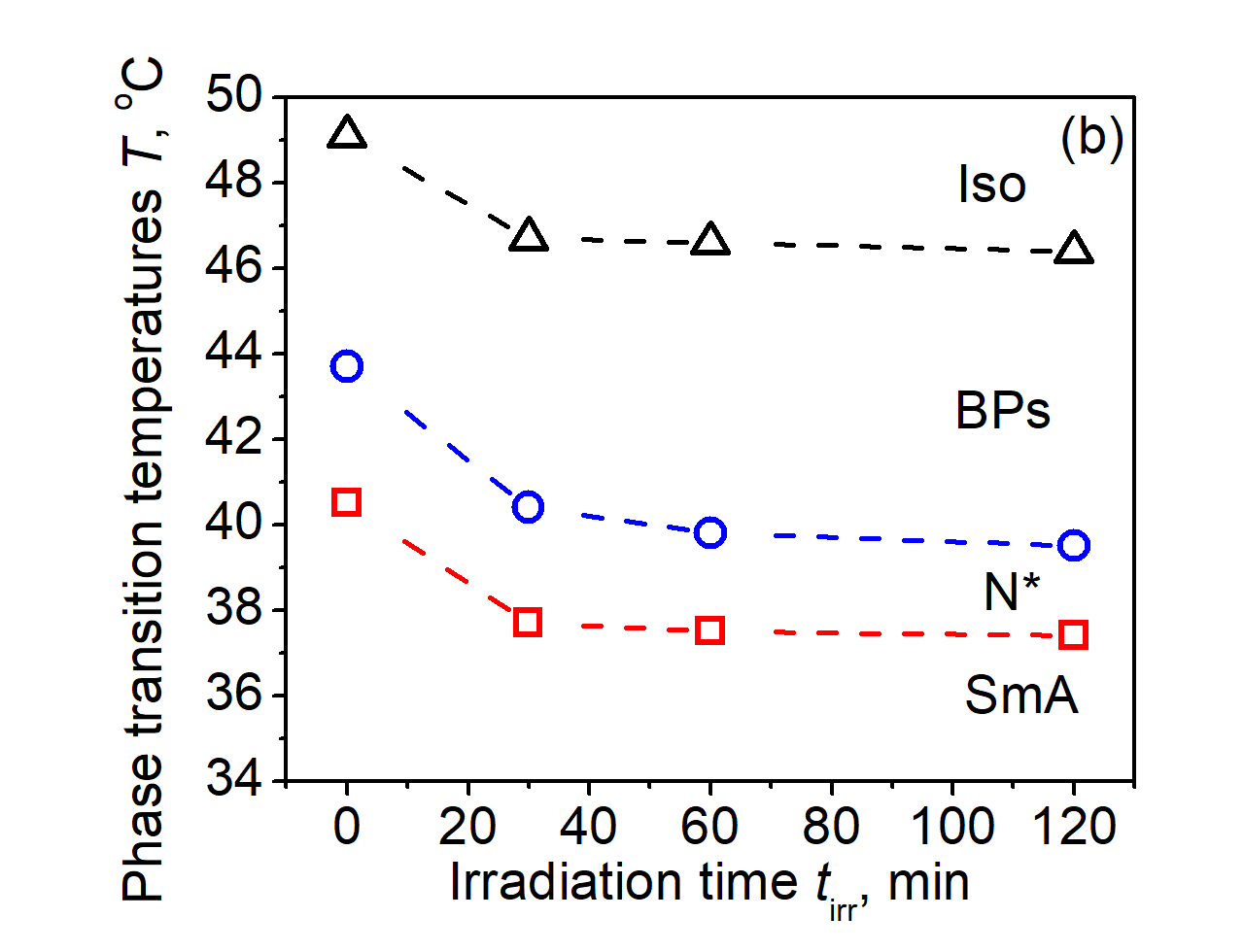} \\
        \hline
        ChD-3805 & 
       \includegraphics[width=4cm, valign=c]{ChD-3805.png} &
        \includegraphics[width=3.5cm, valign=c]{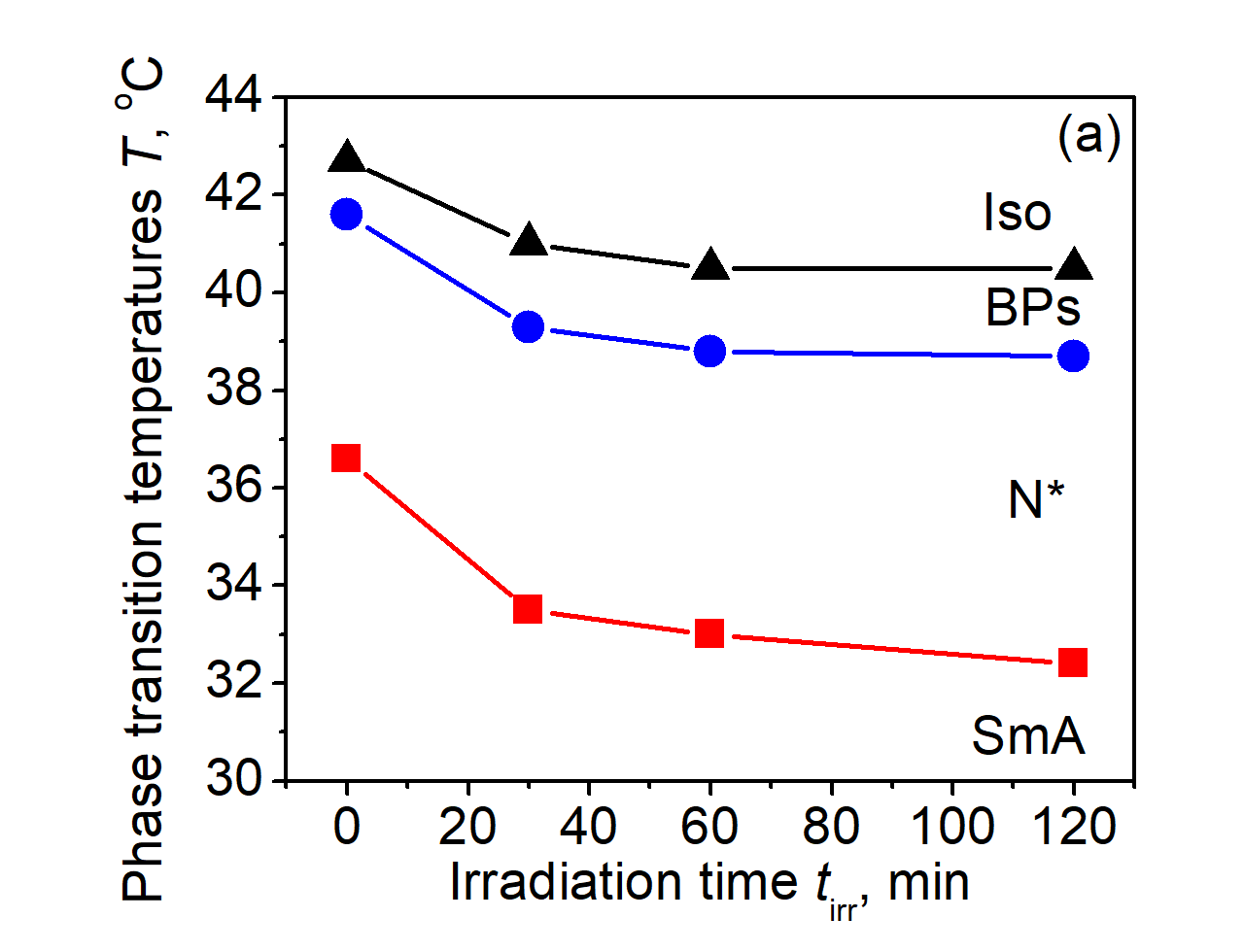} &\includegraphics[width=3.5cm, valign=c]{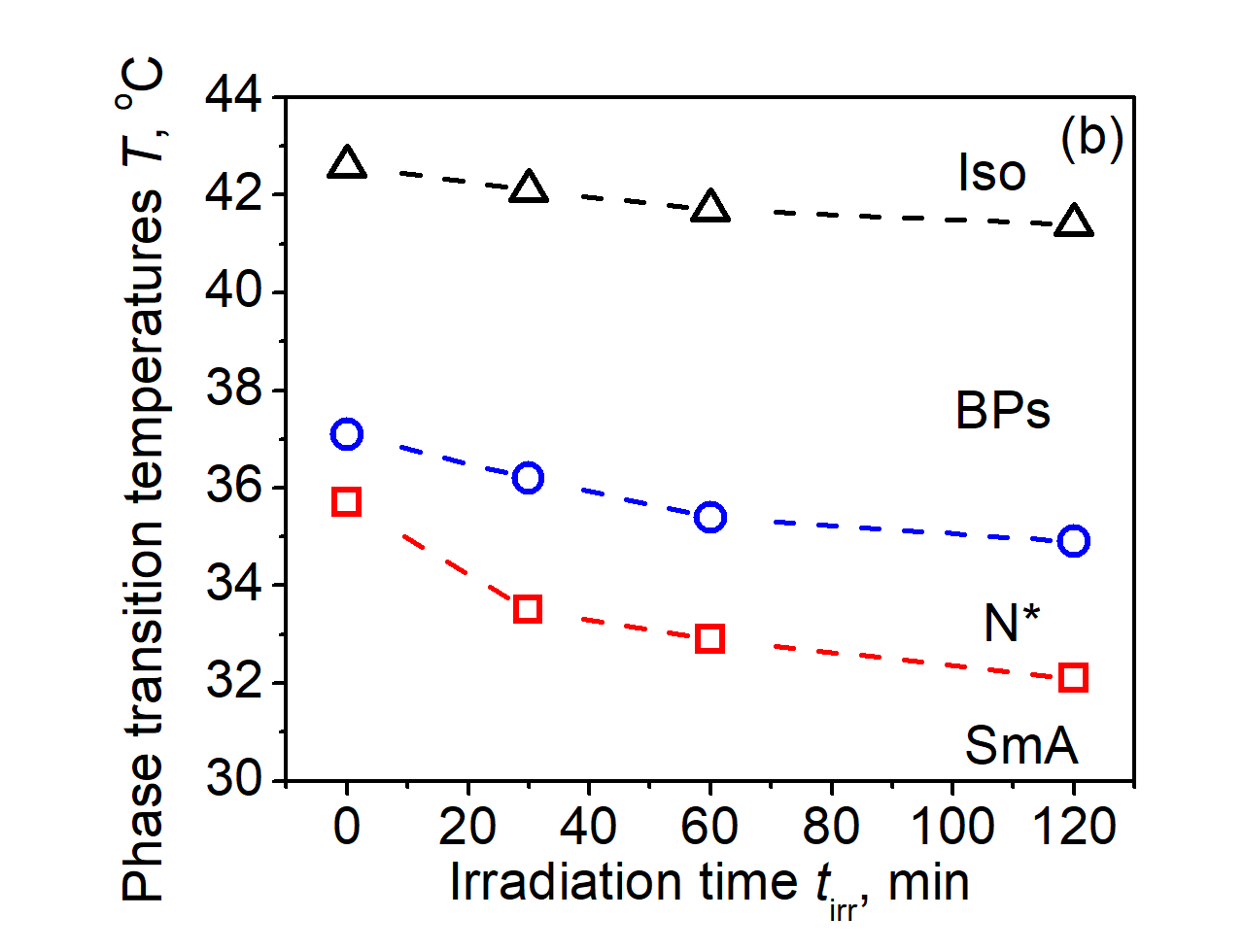} \\
        \hline
       ChD-3816 & 
       \includegraphics[width=3.5cm, valign=c]{ChD-3816.png} &
        \includegraphics[width=3.5cm, valign=c]{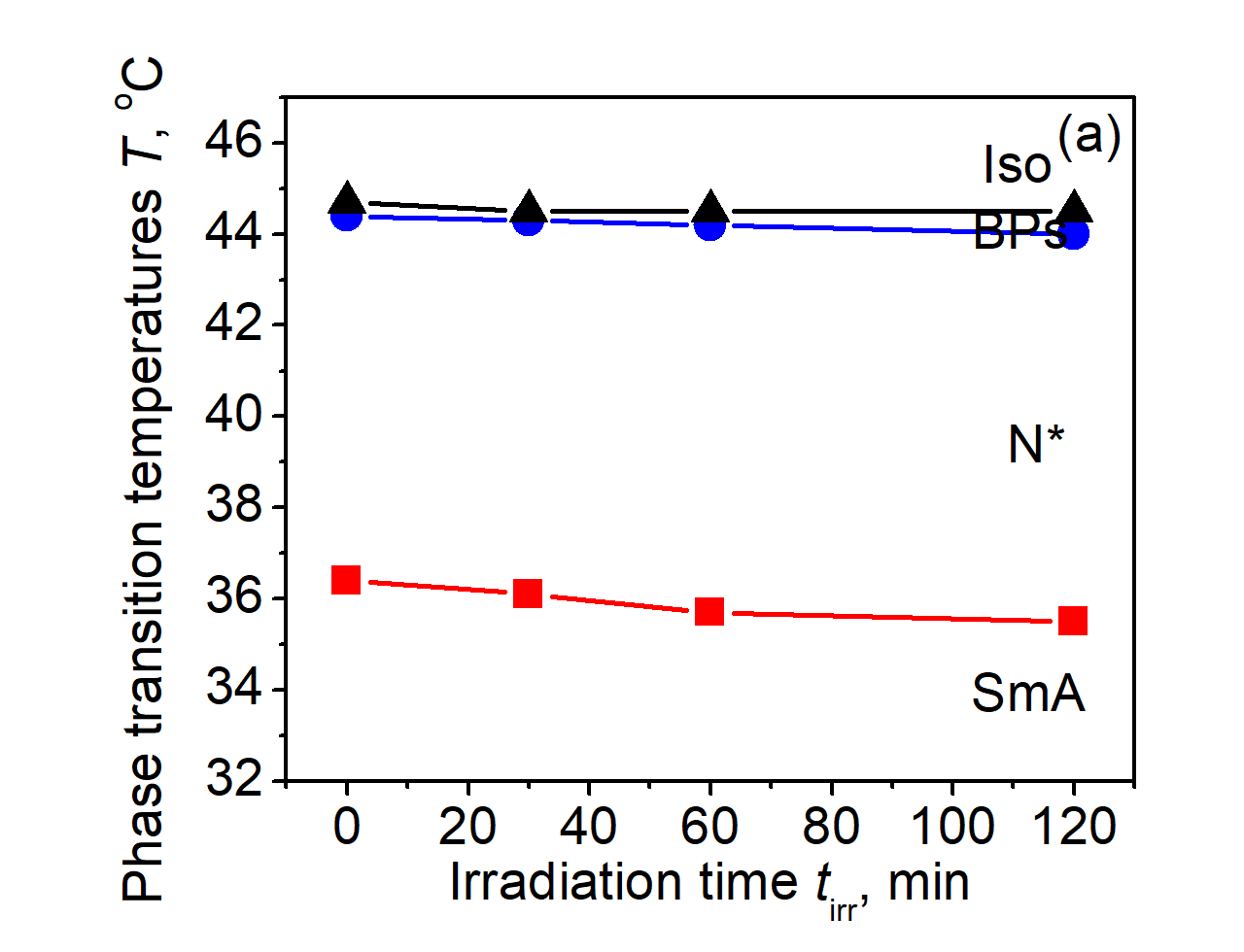} &\includegraphics[width=3.5cm, valign=c]{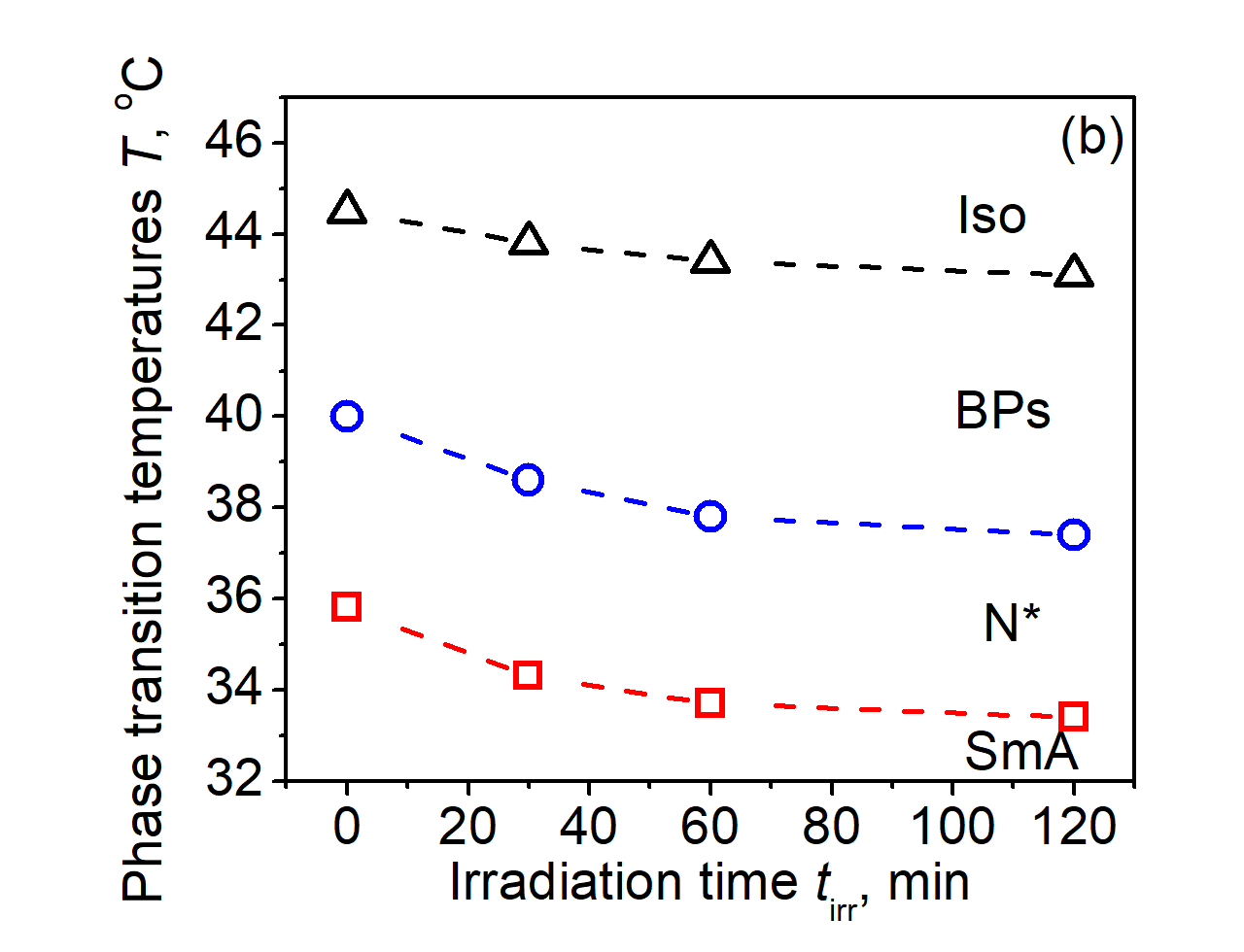} \\
        \hline
        ChD-3501 & 
       \includegraphics[width=4cm, valign=c]{ChD-3501.png} &
        \includegraphics[width=3.5cm, valign=c]{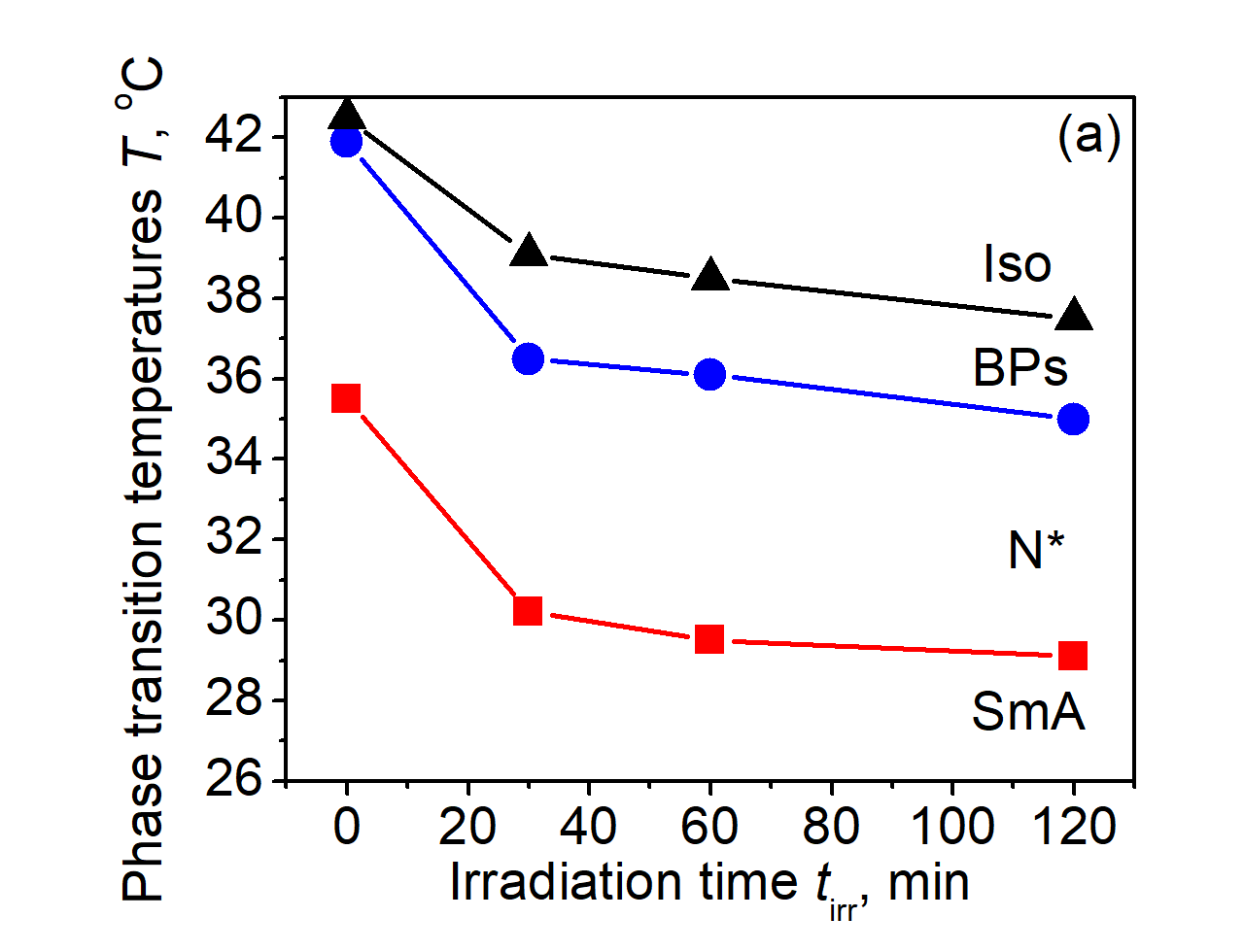} &\includegraphics[width=3.5cm, valign=c]{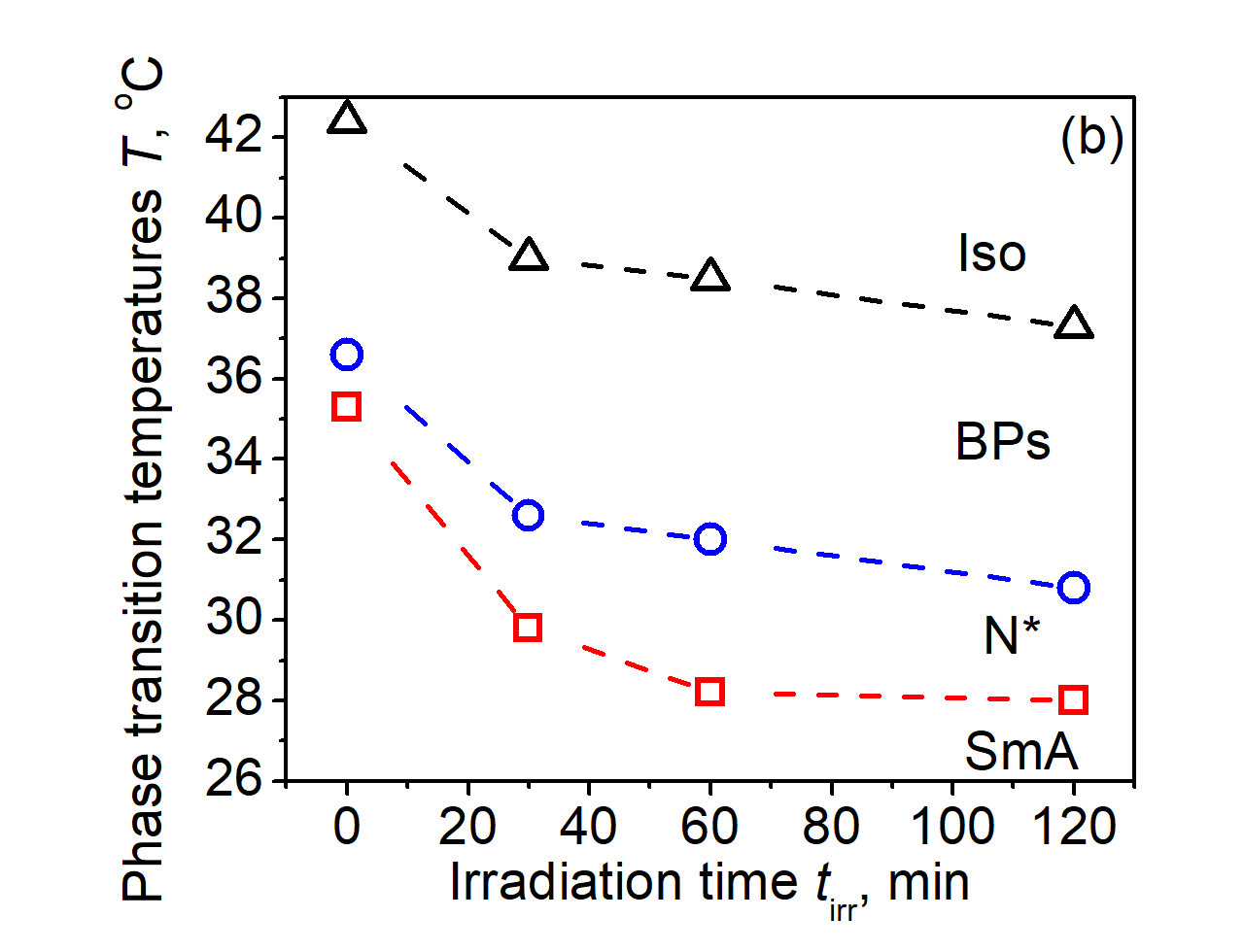} \\
        \hline 
 \end{tabularx}%
}%
\vss%
\end{table}%

\newpage
 
\clearpage

\begin{sidewaystable}[!p]
\centering
\vspace*{-1.5cm}
    \caption{Phase transition temperatures of the base HChM doped with 5~wt\% of achiral \textit{azo} compounds}
    \medskip
    \label{tab:Table4S3}
    \small
    \renewcommand{\arraystretch}{1.8} 
         
    \noindent 
    \setlength{\tabcolsep}{3.5pt}
    \newcolumntype{C}{>{\centering\arraybackslash\hspace{0pt}}X}
    
    \begin{tabularx}{0.99\linewidth}{|>{\hsize=0.6\hsize}C|
    >{\hsize=1.4\hsize}C|c|C|C|C|C|C|}
       \hline
       
        \textbf{\textit{Azo} compound} & \textbf{Chemical formula (\textit{trans}-isomer)} & \textbf{Process} & \multicolumn{5}{c|}{\textbf{Phase transitions ($^\circ$C)}} \\ 
        \cline{4-8} 
        & & & \cellcolor{green!60} \textbf{Before UV} & \cellcolor{magenta!20} \textbf{30~min} & \cellcolor{magenta!50} \textbf{60~min} & \cellcolor{magenta!80} \textbf{120~min} & \cellcolor{green!60}\textbf{Recovery$^a$} \\ 
        \midrule
       
        \multirow{2}{*}{aChD-3490} & 
        \multirow{2}{*}{\includegraphics[width=2cm, valign=c]{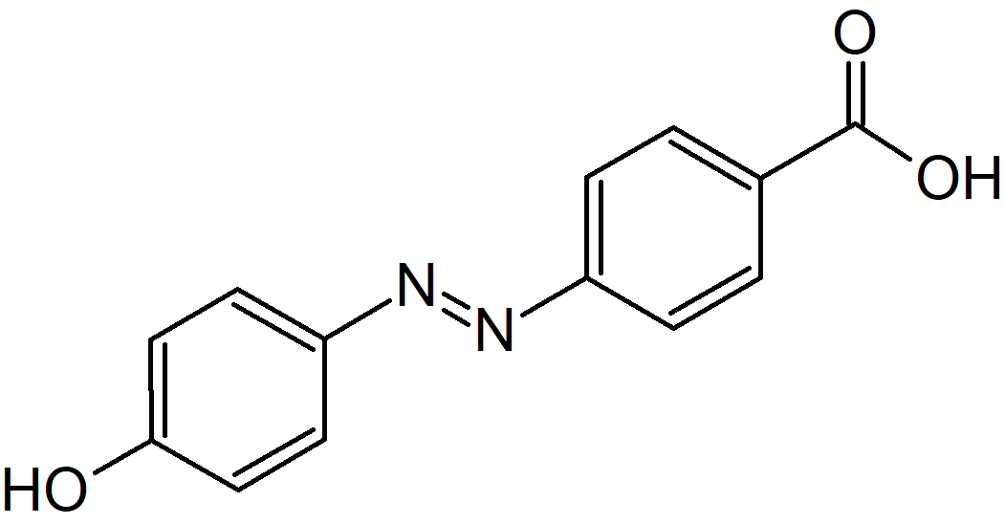}} & \cellcolor{red!20}
        \text{Heating} & SmA 34.4 $\text{N}^*$ 43.4 BPs 43.7 Iso & SmA 33.4 $\text{N}^*$ 42.4 BPs 43.6 Iso & SmA 33.2 $\text{N}^*$ 42.2 BPs 43.6 Iso & SmA 33 $\text{N}^*$ 41.9 BPs 43.2 Iso & SmA 34.4 $\text{N}^*$ 43.4 BPs 43.7 Iso \\
        \cline{3-8}
        & & \cellcolor{cyan!20} \text{Cooling} & Iso 43.7 BPs 39.4 $\text{N}^*$ 34.3 SmA & Iso 43.3 BPs 38.2 $\text{N}^*$ 33.7 SmA & Iso 43 BPs 37.3 $\text{N}^*$ 33.1 SmA & Iso 42.6 BPs 36.7 $\text{N}^*$ 32.9 SmA & Iso 43.7 BPs 39.4 $\text{N}^*$ 34.3 SmA \\
        \hline
        
        \multirow{2}{*}{aChD-3496} & 
        \multirow{2}{*}{\includegraphics[width=3.6cm, valign=c]{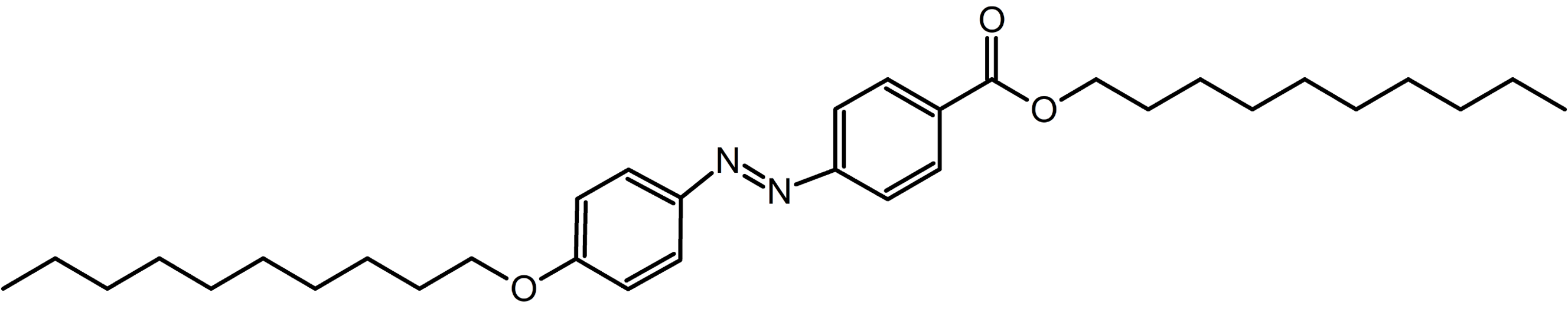}} & \cellcolor{red!20}
        \text{Heating} & SmA 40.3 $\text{N}^*$ 46.4 BPs 47.3 Iso & SmA 33.7 $\text{N}^*$ 40.6 BPs 43.2 Iso & SmA 32 $\text{N}^*$ 40.3 BPs 42.8 Iso & SmA 31.1 $\text{N}^*$ 39.8 BPs 42.7 Iso & SmA 40.3 $\text{N}^*$ 46.4 BPs 47.3 Iso \\
        \cline{3-8}
        & & \cellcolor{cyan!20} \text{Cooling} & Iso 47.2 BPs 41.6 N* 39.7 SmA & Iso 42.9 BPs 36.8 N* 35.2 SmA & Iso 42.5 BPs 36.2 N* 34.5 SmA & Iso 42.1 BPs 35.8 N* 34 SmA & Iso 47.2 BPs 41.6 N* 39.7 SmA \\
        \hline
        
        \multirow{2}{*}{aChD-3497} & 
        \multirow{2}{*}{\includegraphics[width=3.6cm, valign=c]{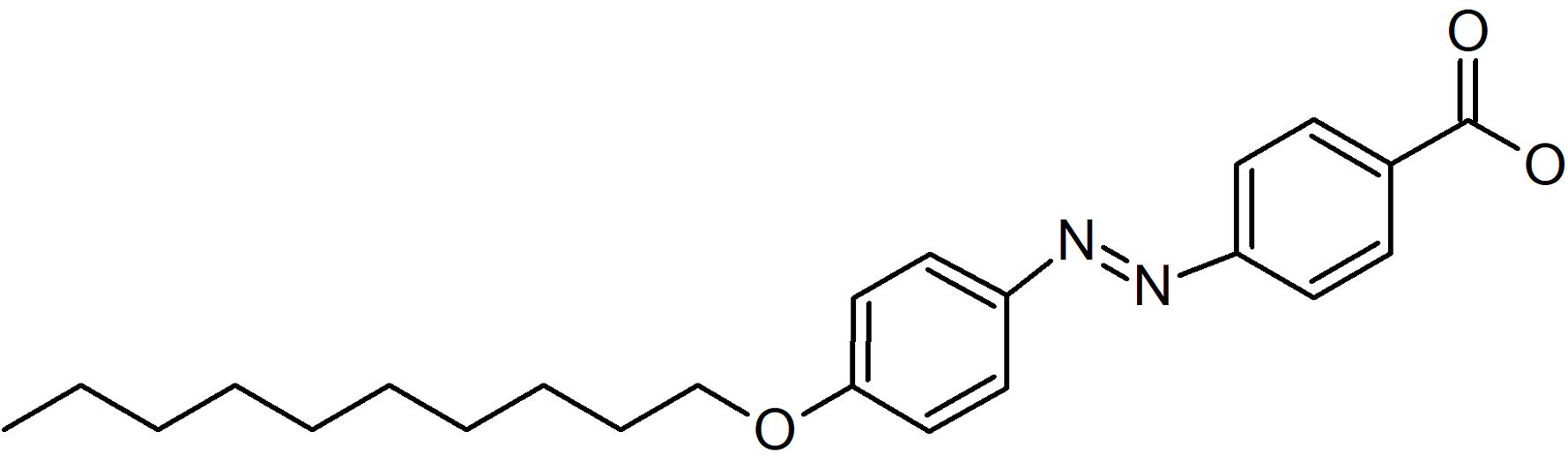}} & \cellcolor{red!20}
        \text{Heating} & SmA 37.2 $\text{N}^*$ 45.7 BPs 46.2 Iso & SmA 36.9 $\text{N}^*$ 45.5 BPs 46 Iso & SmA 36.7 $\text{N}^*$ 45.4 BPs 45.8 Iso & SmA 36.7 $\text{N}^*$ 45.3 BPs 45.4 Iso & SmA 37.2 $\text{N}^*$ 45.7 BPs 46.2 Iso \\
        \cline{3-8}
        & & \cellcolor{cyan!20} \text{Cooling} & Iso 45.8 BPs 41.4 $\text{N}^*$ 37.1 SmA & Iso 45.5 BPs 40.5 $\text{N}^*$ 36.8 SmA & Iso 45.3 BPs 40.1 $\text{N}^*$ 36.4 SmA & Iso 45.1 BPs 39.8 $\text{N}^*$ 36.2 SmA & Iso 45.8 BPs 41.4 $\text{N}^*$ 37.1 SmA \\
        \hline
        
        \multirow{2}{*}{aChD-3862} & 
        \multirow{2}{*}{\includegraphics[width=3.6cm, valign=c]{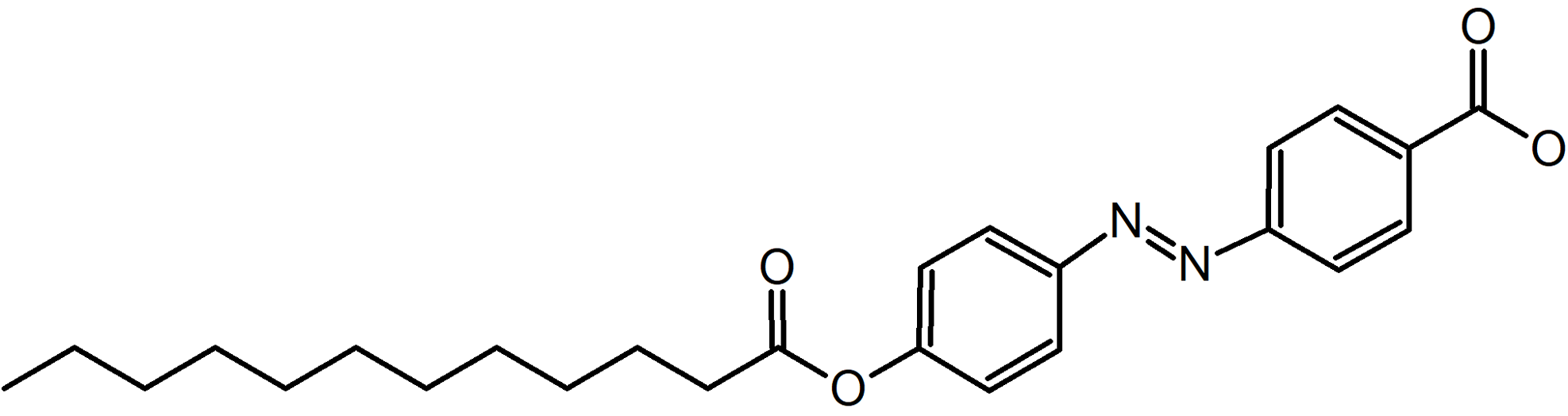}} & \cellcolor{red!20}
        \text{Heating} & SmA 37.2 $\text{N}^*$ 45 BPs 45.6 Iso & SmA 36.8 $\text{N}^*$ 44.8 BPs 46 Iso & SmA 36.4 $\text{N}^*$ 44.3 BPs 45.6 Iso & SmA 36.1 $\text{N}^*$ 44.3 BPs 45.5 Iso & SmA 37.2 $\text{N}^*$ 45 BPs 45.6 Iso \\
        \cline{3-8}
        & & \cellcolor{cyan!20} \text{Cooling} & Iso 45.5 BPs 40.7 $\text{N}^*$ 37.2 SmA & Iso 45.3 BPs 40.1 $\text{N}^*$ 36.7 SmA & Iso 45 BPs 39.5 $\text{N}^*$ 36 SmA & Iso 44.9 BPs 39.3 $\text{N}^*$ 35.7 SmA & Iso 45.5 BPs 40.7 $\text{N}^*$ 37.2 SmA \\
        \hline
        
        \multirow{2}{*}{aChD-4195} & 
        \multirow{2}{*}{\includegraphics[width=2.1cm, valign=c]{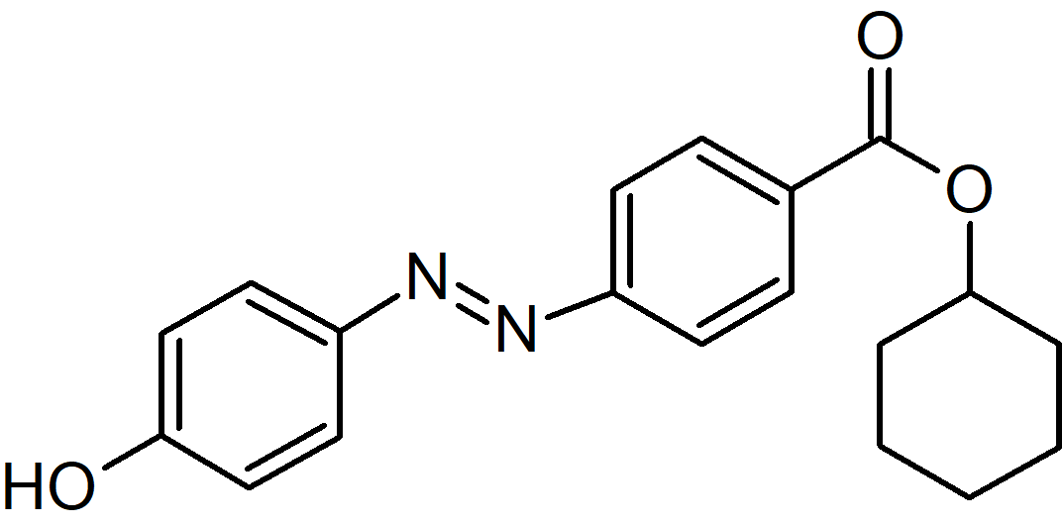}} & \cellcolor{red!20}
        \text{Heating} & SmA 36.5 $\text{N}^*$ 44.8 BPs 45.1 Iso & SmA 35.9 $\text{N}^*$ 44.6 BPs 44.9 Iso & SmA 35.7 $\text{N}^*$ 44.5 BPs 44.7 Iso & SmA 35.7 $\text{N}^*$ 44.4 BPs 44.7 Iso & SmA 36.5 $\text{N}^*$ 44.8 BPs 45.1 Iso \\
        \cline{3-8}
        & & \cellcolor{cyan!20} \text{Cooling} & Iso 45.1 BPs 40.3 $\text{N}^*$ 36.1 SmA & Iso 44.7 BPs 39 $\text{N}^*$ 35.8 SmA & Iso 44.6 BPs 38.6 $\text{N}^*$ 35.5 SmA & Iso 44.3 BPs 38.5 $\text{N}^*$ 35.3 SmA & Iso 45.1 BPs 40.3 $\text{N}^*$ 36.1 SmA \\
        \hline
        
        \multirow{2}{*}{aChD-4197} & 
        \multirow{2}{*}{\includegraphics[width=3cm, valign=c]{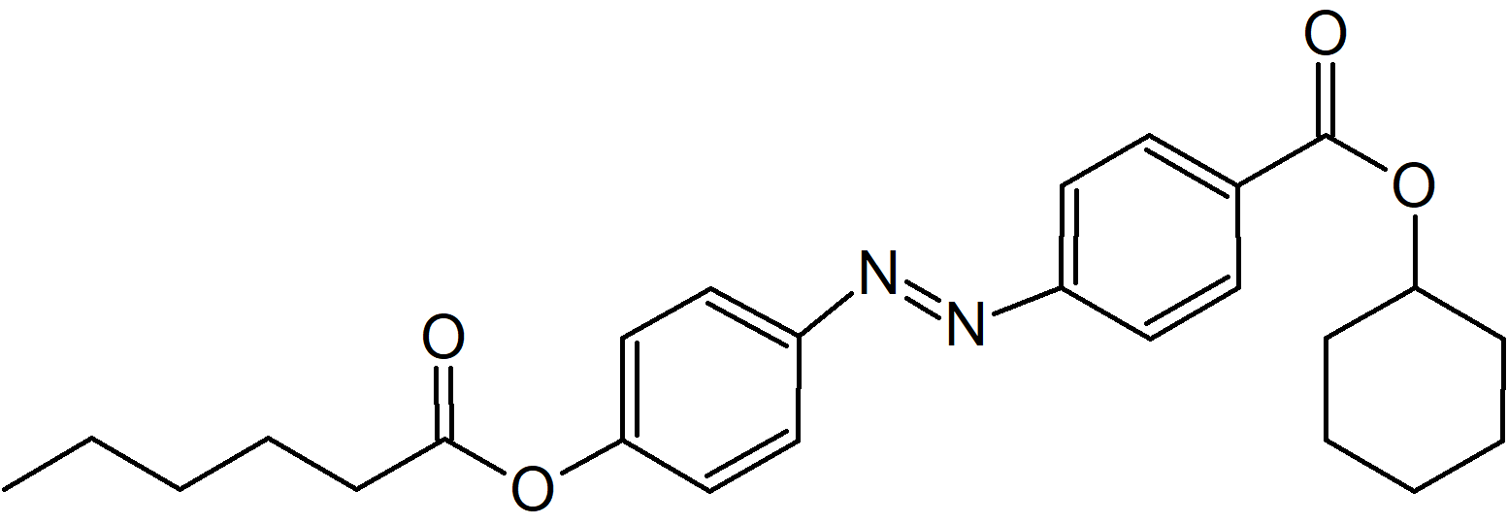}} & \cellcolor{red!20}
        \text{Heating} & SmA 35.4 $\text{N}^*$ 44.8 BPs 45.4 Iso & SmA 33.6 $\text{N}^*$ 40.3 BPs 42.2 Iso & SmA 32.1 $\text{N}^*$ 38.9 BPs 40.7 Iso & SmA 31.7 $\text{N}^*$ 38.3 BPs 40.4 Iso & SmA 35.4 $\text{N}^*$ 44.8 BPs 45.4 Iso \\
        \cline{3-8}
        & & \cellcolor{cyan!20} \text{Cooling} & Iso 45.2 BPs 40.1 $\text{N}^*$ 35.5 SmA & Iso 42 BPs 36.1 $\text{N}^*$ 33 SmA & Iso 40.7 BPs 34.6 $\text{N}^*$ 31.8 SmA & Iso 40.5 BPs 34.3 $\text{N}^*$ 30.5 SmA & Iso 45.2 BPs 40.1 $\text{N}^*$ 35.5 SmA \\
        \hline
        
    \end{tabularx}
    
\vspace{5pt}
    \flushleft \footnotesize $^a$ Reversible \textit{cis}--\textit{trans} isomerization at $80^\circ\text{C}$ for 30~min.
\end{sidewaystable}

\newpage
\clearpage

\begin{table}[H]
   \centering
 \caption{Photo-induced phase transitions of the base HChM doped with 5~wt\% of an achiral \textit{azo} compound}
    \label{tab:Table4S4}
\medskip
   \noindent\centerline{%
   \setlength{\tabcolsep}{2.5pt}
   \renewcommand{\arraystretch}{1.5}
\begin{tabularx}{0.99\textwidth}{|
    >{\centering\arraybackslash}m{1.8cm}|
    >{\centering\arraybackslash}m{4cm}|
    >{\centering\arraybackslash}X|
    >{\centering\arraybackslash}X|}
        \hline
       \vspace{4mm}\textbf{\shortstack{\textit{Azo}\\compound}} & 
        \vspace{3mm}\textbf{Chemical formula (\textit{trans}-isomer)} & 
        \multicolumn{2}{c|}{\textbf{Photo-induced phase diagrams}} 
        \\ 
        \cline{3-4}
       & & \cellcolor{red!20}\textbf{Upon heating} & \cellcolor{cyan!20}\textbf{Upon cooling} \\ 
        \hline
        aChD-3490 & 
        \includegraphics[width=2.2cm, valign=c]{aChD3490.png} &
        \includegraphics[width=3.5cm, valign=c]{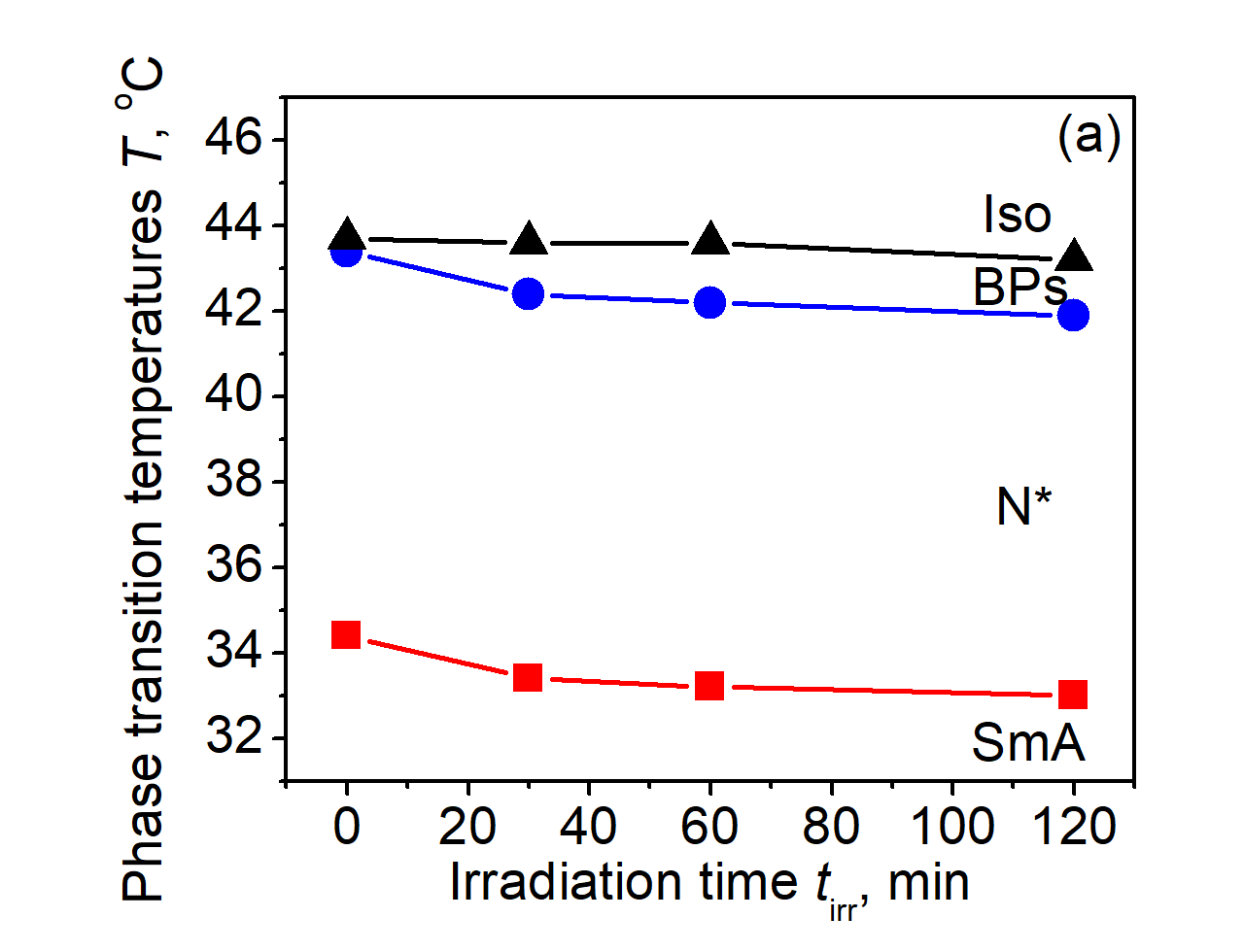} &\includegraphics[width=3.5cm, valign=c]{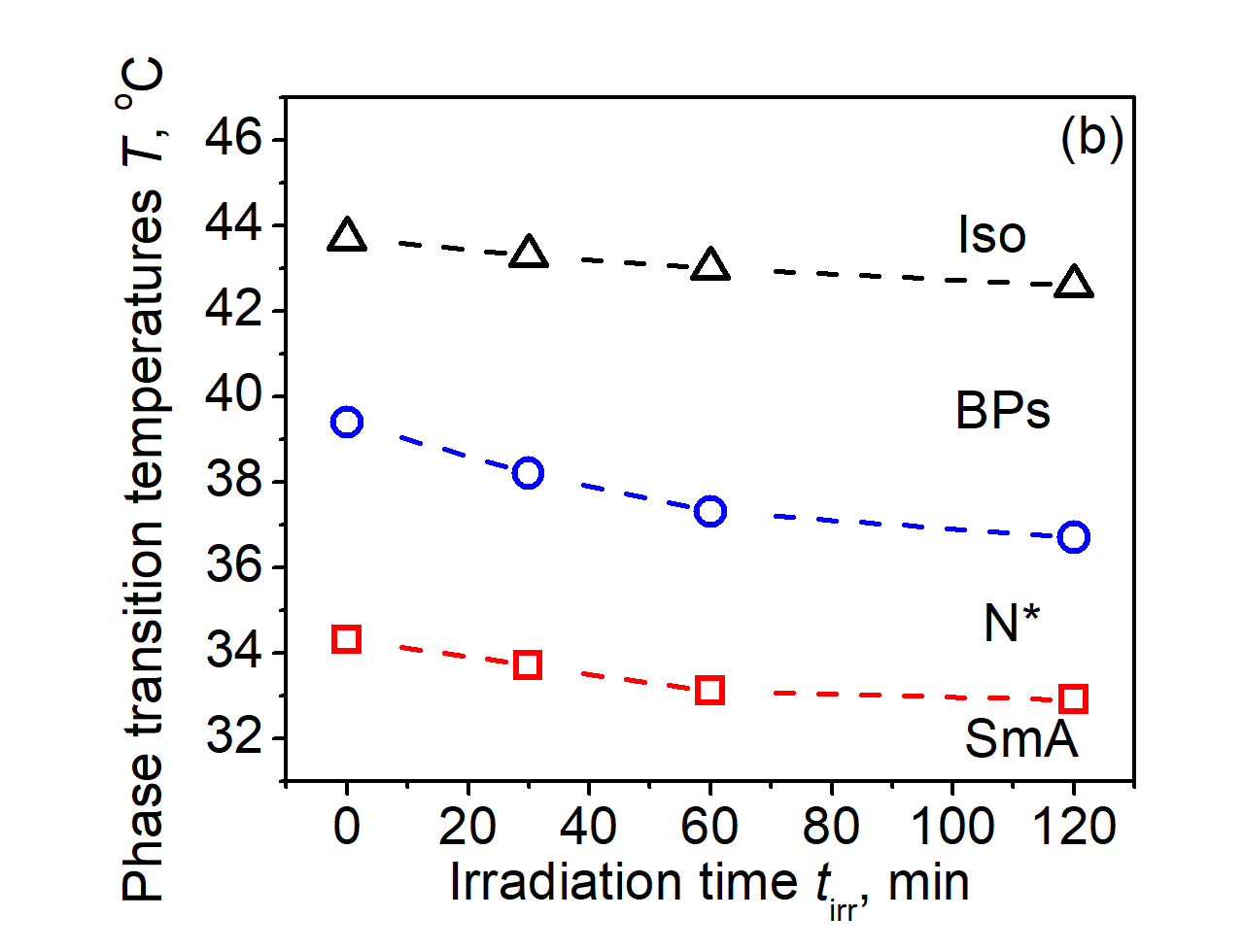} \\
       \hline
      aChD-3496 & 
      \includegraphics[width=4cm, valign=c]{aChD3496.png} &
        \includegraphics[width=3.5cm, valign=c]{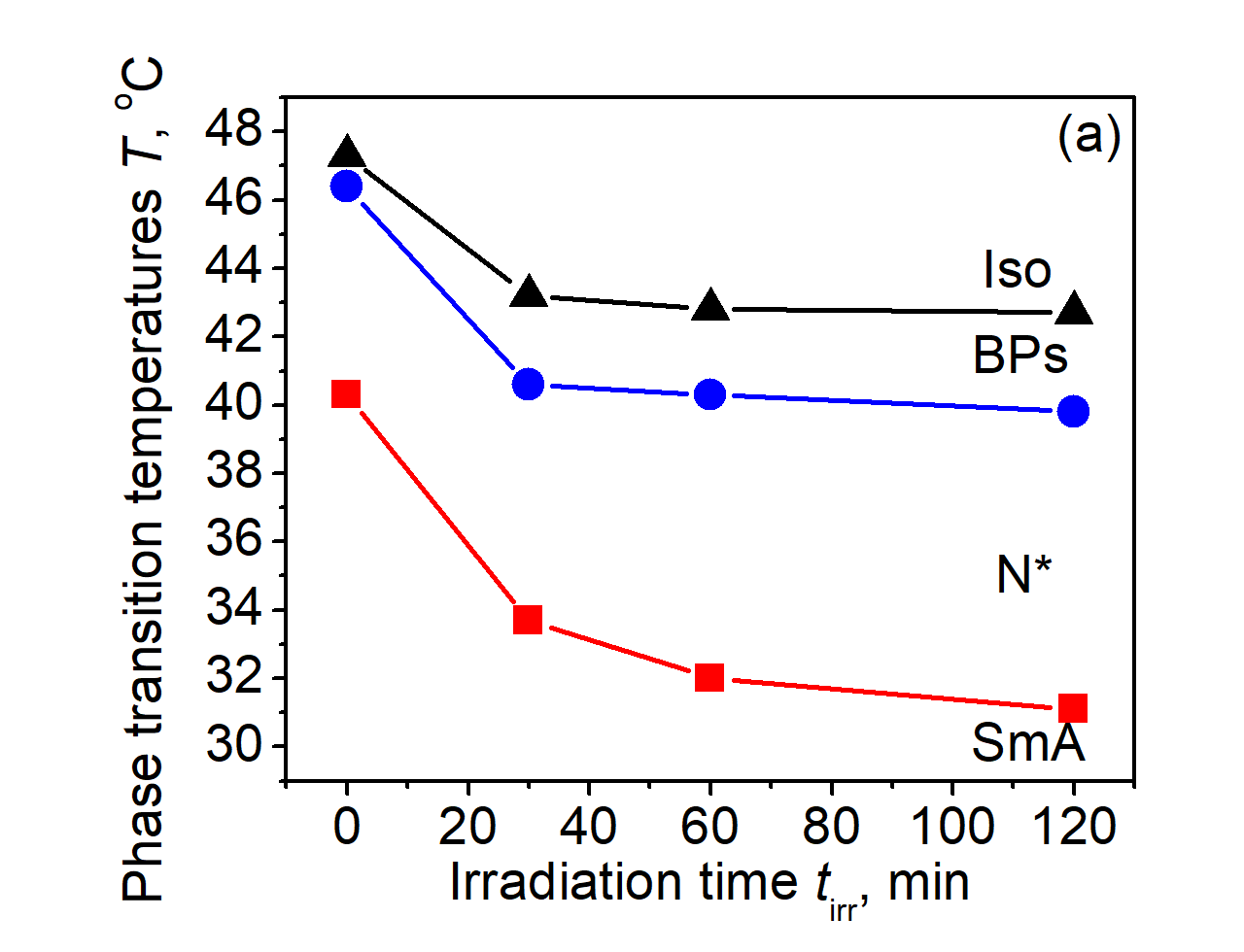} &\includegraphics[width=3.5cm, valign=c]{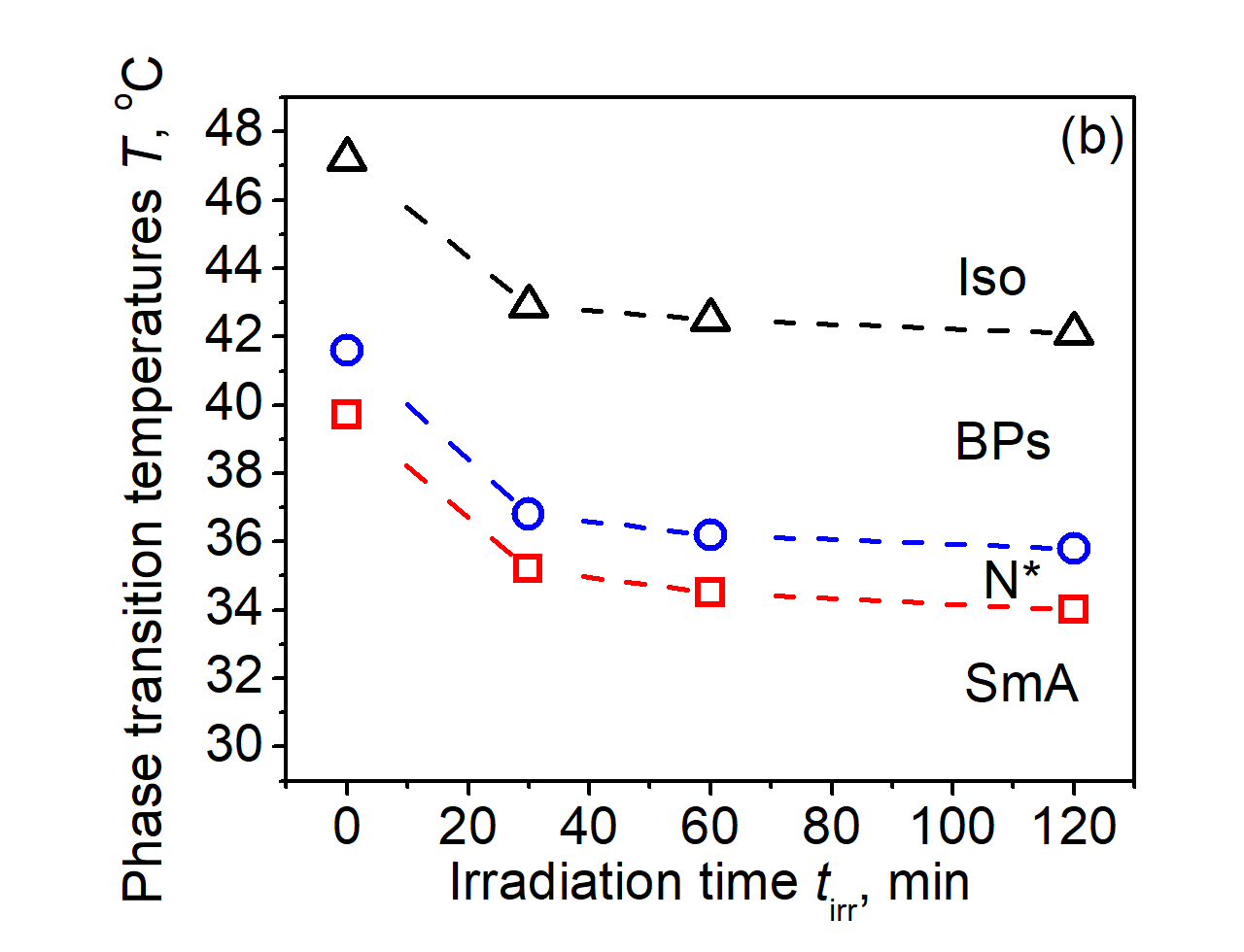} \\
        \hline
       aChD-3497 & 
       \includegraphics[width=3.7cm, valign=c]{aChD3497.png} &
        \includegraphics[width=3.5cm, valign=c]{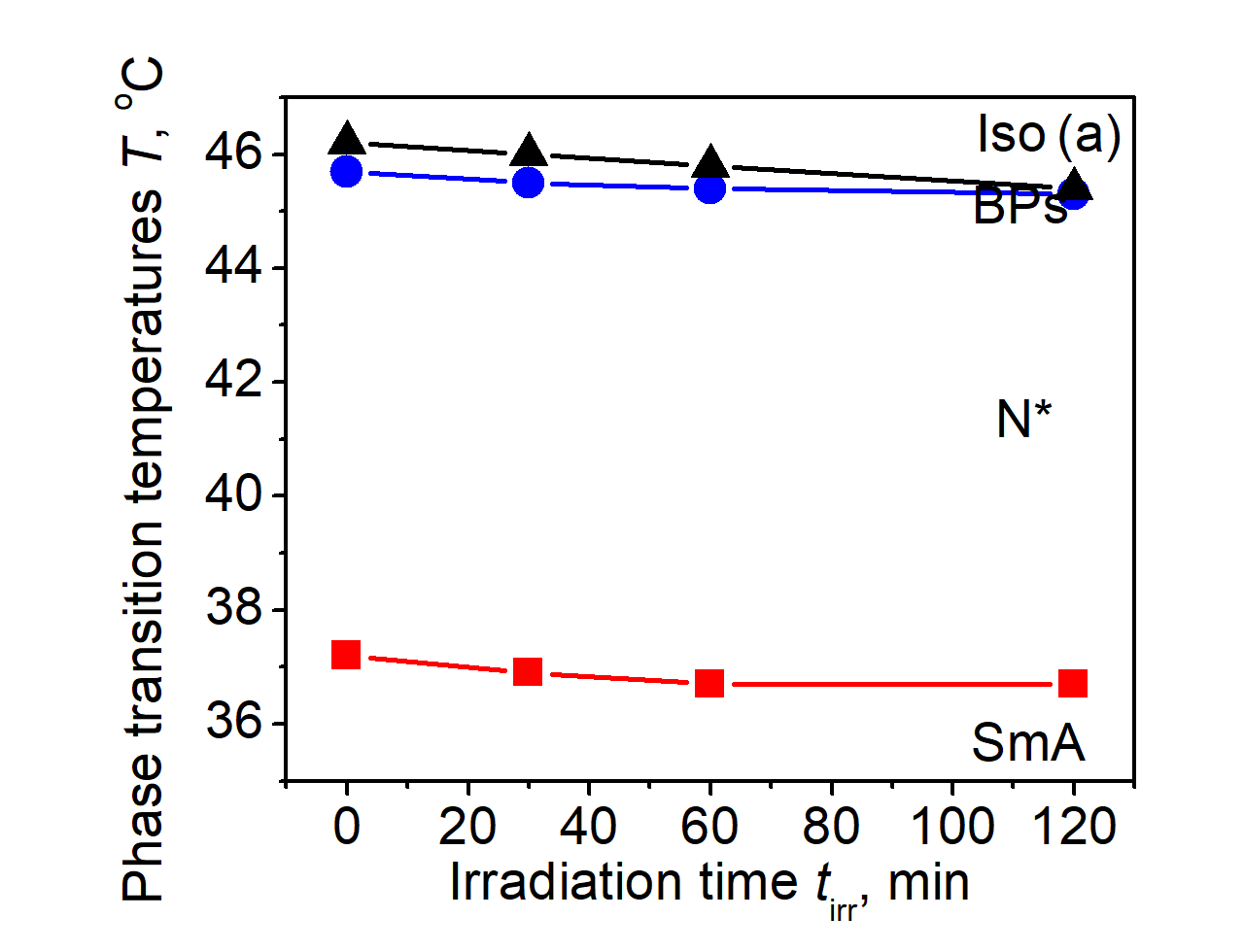} &\includegraphics[width=3.5cm, valign=c]{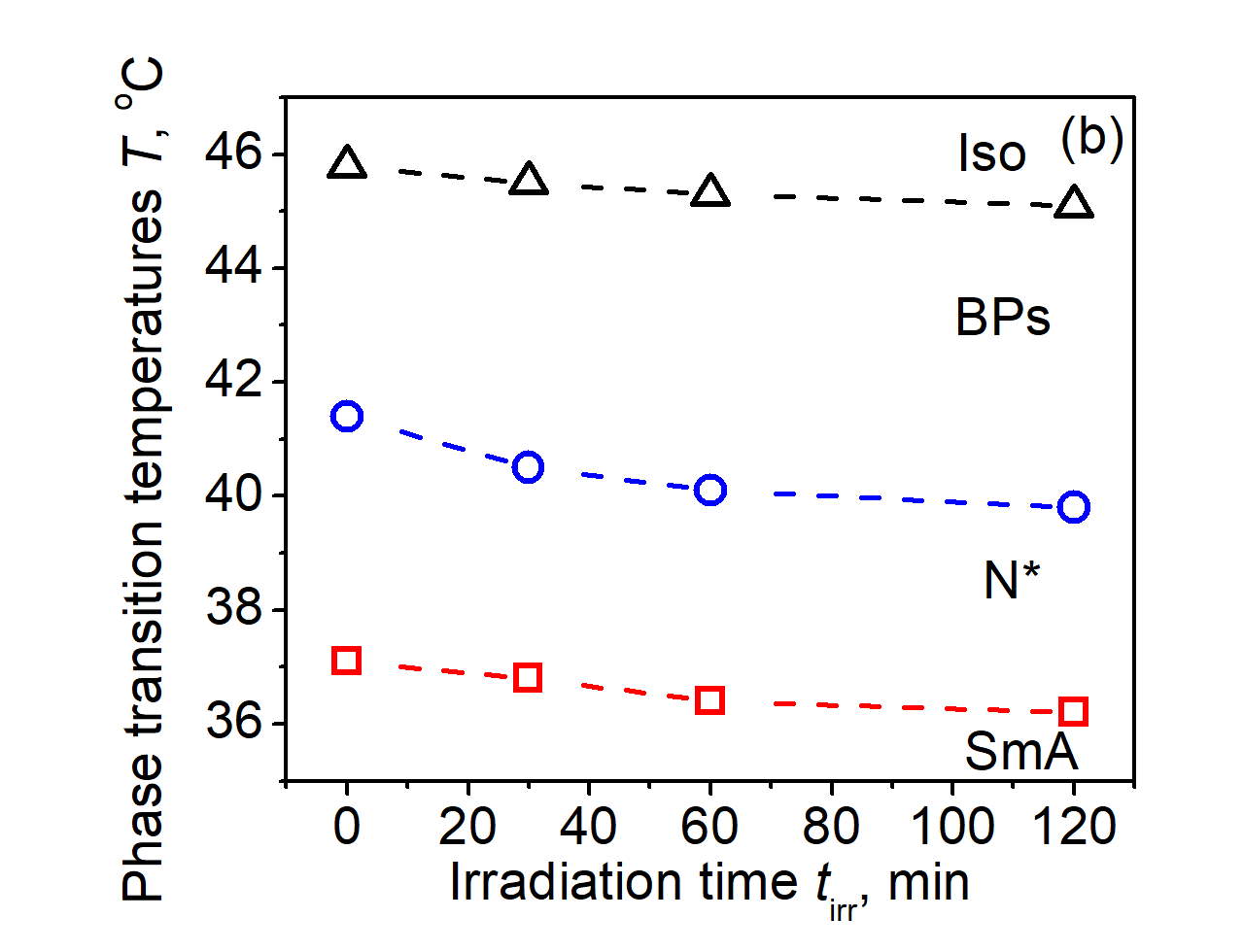} \\
        \hline
       aChD-3862 & 
       \includegraphics[width=4cm, valign=c]{aChD3862.png} &
        \includegraphics[width=3.5cm, valign=c]{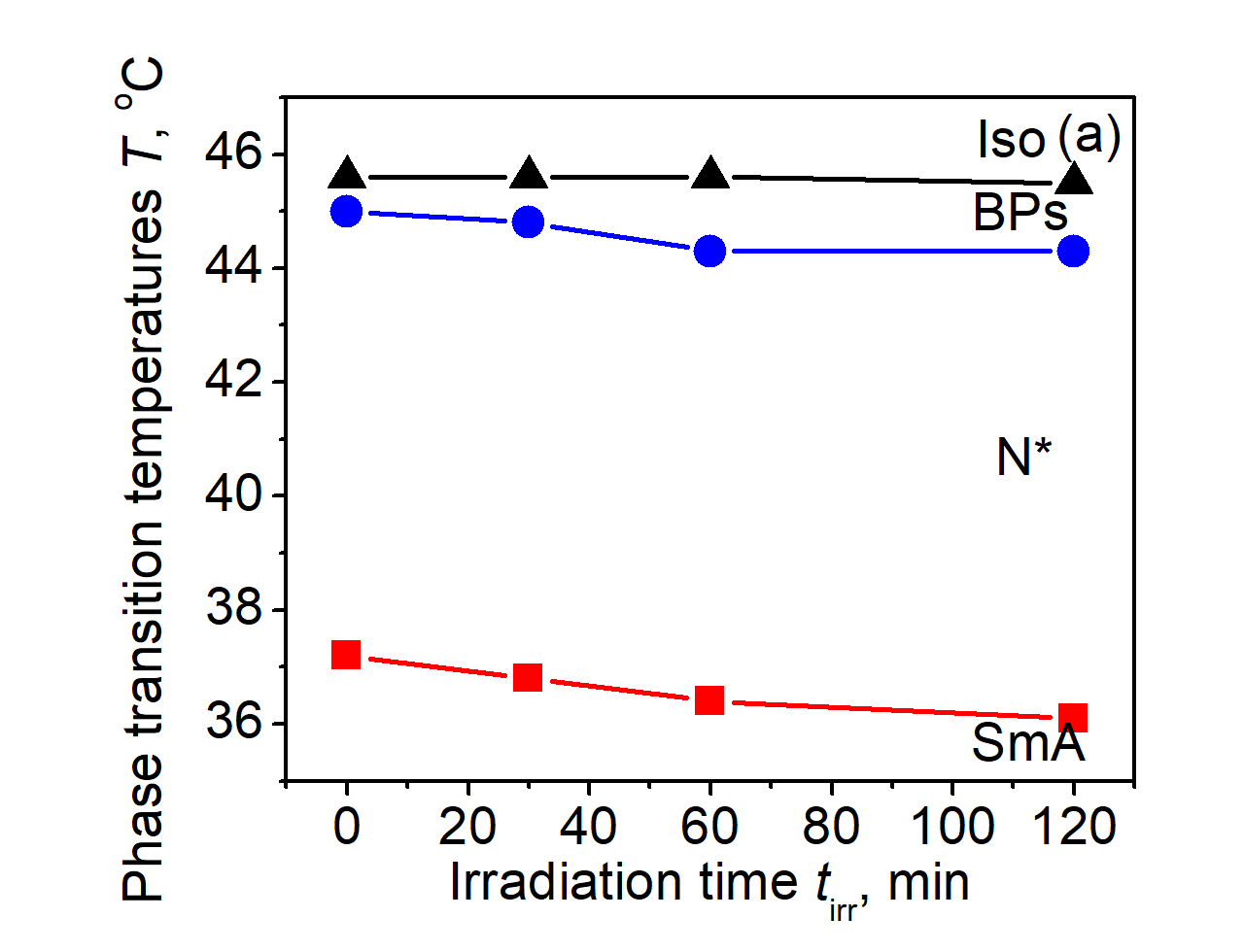} &\includegraphics[width=3.5cm, valign=c]{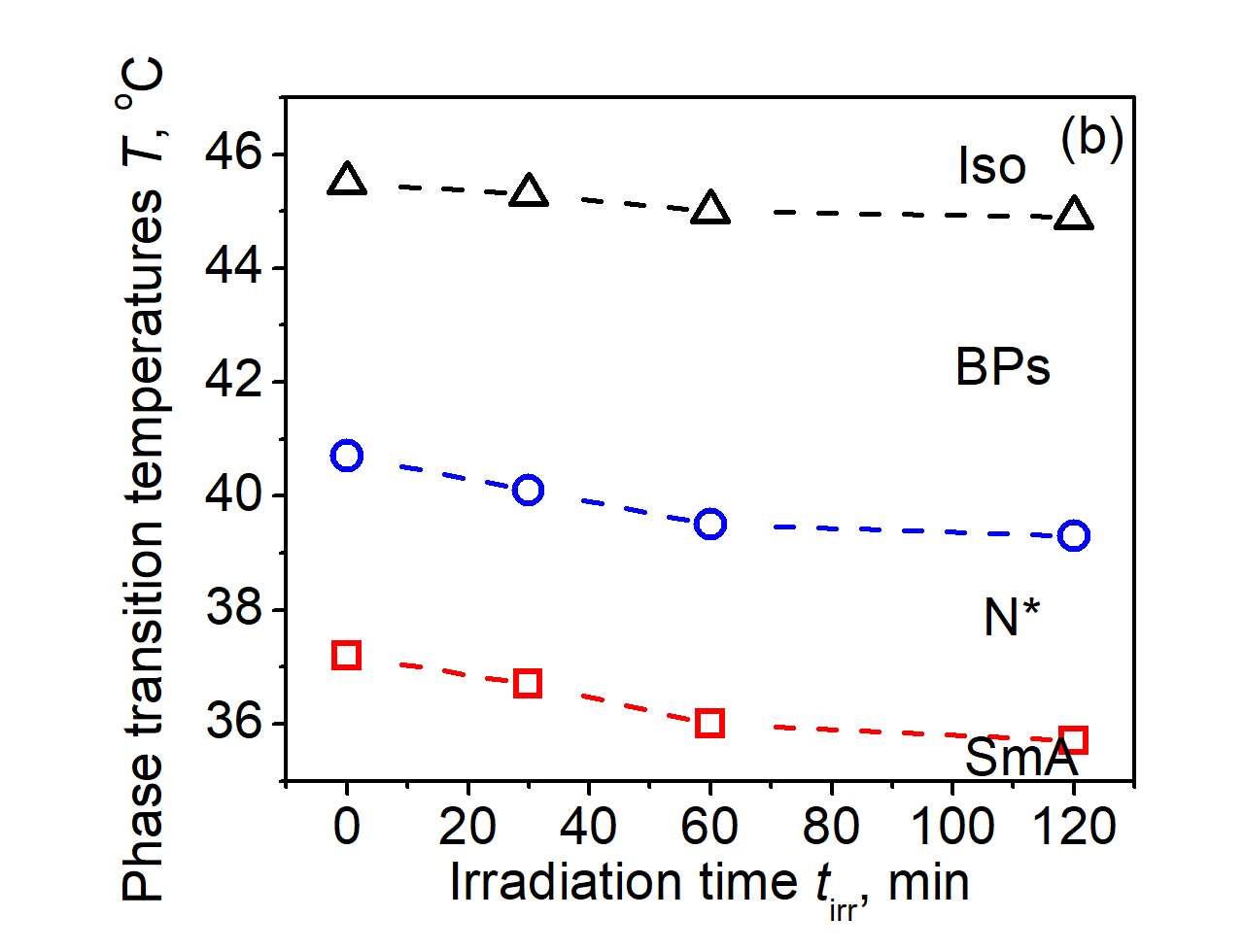} \\
        \hline
       aChD-4195 & 
       \includegraphics[width=2.3cm, valign=c]{aChD4195.png} &
        \includegraphics[width=3.5cm, valign=c]{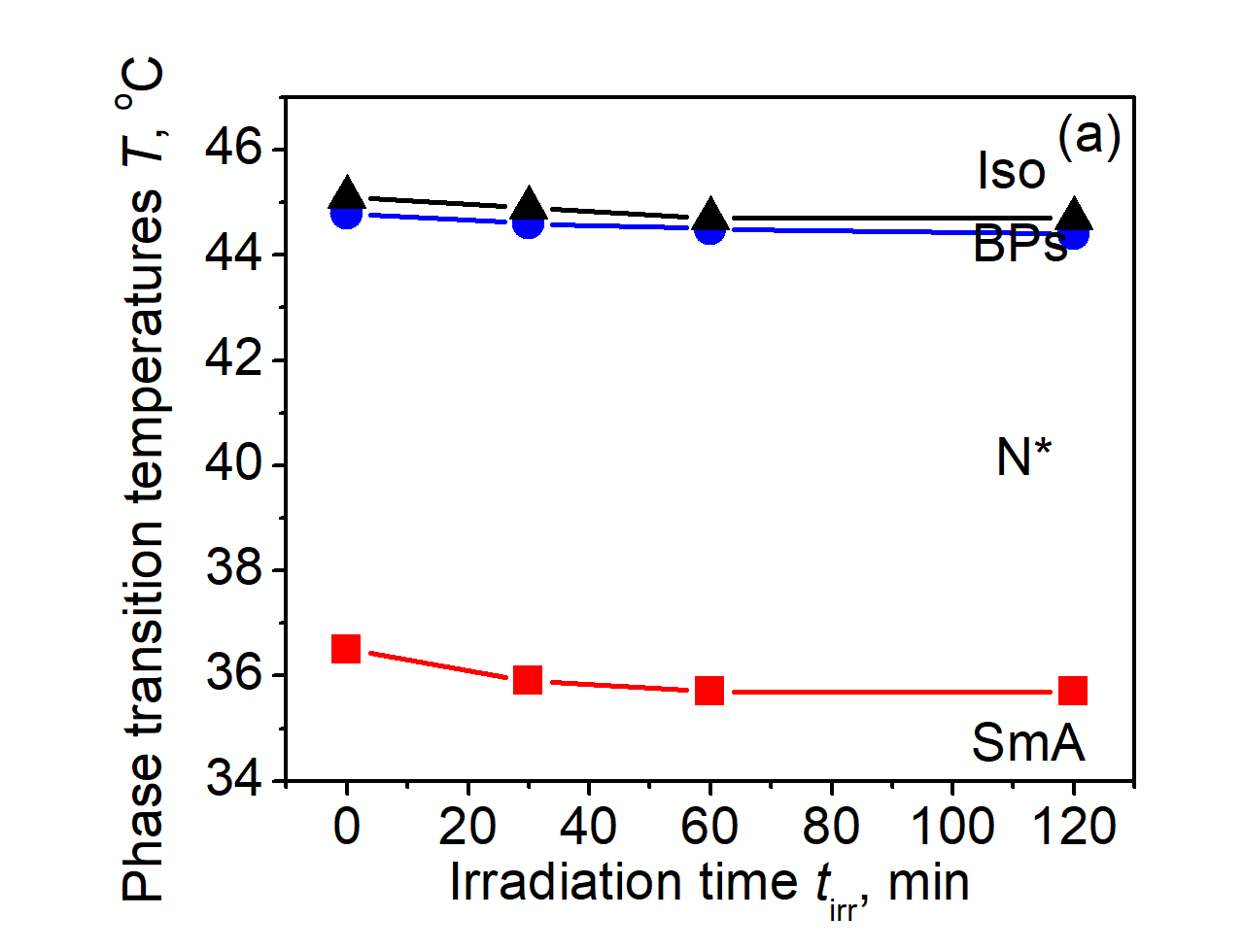} &\includegraphics[width=3.5cm, valign=c]{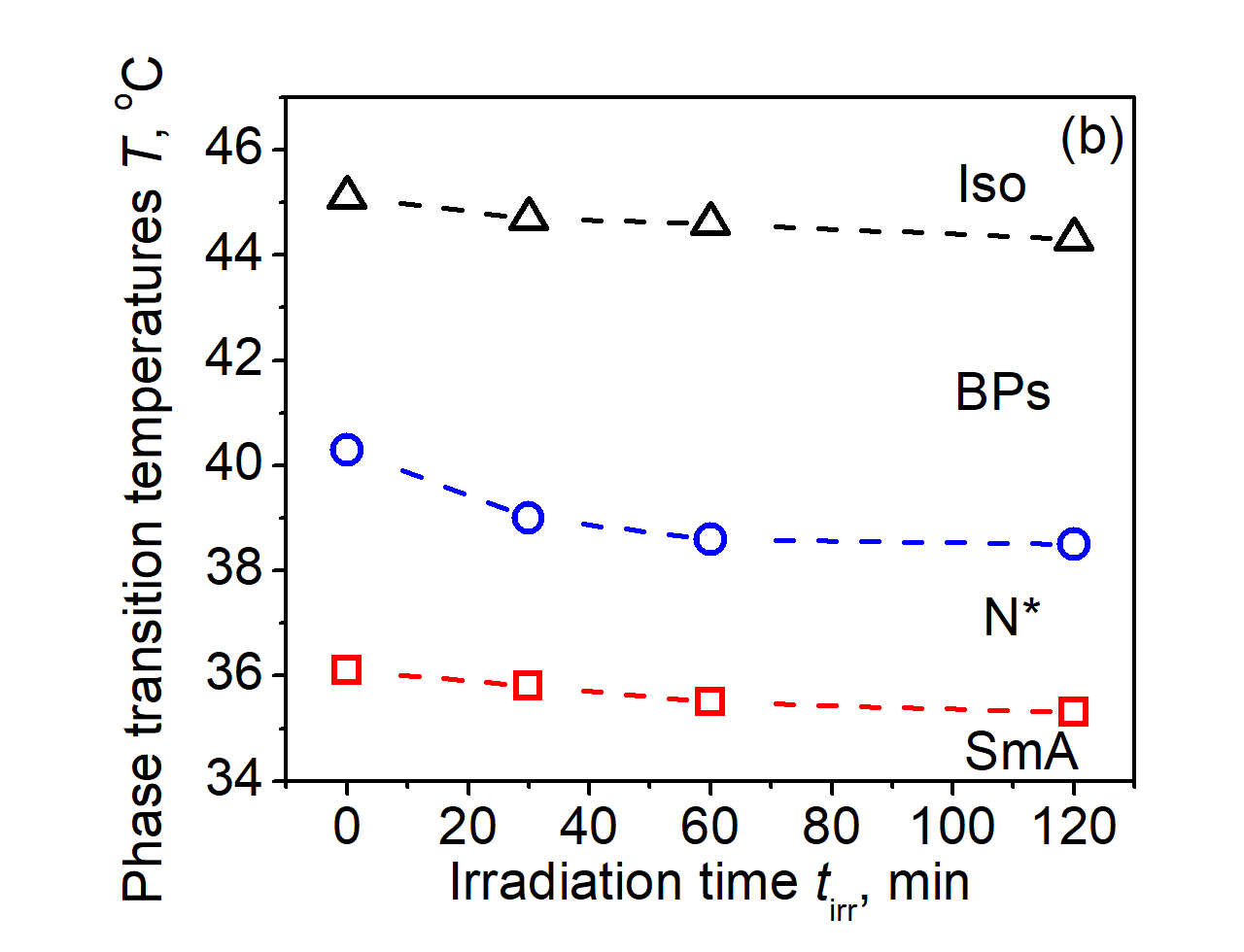} \\
        \hline 
       aChD-4197 & 
       \includegraphics[width=3.3cm, valign=c]{aChD4197.png} &
        \includegraphics[width=3.5cm, valign=c]{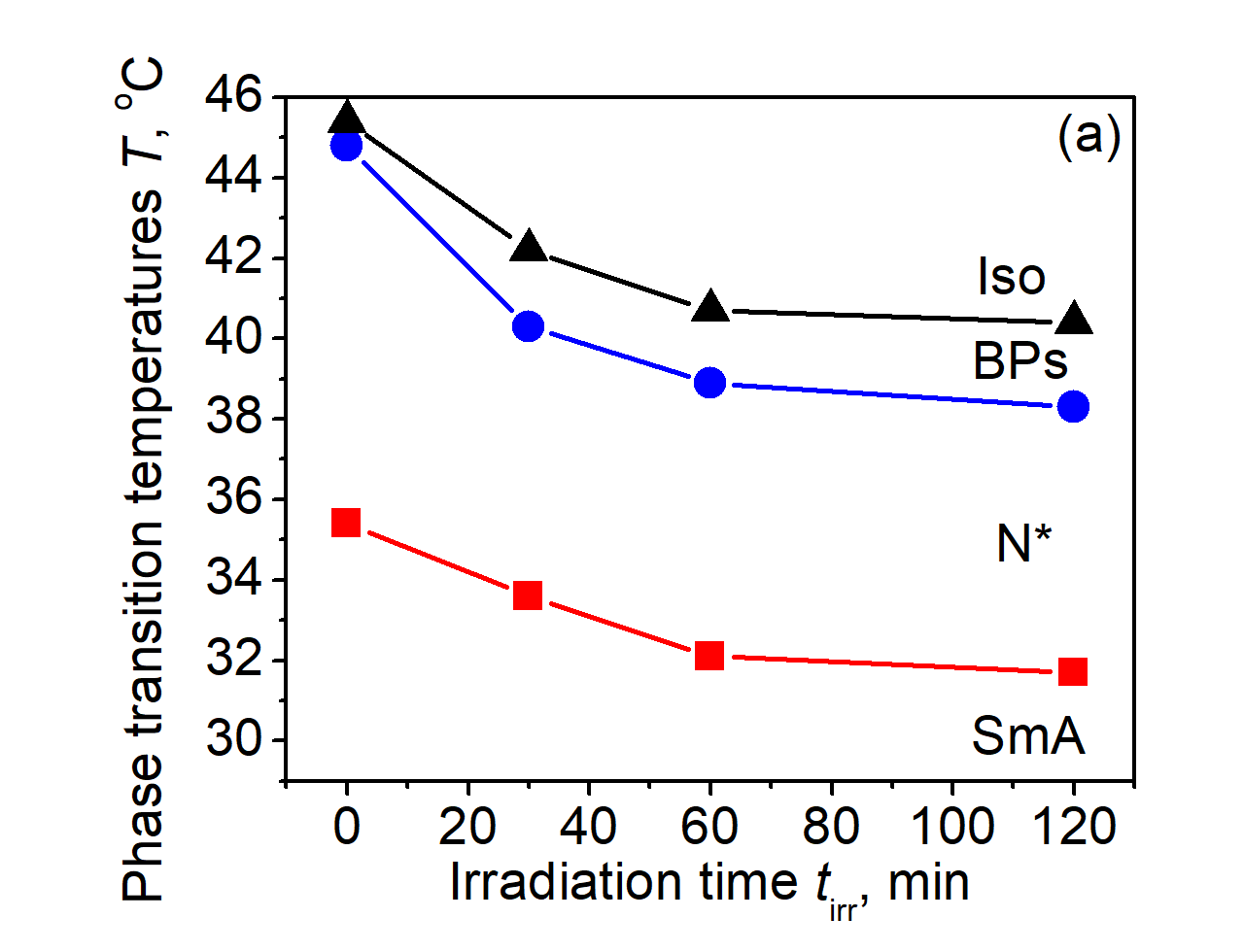} &\includegraphics[width=3.5cm, valign=c]{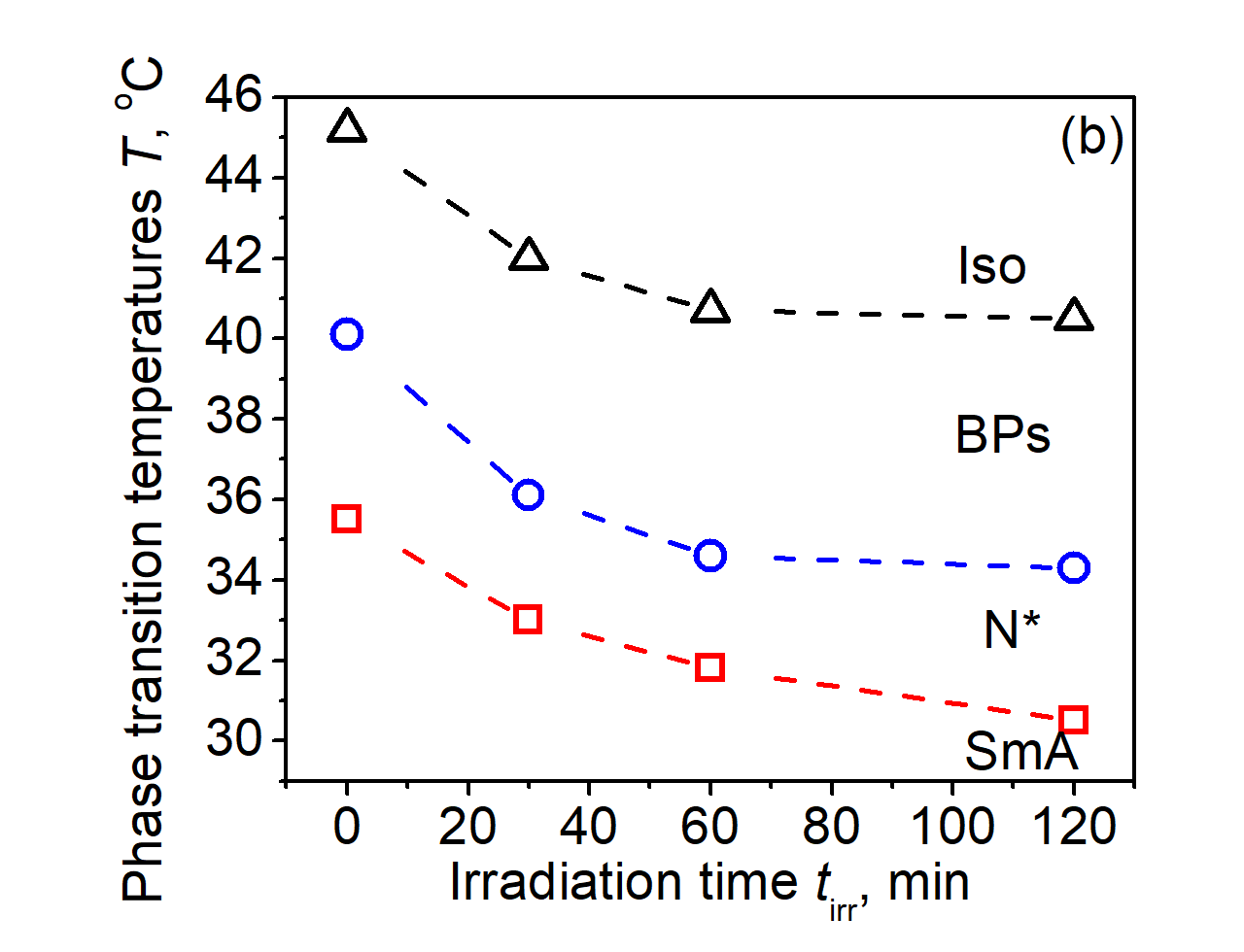} \\
        \hline
 \end{tabularx}%
}%
\vss%
\end{table}%
\newpage
\clearpage

\begin{sidewaystable}[!p]
\centering
\vspace*{-1.5cm}
    \caption{Phase transition temperatures of the base HChM doped with 5~wt\% of homologous series of chiral \textit{azo} compounds}
    \medskip
    \label{tab:Table4S5}
    \small
    \renewcommand{\arraystretch}{1.8} 
         
    \noindent 
    \setlength{\tabcolsep}{3.5pt}
    \newcolumntype{C}{>{\centering\arraybackslash\hspace{0pt}}X}
    
    \begin{tabularx}{0.99\linewidth}{|>{\hsize=0.6\hsize}C|
    >{\hsize=1.4\hsize}C|c|C|C|C|C|C|}
       \hline
       
        \textbf{\textit{Azo} compound} & \textbf{Chemical formula (\textit{trans}-isomer)} & \textbf{Process} & \multicolumn{5}{c|}{\textbf{Phase transitions ($^\circ$C)}} \\ 
        \cline{4-8} 
        & & & \cellcolor{green!60} \textbf{Before UV} & \cellcolor{magenta!20} \textbf{30~min} & \cellcolor{magenta!50} \textbf{60~min} & \cellcolor{magenta!80} \textbf{120~min} & \cellcolor{green!60}\textbf{Recovery$^a$} \\ 
        \midrule
       
        \multirow{2}{*}{ChD-4185} & 
        \multirow{2}{*}{\includegraphics[width=2.4cm, valign=c]{ChD-4185.png}} & \cellcolor{red!20}
        \text{Heating} & SmA 30.2 $\text{N}^*$ 37.4 BPs 38 Iso & SmA 27.9 $\text{N}^*$ 34.7 BPs 36.7 Iso & SmA 26.7 $\text{N}^*$ 33.8 BPs 36.4 Iso & SmA 25.7 $\text{N}^*$ 33.4 BPs 36.2 Iso & SmA 30.2 $\text{N}^*$ 37.4 BPs 38 Iso \\
        \cline{3-8}
        & & \cellcolor{cyan!20} \text{Cooling} & Iso 37.9 BPs 33.6 $\text{N}^*$ 29.2 SmA & Iso 36.6 BPs 31.6 $\text{N}^*$ 27.4 SmA & Iso 36.2 BPs 30.9 $\text{N}^*$ 26.1 SmA & Iso 36.1 BPs 30.8 $\text{N}^*$ 25.6 SmA & Iso 37.9 BPs 33.6 $\text{N}^*$ 29.2 SmA \\
        \hline
        \multirow{2}{*}{ChD-4187} & 
        \multirow{2}{*}{\includegraphics[width=2.8cm, valign=c]{ChD-4187.png}} & \cellcolor{red!20}
        \text{Heating} & SmA 29.8 $\text{N}^*$ 38.4 BPs 38.9 Iso & SmA 28.1 $\text{N}^*$ 36 BPs 37.8 Iso & SmA 27.8 $\text{N}^*$ 35.3 BPs 37.5 Iso & SmA 27.6 $\text{N}^*$ 35.1 BPs 37.3 Iso & SmA 29.8 $\text{N}^*$ 38.4 BPs 38.9 Iso\\
        \cline{3-8}
        & & \cellcolor{cyan!20} \text{Cooling} & Iso 38.8 BPs 34.4 $\text{N}^*$ 29.7 SmA & Iso 37.5 BPs 32.4 $\text{N}^*$ 27.8 SmA & Iso 37.3 BPs 32.1 $\text{N}^*$ 27.3 SmA & Iso 37.1 BPs 31.5 $\text{N}^*$ 27.3 SmA & Iso 38.8 BPs 34.4 $\text{N}^*$ 29.7 SmA \\
        \hline
        \multirow{2}{*}{ChD-3816} & 
        \multirow{2}{*}{\includegraphics[width=3.2cm, valign=c]{ChD-3816.png}} & \cellcolor{red!20}
        \text{Heating} & SmA 36.4 $\text{N}^*$ 44.4 BPs 44.7 Iso & SmA 36.1 $\text{N}^*$ 44.3 BPs 44.5 Iso & SmA 35.7 $\text{N}^*$ 44.2 BPs 44.5 Iso & SmA 35.5 $\text{N}^*$ 44 BPs 44.5 Iso & SmA 36.4 $\text{N}^*$ 44.4 BPs 44.7 Iso \\
        \cline{3-8}
        & & \cellcolor{cyan!20} \text{Cooling} & Iso 44.5 BPs 40 $\text{N}^*$ 35.8 SmA & Iso 43.8 BPs 38.6 $\text{N}^*$ 34.3 SmA & Iso 43.4 BPs 37.8 $\text{N}^*$ 33.7 SmA & Iso 43.1 BPs 37.4 $\text{N}^*$ 33.4 SmA & Iso 44.5 BPs 40 $\text{N}^*$ 35.8 SmA \\
        \hline
        \multirow{2}{*}{ChD-4211} & 
        \multirow{2}{*}{\includegraphics[width=3.6cm, valign=c]{ChD-4211.png}} & \cellcolor{red!20}
        \text{Heating} & SmA 35.4 $\text{N}^*$ 41.9 BPs 43.3 Iso & SmA 33.1 $\text{N}^*$ 39.8 BPs 42 Iso & SmA 31.5 $\text{N}^*$ 39.8 BPs 41.2 Iso & SmA 31.2 $\text{N}^*$ 39.2 BPs 41 Iso & SmA 35.4 $\text{N}^*$ 41.9 BPs 43.3 Iso \\
        \cline{3-8}
        & & \cellcolor{cyan!20} \text{Cooling} & Iso 42.9 BPs 37.6 $\text{N}^*$ 36 SmA & Iso 41.8 BPs 36.1 $\text{N}^*$ 32.9 SmA & Iso 41.4 BPs 35.5 $\text{N}^*$ 32.3 SmA & Iso 40.8 BPs 34.7 $\text{N}^*$ 31.8 SmA & Iso 42.9 BPs 37.6 $\text{N}^*$ 36 SmA \\
        \hline
        \multirow{2}{*}{ChD-4212} & 
        \multirow{2}{*}{\includegraphics[width=3.8cm, valign=c]{ChD-4212.png}} & \cellcolor{red!20}
        \text{Heating} & SmA 36.1 $\text{N}^*$ 41.5 BPs 42.2 Iso & SmA 33.2 $\text{N}^*$ 38.6 BPs 40.7 Iso & SmA 32.2 $\text{N}^*$ 38.1 BPs 40.3 Iso & SmA 31.4 $\text{N}^*$ 37.6 BPs 39.7 Iso & SmA 36.1 $\text{N}^*$ 41.5 BPs 42.2 Iso \\
        \cline{3-8}
        & & \cellcolor{cyan!20} \text{Cooling} & Iso 42.1 BPs 36.7 $\text{N}^*$ 35.6 SmA & Iso 40.6 BPs 34.8 $\text{N}^*$ 31.9 SmA & Iso 40.1 BPs 34.1 $\text{N}^*$ 31.2 SmA & Iso 39.5 BPs 33.2 $\text{N}^*$ 30.3 SmA & Iso 42.1 BPs 36.7 $\text{N}^*$ 35.6 SmA \\
        \hline
        \multirow{2}{*}{ChD-3805} & 
        \multirow{2}{*}{\includegraphics[width=3.6cm, valign=c]{ChD-3805.png}} & \cellcolor{red!20}
        \text{Heating} & SmA 36.3 $\text{N}^*$ 41.6 BPs 42.7 Iso & SmA 33.5 $\text{N}^*$ 39.3 BPs 41 Iso & SmA 33 $\text{N}^*$ 38.8 BPs 40.5 Iso & SmA 32.4 $\text{N}^*$ 38.7 BPs 40.5 Iso & SmA 36.3 $\text{N}^*$ 41.6 BPs 42.7 Iso \\
        \cline{3-8}
        & & \cellcolor{cyan!20} \text{Cooling} & Iso 42.6 BPs 37.1 $\text{N}^*$ 35.7 SmA & Iso 42.1 BPs 36.2 $\text{N}^*$ 33.5 SmA & Iso 41.7 BPs 35.4 $\text{N}^*$ 32.9 SmA & Iso 41.4 BPs 34.9 $\text{N}^*$ 32.1 SmA & Iso 42.6 BPs 37.1 $\text{N}^*$ 35.7 SmA \\
        \hline
        
    \end{tabularx}
\vspace{5pt}
    \flushleft \footnotesize $^a$ Reversible \textit{cis}--\textit{trans} isomerization at 80\,$^\circ$C for 30~min.
\end{sidewaystable}
\newpage
\clearpage

\begin{table}[H]
   \centering
 \caption{Photo-induced phase transitions of the base HChM doped with 5~wt\% of a homologous series of chiral \textit{azo} compound}
    \label{tab:Table4S6}
\medskip
     \noindent\centerline{%
   \setlength{\tabcolsep}{2.5pt}
   \renewcommand{\arraystretch}{1.5}
\begin{tabularx}{0.99\textwidth}{|
    >{\centering\arraybackslash}m{1.8cm}|
    >{\centering\arraybackslash}m{4cm}|
    >{\centering\arraybackslash}X|
    >{\centering\arraybackslash}X|}
        \hline
  \textbf{\shortstack{\textit{Azo}\\compound}} & 
  \textbf{Chemical formula (\textit{trans}-isomer)} & 
  \multicolumn{2}{c|}{\textbf{Photo-induced phase diagrams}} 
        \\ 
  & & \cellcolor{red!20}\textbf{Upon heating} & \cellcolor{cyan!20}\textbf{Upon cooling} \\ 
        \hline
        ChD-4185 & 
        \includegraphics[width=2.8cm, valign=c]{ChD-4185.png} &
        \includegraphics[width=3.5cm, valign=c]{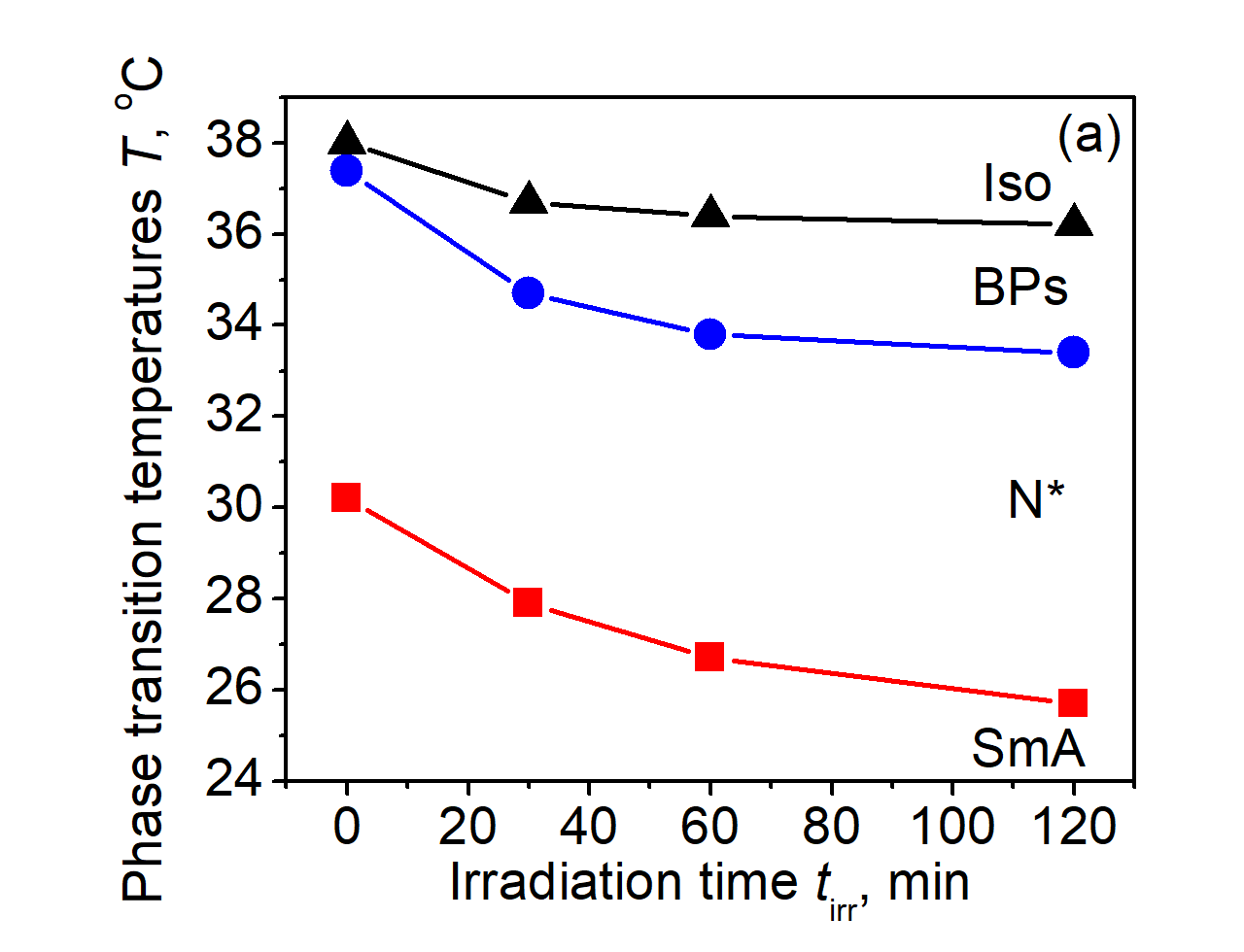} &\includegraphics[width=3.5cm, valign=c]{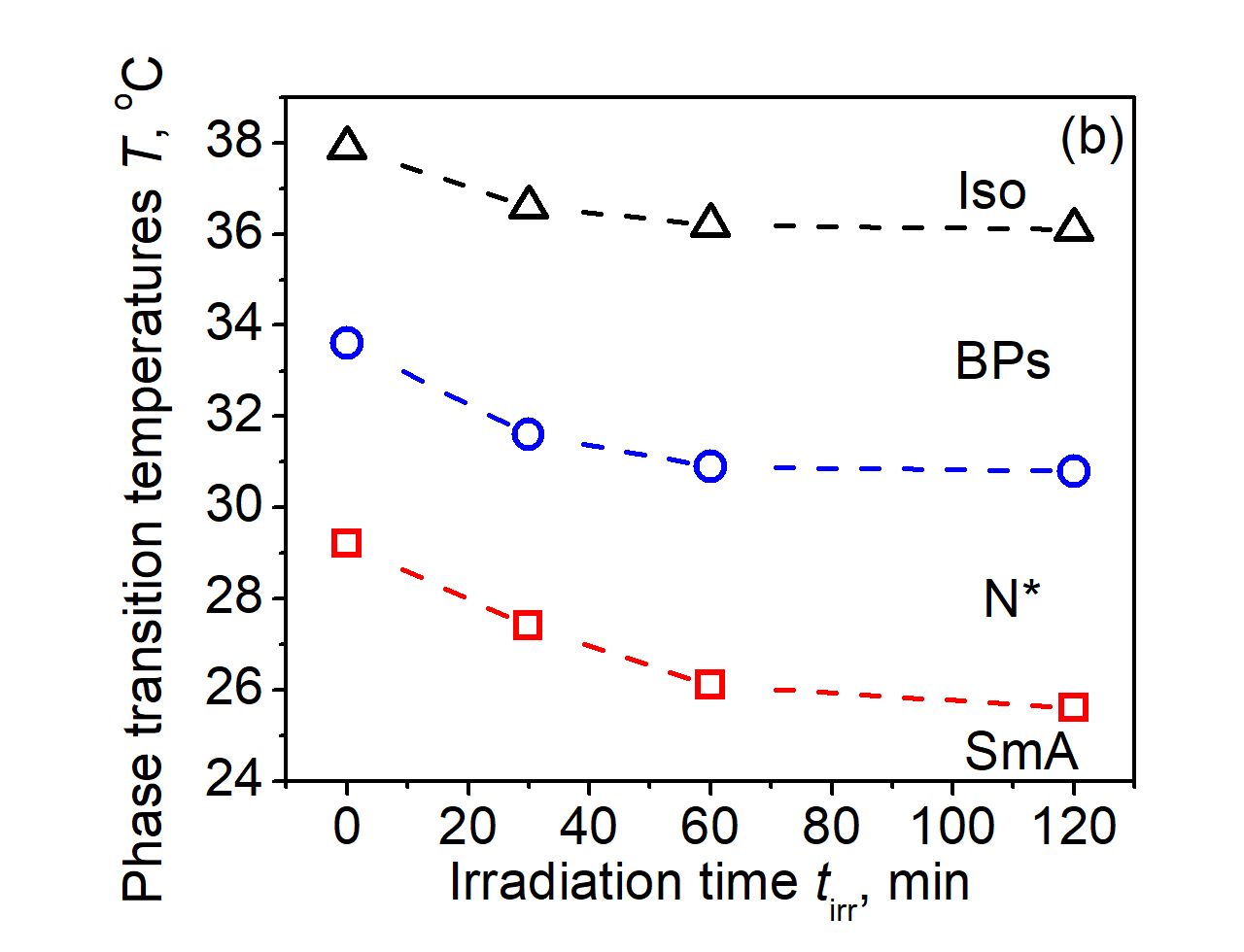} \\
       \hline
      ChD-4187 & 
      \includegraphics[width=3.2cm, valign=c]{ChD-4187.png} &
        \includegraphics[width=3.5cm, valign=c]{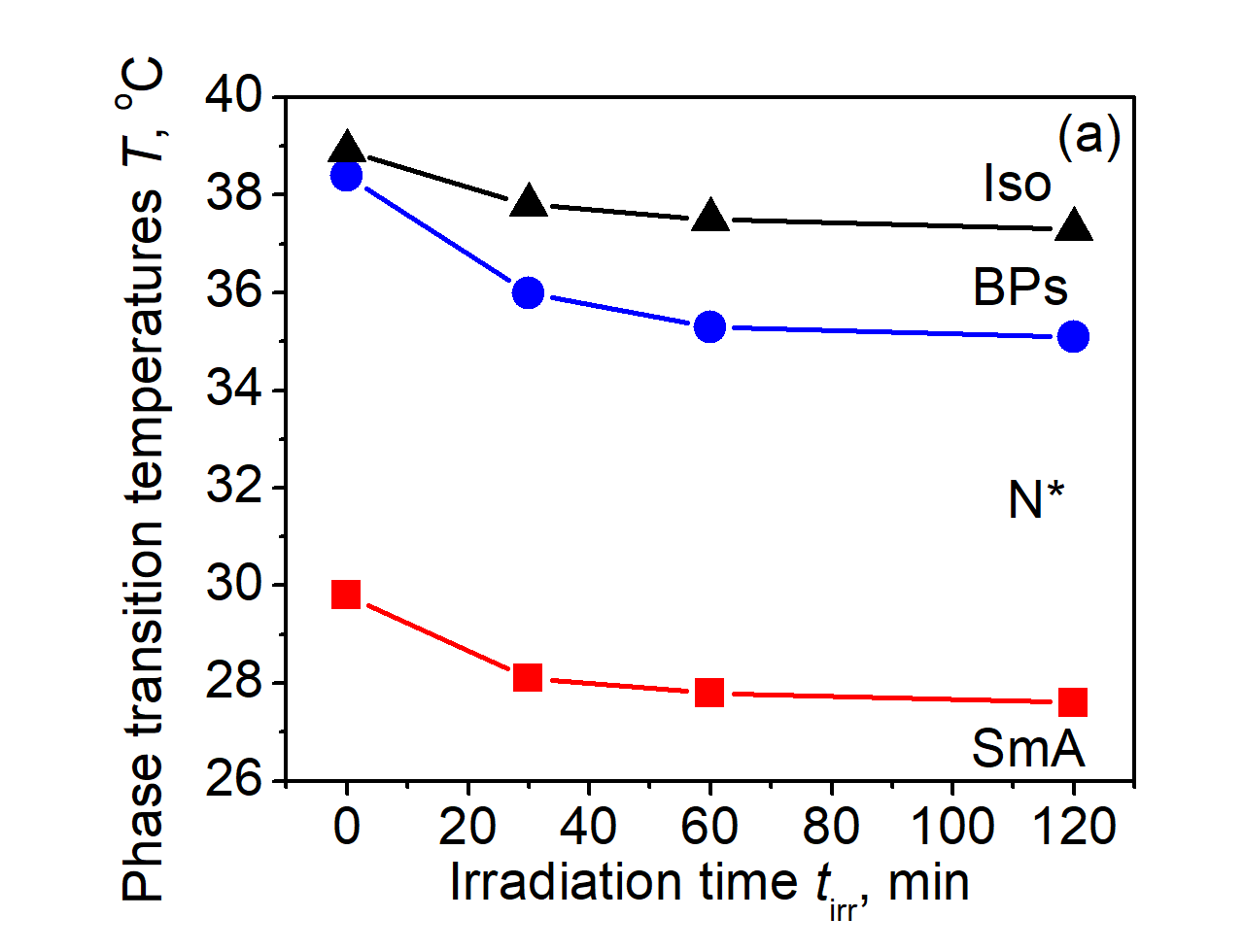} &\includegraphics[width=3.5cm, valign=c]{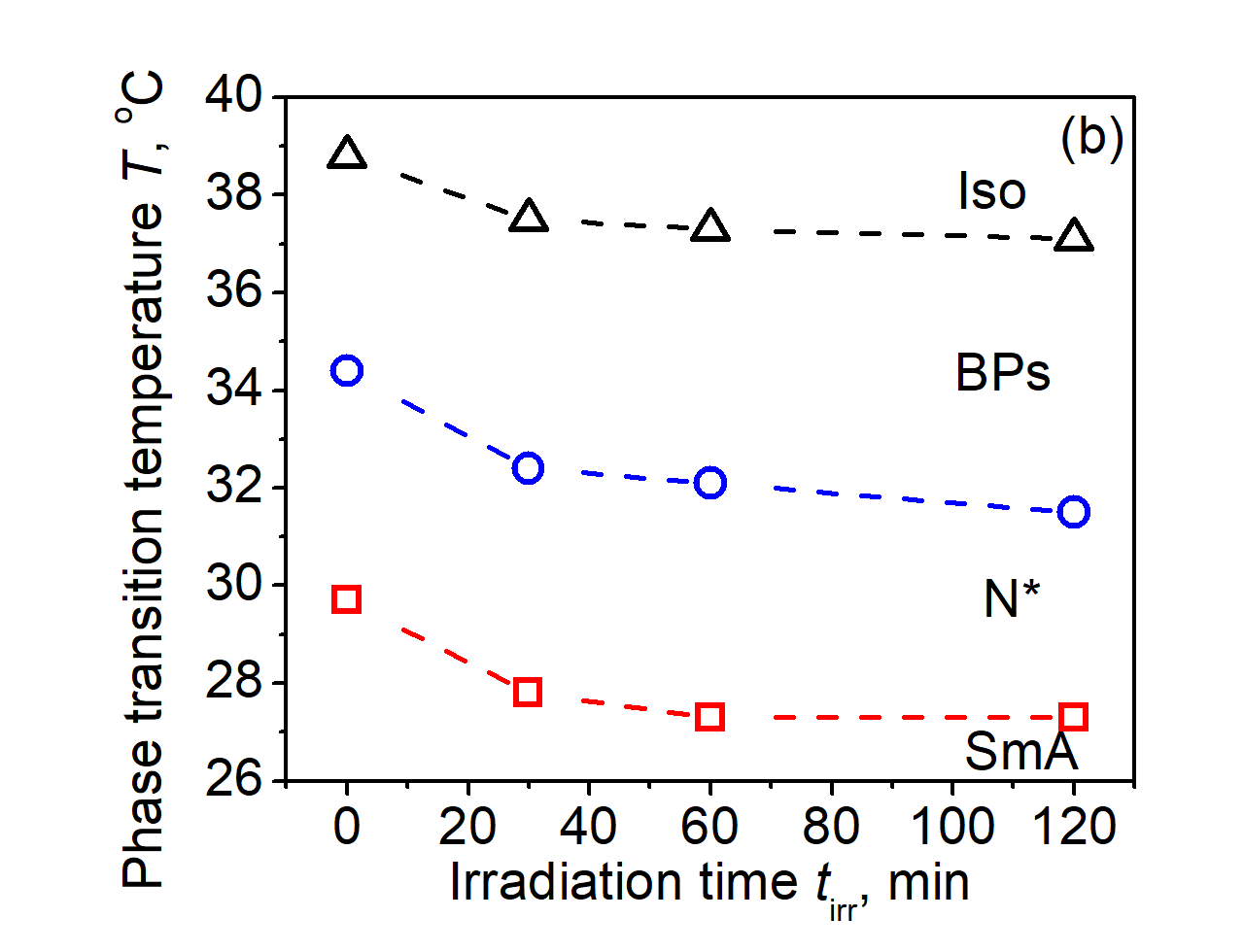} \\
        \hline
       ChD-3816 & 
       \includegraphics[width=3.5cm, valign=c]{ChD-3816.png} &
        \includegraphics[width=3.5cm, valign=c]{Table_4S2_ChD-3816_heating.png} &\includegraphics[width=3.5cm, valign=c]{Table_4S2_ChD-3816_cooling.png} \\
        \hline
       ChD-4211 & 
       \includegraphics[width=3.8cm, valign=c]{ChD-4211.png} &
        \includegraphics[width=3.5cm, valign=c]{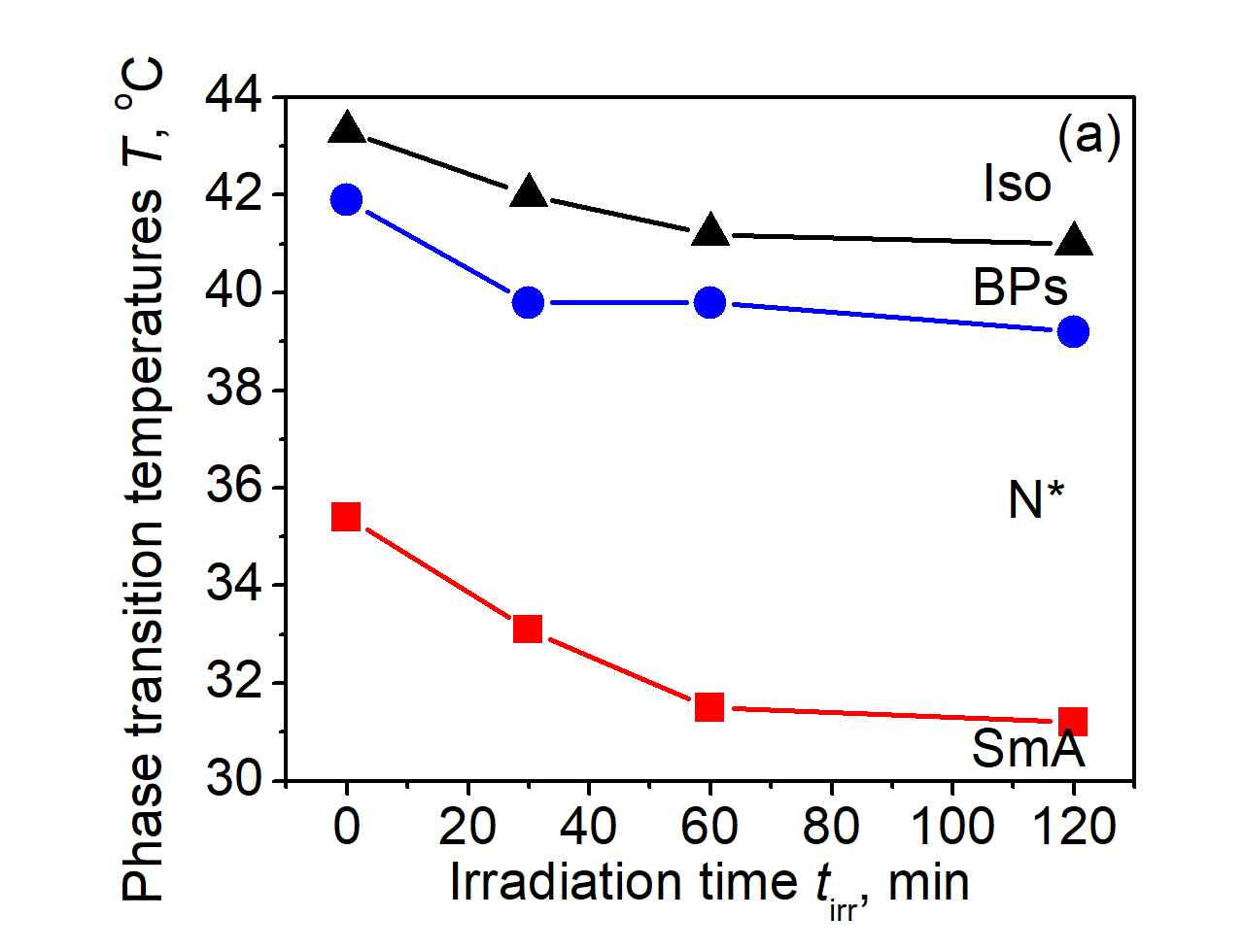} &\includegraphics[width=3.5cm, valign=c]{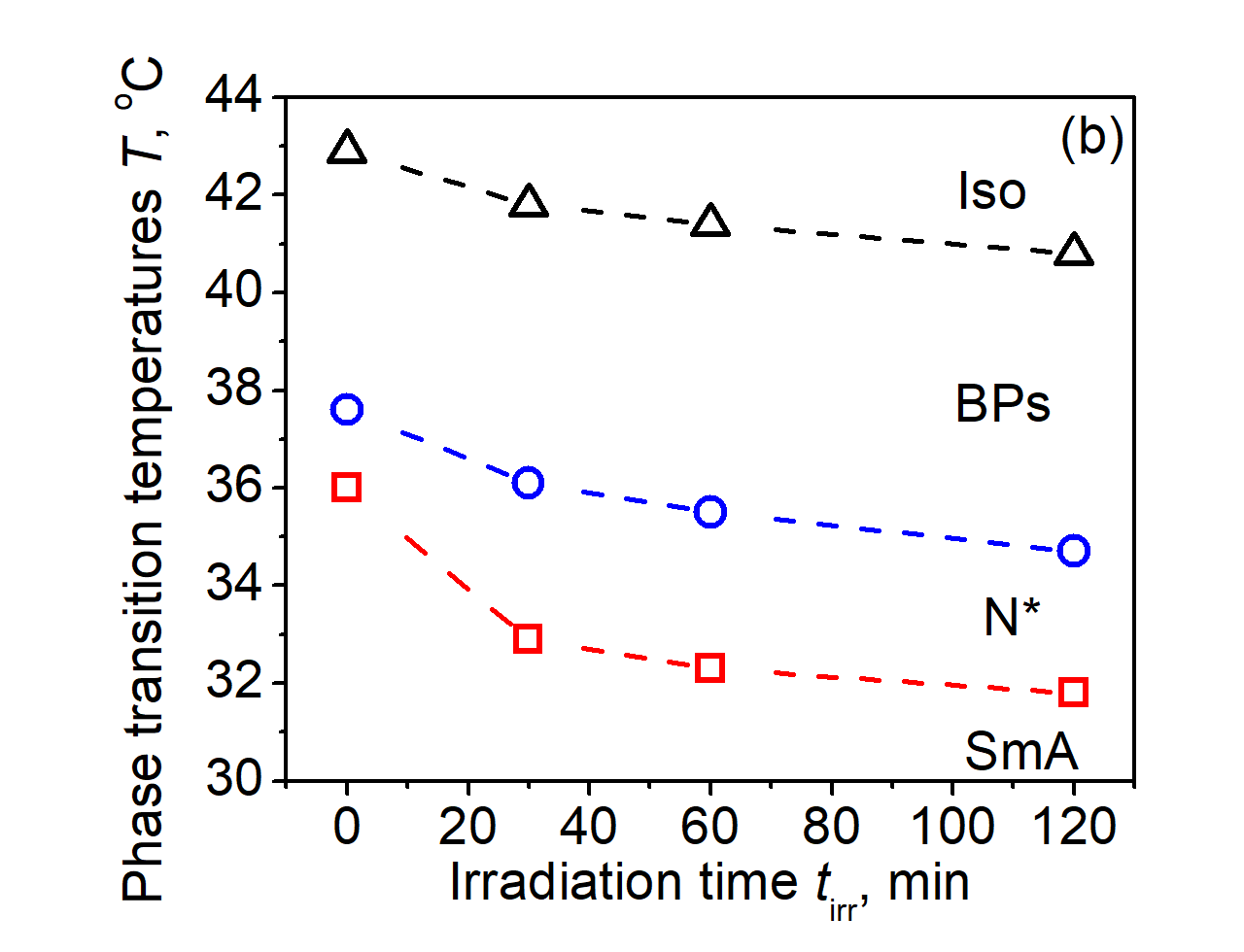} \\
        \hline
       ChD-4212 & 
       \includegraphics[width=4cm, valign=c]{ChD-4212.png} &
        \includegraphics[width=3.5cm, valign=c]{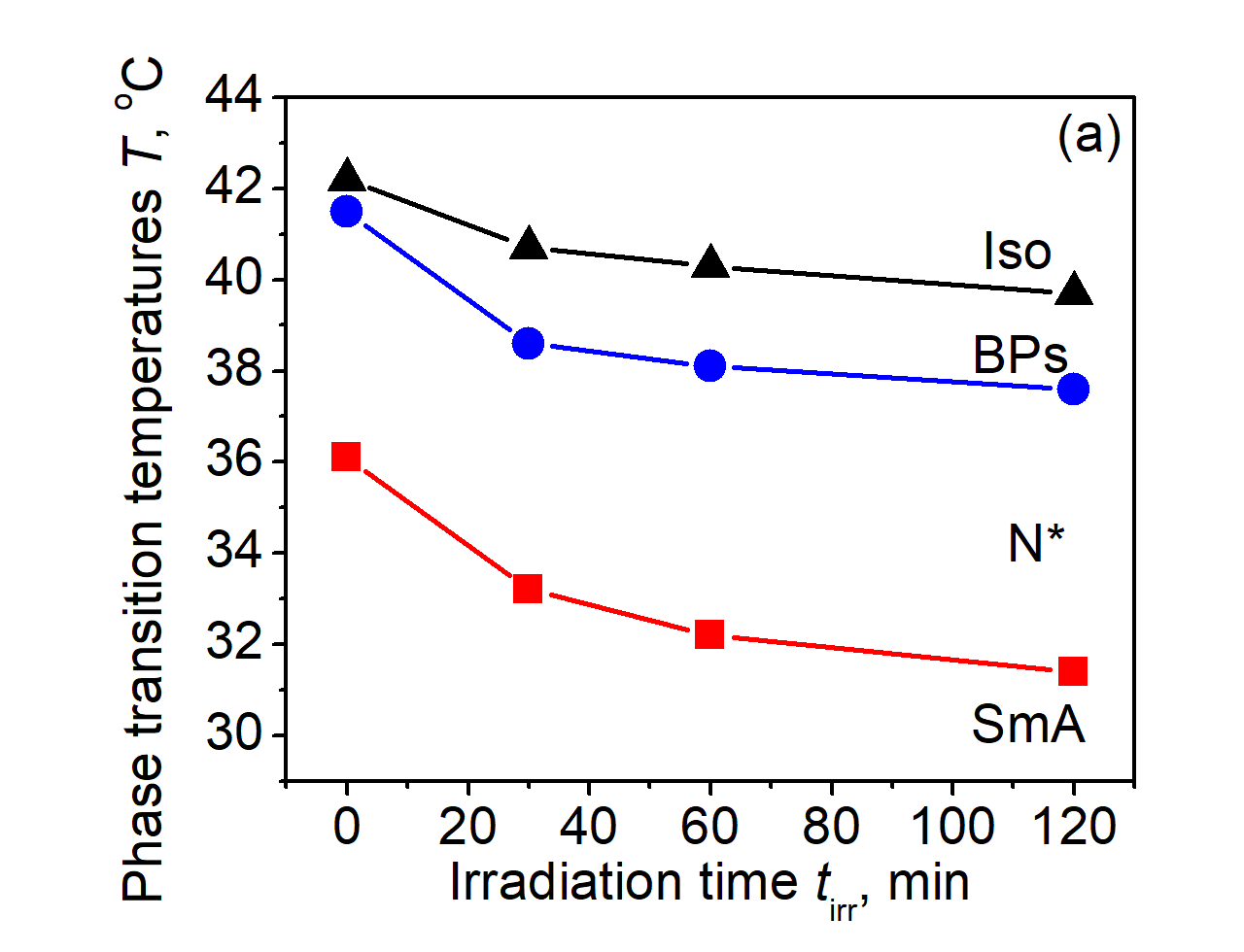} &\includegraphics[width=3.5cm, valign=c]{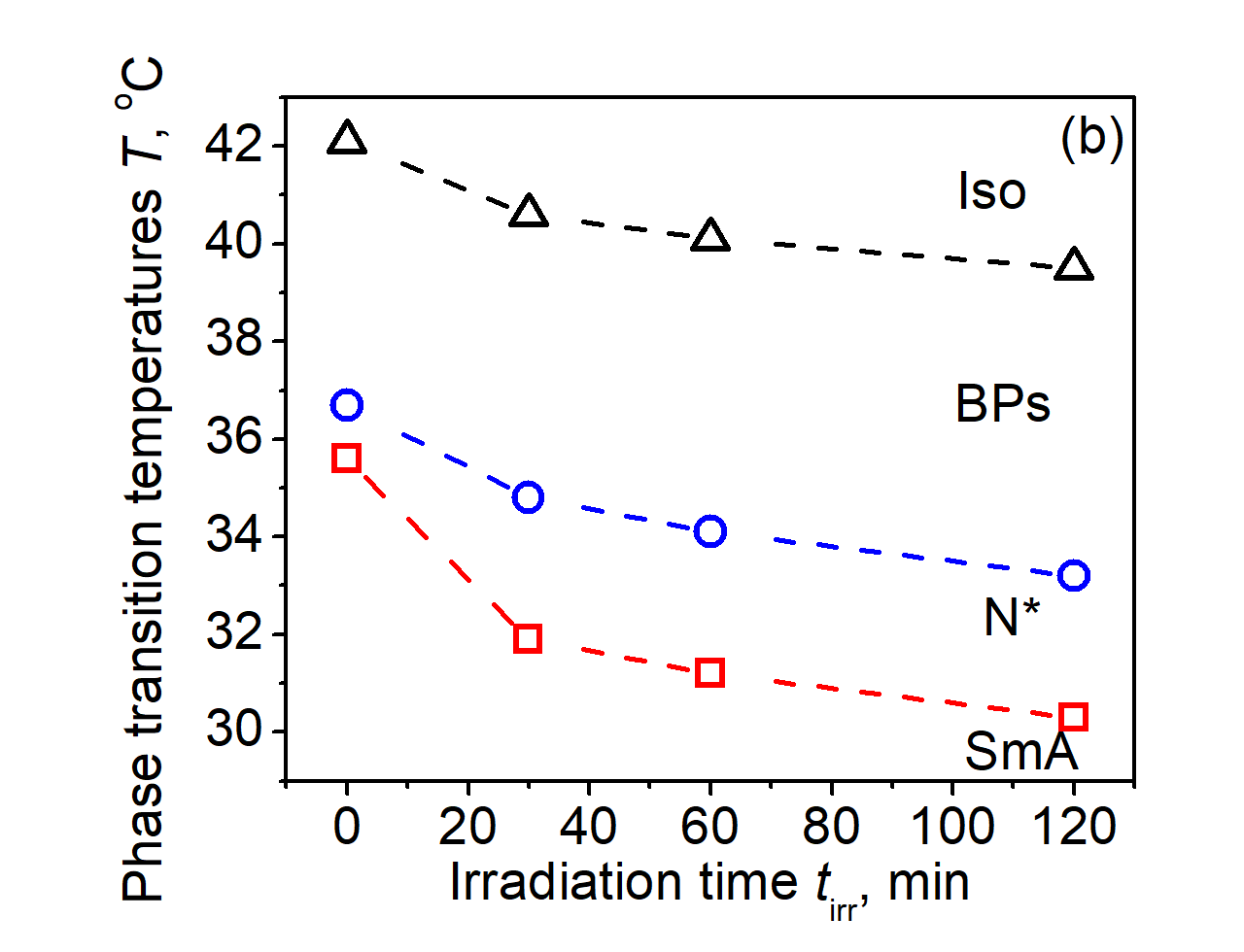} \\
        \hline 
       ChD-3805 & 
      \includegraphics[width=4cm, valign=c]{ChD-3805.png} &
        \includegraphics[width=3.5cm, valign=c]{Table_4S2_ChD-3805_heating.png} &\includegraphics[width=3.5cm, valign=c]{Table_4S2_ChD-3805_cooling.png} \\
        \hline
\end{tabularx}%
}%
\vss%
\end{table}%
\newpage
\clearpage


\section{\textit{Cis}-isomer efficiency as a parameter for estimating the broadening of the BP temperature range}

To compare the influence of UV irradiation --- which leads to the formation of \textit{cis}-isomer of the \textit{azo} compound within the base HChM (65~wt\% COC and 35~wt\% E7) --- on the broadening of the BP temperature range, we introduce a \textit{cis}-isomer efficiency coefficient, $\xi$, defined as follows:

\begin{equation}
  \xi = \frac{\Delta T^{t_{\text{irr}}=i}_{\text{BP}}}{\Delta T^{t_{\text{irr}}=0}_{\text{BP}}} 
\end{equation}

where $\Delta T^{t_{\text{irr}}=0}_{\text{BP}}$ and $\Delta T^{t_{\text{irr}}=i}_{\text{BP}}$ are the BP temperature ranges (including BP-I, BP-II and BP-III) before ($t_{\text{irr}} = 0$~min) and after ($t_{\text{irr}} = i$~min, where $0 < i \leq 30$) UV irradiation of the BPM, respectively.

Upon UV irradiation, the concentration of the \textit{cis}-isomer increases. During short exposure time $t_{\text{irr}}$, an almost linear dependence of the BP temperature range ($\Delta{T}_{\text{BP}}$) on $t_{\text{irr}}$ is observed. Fig.~S5.1 shows near-linear dependences for various BPMs, containing either the chiral or achiral \textit{azo} compound at irradiation times $t_{\text{irr}} \leq  30$~min. 

\begin{figure}[ht]
\begin{adjustwidth}{-1in}{0in}
\centering
  \includegraphics[width=13cm]{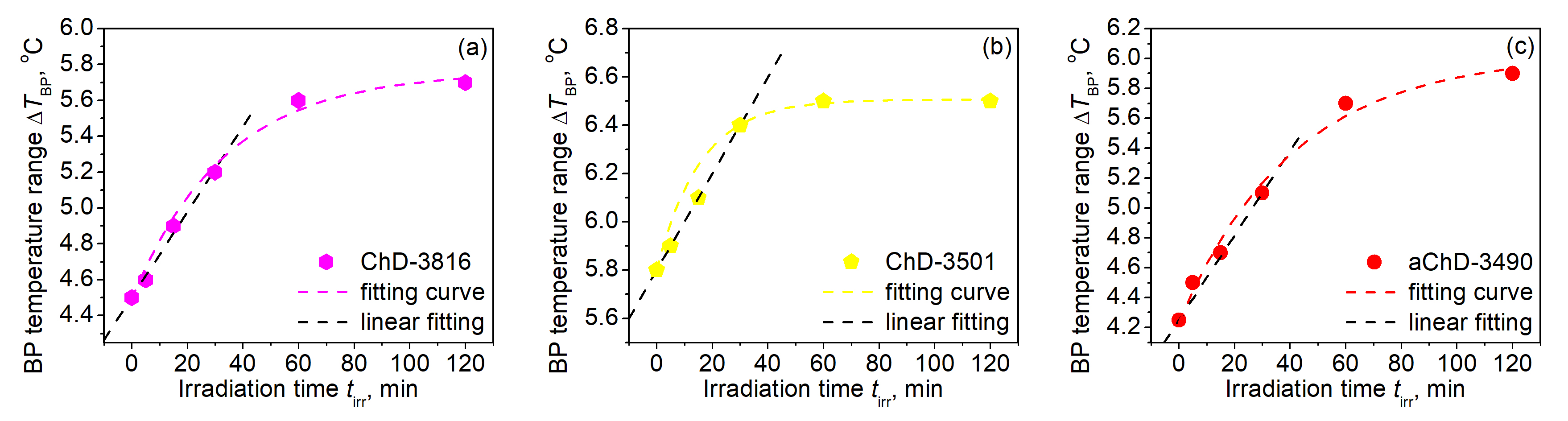}
  \caption{\textbf{Dependence of the BP temperature range on irradiation time (\(t_{\text{irr}}\)) for BPMs containing: (a)~ChD-3816 (solid magenta hexagons), (b)~ChD-3501 (solid yellow pentagons), and (c)~aChD-3490 (solid red circles) \textit{azo} compounds.}}
  \label{fgr:Figure5S1}
  \end{adjustwidth}
\end{figure} 

Notably, this approach provides an estimate of the BP range for various \textit{azo}-doped HChMs after UV exposure within the irradiation time interval of 0--30~min.

\section{Concentration dependence of the BP temperature range: the ``\textit{cinema hall}'' model}

Using the chiral dopant S-811 (Merck, Germany, Darmstadt) as an example, we demonstrate the concentration dependence of the BP temperature range for a BPM based on 95~wt\% of the base HChM doped with 5~wt\% of S-811. We chose S-811 because, first, this chiral dopant lacks an \textit{azo} moiety and, second, it induces a left-handed helix that coincides with the handedness of the base HChM. Analogous to the data described in the main text (Fig.~7(a)), an unequal distribution of the BP temperature range ($\Delta T_{\text{BP}}$) is observed upon varying the S-811 concentration (Fig.~S6.1).
These results are qualitatively similar for various ChDs and coincide with those reported for BPMs containing \ce{BaTiO3} nanoparticles (NPs).\cite{Wang2012,KasianGvozd2025} The observed distribution can be interpreted by the ``\textit{cinema hall}'' model (Fig.~S6.2).

\begin{figure}[ht]
\begin{adjustwidth}{-1in}{0in}
\centering
  \includegraphics[width=7.5cm]{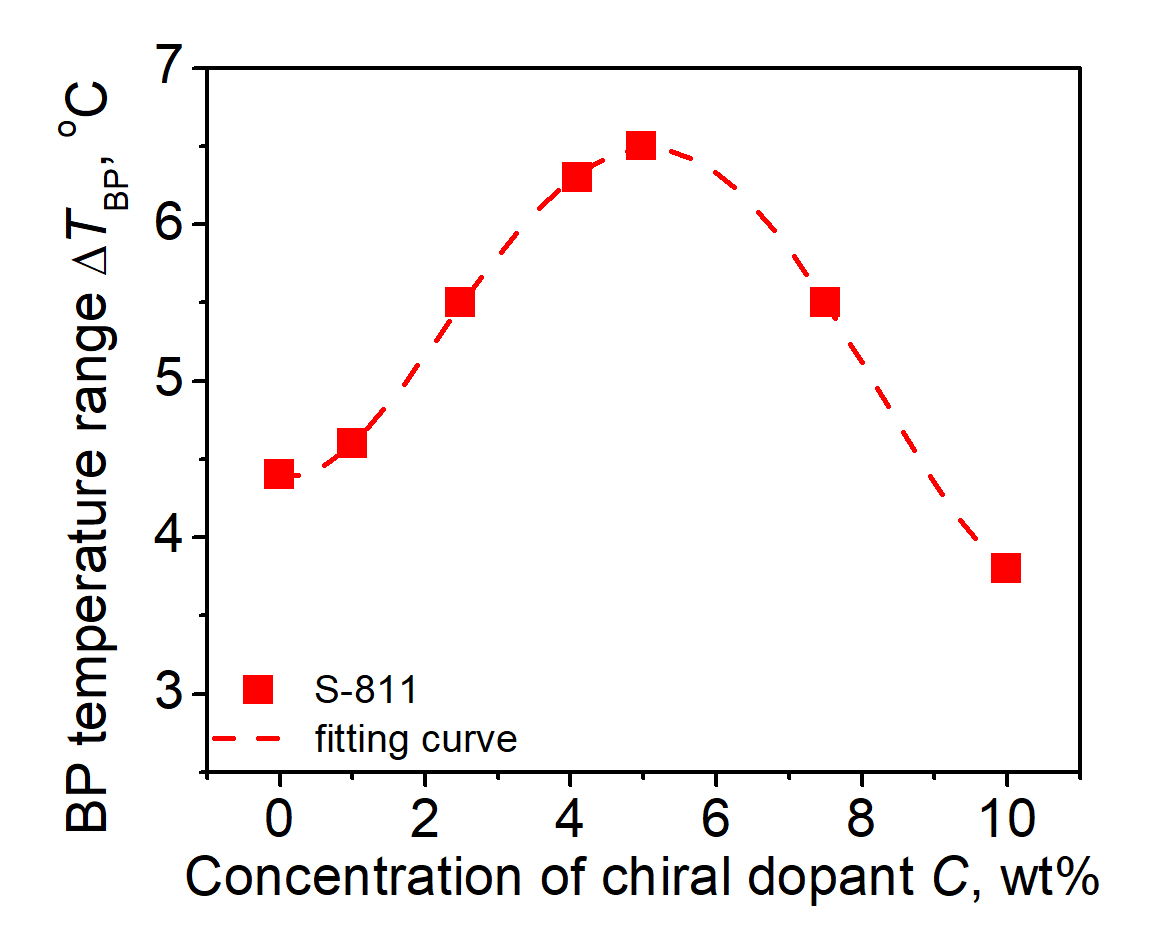}
  \caption{\textbf{Concentration dependence of the BP temperature range $\Delta T_{\text{BP}}$ for the BPM containing left-handed chiral compound S-811.}}
  \label{fgr:Figure6S1}
  \end{adjustwidth}
\end{figure} 

In terms of this model, low concentrations of chiral dopant (ChD) molecules correspond to a situation when all spectators have taken their seats. At a concentration of ChD higher than 5~wt\% the decrease of the BP temperature range $\Delta T_{\text{BP}}$) is observed (Figs.~7(a) and S6.1).

\begin{figure}[!ht]
\begin{adjustwidth}{-1in}{0in}
\centering
  \includegraphics[width=15cm]{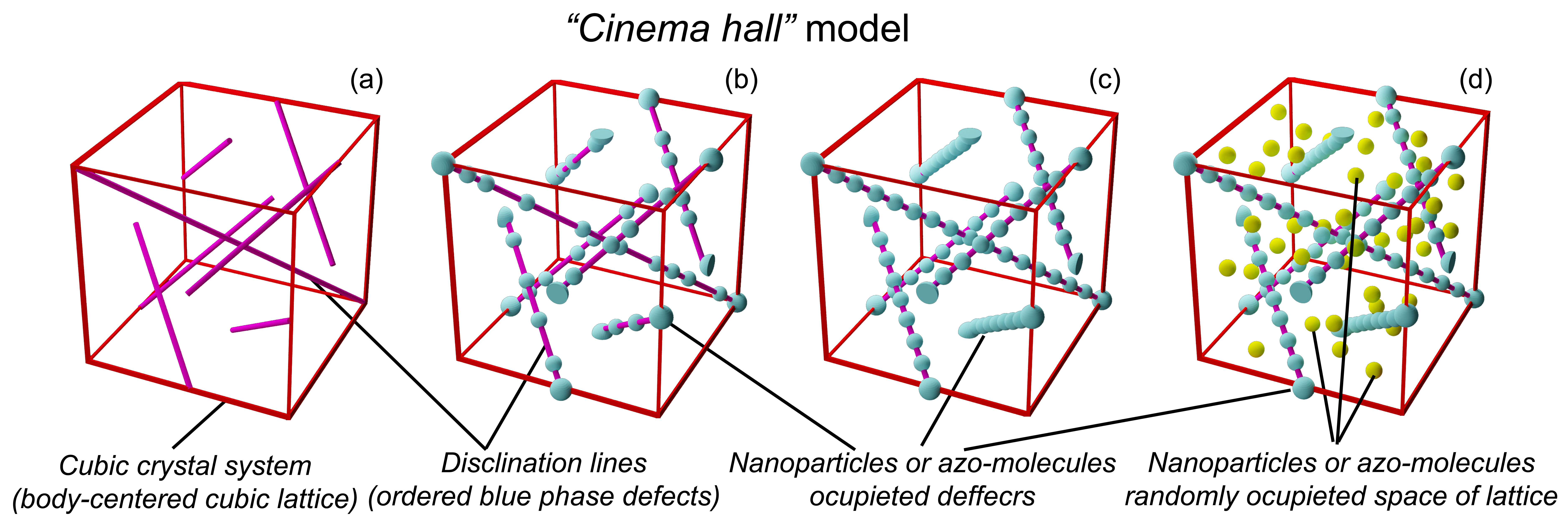}
  \caption{\textbf{Schematic illustration of the ``\textit{cinema hall}'' model explaining the concentration dependence of the BP temperature range for systems doped with NPs or \textit{azo} molecules (body-centered cubic lattice): (a)~base HChM; (b)~low dopant concentration; (c)~threshold concentration with fully occupied defects; and (d)~excessive dopant concentration with molecules (yellow spheres) occupying the free volume of the lattice.}}
  \label{fgr:Figure6S2}
  \end{adjustwidth}
\end{figure} 

Such a dependence can be explained by the fact that the ChD molecules occupy both the disclination lines of the BP lattice and the free volume within the BPM. Within the ``\textit{cinema hall}'' analogy, the disclination lines act as ``\textit{seats}'' that are energetically favorable for the ChD molecules. Once these ``\textit{seats}'' are fully occupied, any additional molecules are forced into the ``\textit{aisles}'' (the free volume of the BPM), which disrupts the BP lattice stability. This leads to a noticeable decrease in $\Delta T_{\text{BP}}$ as observed in Figs.~6S1 and 7(a). 

In contrast to the disclination lines, the occupancy of the free volume introduces elastic distortions that destabilize the blue phase. Within the ``\textit{cinema hall}'' analogy, one can imagine a situation where, during a compelling movie, the hall is not only filled to capacity, but extra spectators also occupy the aisles. These additional viewers randomly obstruct the view, creating discomfort for the audience in their assigned seats. Analogously, ChD molecules residing in the free volume act as these ``\textit{extra spectators},'' inducing elastic distortions and steric hindrance within the BP lattice. This leads to the disruption of the double-twist cylinder arrangement, eventually causing the observed destabilization and narrowing of the BP temperature range at high dopant concentrations.

\end{document}